\documentclass[traditabstract]{aa} 
\usepackage{graphicx}
\usepackage{supertabular}
\usepackage{txfonts}
\begin{document}
\title{The VMC Survey}
\subtitle{II. A multi-wavelength study of LMC planetary nebulae and their mimics\thanks{Based on observations made with VISTA at Paranal Observatory under program ID 179.B-2003 and the Wide Field Imager of the Max-Planck-ESO 2.2m telescope at La Silla Observatory under programs 066.B-0553, 68.C-0019(A) and 076.C-0888.}}

   \author{B. Miszalski
           \inst{1,2,3}
           \and
           R. Napiwotzki
           \inst{1}
           \and
           M. -R. L. Cioni
           \inst{1,4}\thanks{Research Fellow of the Alexander von Humboldt Foundation}
           \and
           M. A. T. Groenewegen\inst{5}
           \and
           J. M. Oliveira\inst{6}
           \and 
           A. Udalski
           \inst{7}
          }

\institute{Centre for Astrophysics Research, STRI, University of Hertfordshire, College Lane Campus, Hatfield AL10 9AB, UK
\and
South African Astronomical Observatory, PO Box 9, Observatory, 7935, South Africa
\and
Southern African Large Telescope Foundation, PO Box 9, Observatory, 7935, South Africa\\
\email{brent@saao.ac.za}
         \and
         University Observatory Munich, Scheinerstrasse 1, D-81679, M\"unchen, Germany
         \and
         Royal Observatory of Belgium, Ringlaan 3, 1180 Ukkel, Belgium
         \and
School of Physical \& Geographical Sciences, Lennard-Jones Laboratories, Keele University, Staffordshire ST5 5BG, UK
         \and
         Warsaw University Observatory, Al. Ujazdowskie 4, PL-00-478, Warsaw, Poland\\
         }
   \date{Received -; accepted -}

   \abstract{
   The VISTA Magellanic Cloud (VMC) survey is assembling a deep, multi-epoch atlas of $YJK_s$ photometry across the Magellanic Clouds. Prior to the VMC survey only the brightest Magellanic Cloud PNe (MCPNe) were accessible at near-infrared (NIR) wavelengths. It is now possible for the first time to assemble the NIR properties of MCPNe and to identify contaminating non-PNe mimics which are best revealed at NIR wavelengths (e.g. HII regions and symbiotic stars). To maintain the unique scientific niche that MCPNe occupy these contaminants must be removed. Here we conduct a VMC-led, multi-wavelength study of 102 objects previously classified as PNe that are located within the first six VMC tiles observed. We present images, photometry, lightcurves, diagnostic colour-colour diagrams and spectral energy distributions used to analyse the entire sample. At least five PNe have newly resolved nebula morphologies, a task previously only possible with the \emph{HST}. A total 45/67 (67\%) of Reid \& Parker (RP) catalogued objects were reclassified as non-PNe, most of which were located in the vicinity of 30 Doradus. This sample included 16 field stars, 5 emission line stars, 19 HII regions, 4 symbiotic star candidates and 1 young stellar object. We discuss possible selection effects responsible for their inclusion in the RP catalogue and the implications for binary central star surveys targeting LMC PNe. A total of five new LMC symbiotic star candidates identified, compared to eight previously known, underlines the important role the VMC survey will have in advancing Magellanic symbiotic star studies.
   }
   \keywords{planetary nebulae: general - stars: AGB and post-AGB - binaries: symbiotic - Magellanic Clouds}
   \maketitle
   \section{Introduction}

   The gaseous shells of Planetary Nebulae (PNe) shine for a brief $\sim$$10^4$ yr after being ejected by low-intermediate mass stars during the asymptotic giant branch (AGB) phase. Most surveys for PNe have so far focused on detecting their strongest [O~III] and H$\alpha$ emission lines and they can be routinely detected in the Milky Way and the Magellanic Clouds (Parker et al. 2006; Shaw 2006; Stanghellini 2009; Parker \& Frew 2011). As these lines are not unique to PNe, catalogues of PN candidates are built and entries often have an associated PN-likelihood informed by available morphological, photometric and spectroscopic properties. With the advent of modern, large-scale, multi-wavelength surveys these candidate lists are being subjected to increasingly greater scrutiny than previously possible. As a result, many previous PNe candidates can now be confidently reclassified as one of many possible other objects or `mimics' (Frew \& Parker 2010). Classification is now becoming more routine and can confidently be performed with a sufficiently large variety of multi-wavelength data, reducing and in some cases even eliminating the dependence on more time-consuming spectroscopy. 

   Establishing a verified population of bona-fide Magellanic Cloud PNe (MCPNe) has added significance over Galactic PNe because of their location at a known distance. The nearest large extragalactic population of PNe are located in the LMC and, unlike the Galactic population, can be assembled to very high completeness thanks to the low reddening and low inclination angle (van der Marel \& Cioni 2001). This facilitates population-wide studies that would otherwise be unfeasible or fundamentally biased with Galactic PNe. Particularly dependent on the completeness and reliability of the PN population is the [O~III] planetary nebula luminosity function (PNLF) which was formulated in the LMC (Jacoby 1980, 1989). Ciardullo (2010) reviewed our current understanding of the PNLF whose bright-end cut-off acts as a standard candle distance indicator to distant galaxies. There are aspects of the PNLF that are poorly understood and studying in detail its constituent PNe for an entire population is a high priority (Ciardullo et al. 2010; Cioni et al. 2011). The ability to reliably identify MCPNe at the intermediate distance of the Magellanic Clouds will also provide essential training before more distant galaxies become similarly accessible with larger telescopes (Parker \& Shaw 2006). MCPNe may also play a critical role in measuring the binary fraction of PNe (Shaw et al. 2007c) to compare against Galactic estimates (Miszalski et al. 2009). If this is to be accomplished any binary non-PN contaminants such as symbiotic stars must be removed first (see e.g. Miszalski et al. 2009). Further applications of MCPNe are reviewed by Shaw (2006) and Stanghellini (2009).

   Multi-wavelength analyses of PNe in the near-infrared (NIR) and mid-infrared (MIR) are particularly useful for assessing the veracity of PN candidates (e.g. Schmeja \& Kimeswenger 2001; Cohen et al. 2007, 2011; Corradi et al. 2008). Hora et al. (2008) commenced a NIR and MIR study of a subset of LMC PNe, but because previous NIR surveys were only deep enough to detect the brightest PNe, their main focus was at MIR wavelengths where the \emph{Spitzer} Surveying the Agents of a Galaxy's Evolution (SAGE, Meixner et al. 2006) survey provides deep enough coverage to detect faint PNe. The sample assessed by Hora et al. (2008) consisted of mostly bright PNe found from a variety of surveys (Leisy et al. 1997) and did not include the significant new discoveries from the H$\alpha$-selected Reid \& Parker (RP) catalogue of 291 `true', 54 `likely' and 115 `possible' PNe (Reid \& Parker 2006a, 2006b; hereafter RP2006a, RP2006b). 
   
   The VISTA Magellanic Cloud (VMC) survey\footnote{http://star.herts.ac.uk/$\sim$mcioni/vmc/} is obtaining deep $YJK_s$ photometry of the Magellanic Clouds and Bridge at sub-arcsecond resolution (Cioni et al. 2011). The VMC observations therefore offer a unique opportunity to refine the catalogued population of MCPNe in conjunction with comparable optical (Zaritsky et al. 2004) and MIR observations (Meixner et al. 2006). Six LMC tiles were observed during the first year of VMC operations and include a total of 102 objects catalogued as true, likely or possible PNe from Leisy et al. (1997), RP2006b and Miszalski et al. (2011b). In this first paper on PNe in the VMC survey we expand upon our first results summarised in Cioni et al. (2011). 

This paper is structured as follows. Sect. \ref{sec:sample} describes our sample selection from the available MCPNe catalogues. Sect. \ref{sec:obs} describes the multi-wavelength observations at our disposal. Sect. \ref{sec:results} presents lightcurves, spectral energy distributions and diagnostic diagrams which summarise our new classifications of the sample. Sect. \ref{sec:indiv} describes the classifications of a subset of the sample in detail and Sect. \ref{sec:discussion} discusses the range of possible selection effects encountered during the study and the impact of our new classifications on searches for binary central stars. We conclude in Sect. \ref{sec:conclusion}.

   \section{Previous reclassifications and sample selection}
   \label{sec:sample}
   As the newest and least studied LMC PNe, the RP sample may be more susceptible to containing a larger fraction of contaminating non-PNe than the Leisy et al. (1997) sample. The extent of contamination is not yet known as only a small fraction of RP objects have been asssesed for non-PNe members (Shaw, Reid \& Parker 2007a; Cohen et al. 2009; Van Loon et al. 2010; Woods et al. 2011; Reid \& Parker 2010, hereafter RP2010). When a non-PN classification is made in these studies it is usually done so without the presentation of images and spectra, making it difficult to independently judge the reliability of the classifications. For this reason we incorporate objects in our sample even if they may have already been classified or reclassified as non-PNe. 

\begin{figure*}
      \begin{center}
         \includegraphics[scale=1.0,angle=270]{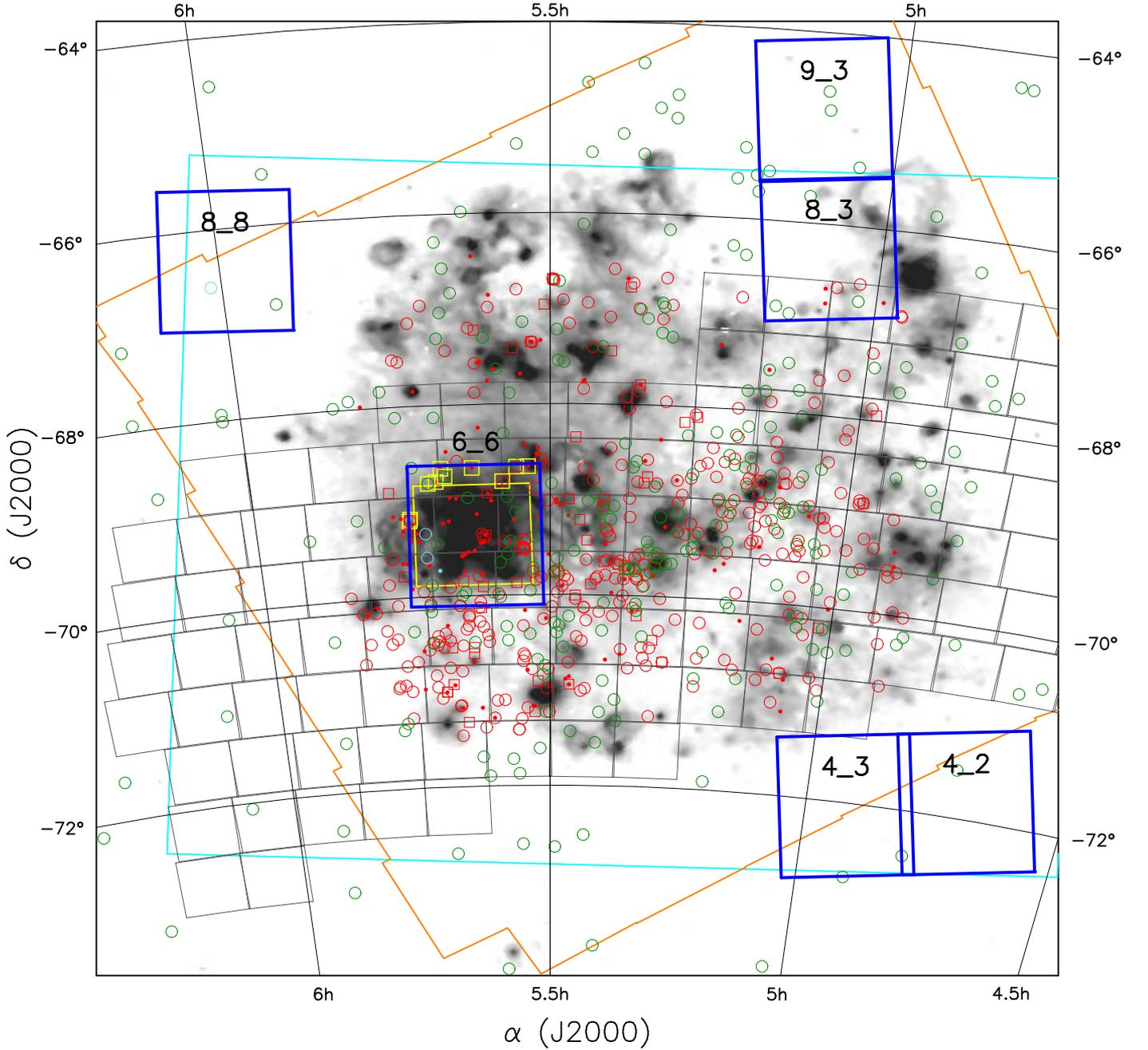}
      \end{center}
      \caption{SHASSA H$\alpha$ mosaic of the LMC (Gaustad et al. 2001) in a lambert azimuthal equal-area projection overlaid with catalogued LMC PNe (Leisy et al. 1997, green symbols; RP2006b, red symbols; Miszalski et al. 2011b, light blue symbols). Different symbols correspond to true (circles), likely (squares) and possible (dots) PNe. VMC tiles observed in the first year of operations are shown in blue (see Cioni et al. 2011). Complementary survey footprints are also shown including $UBVi$ photometry from Zaritsky et al. (2004) in cyan, \emph{Spitzer} MIR photometry from Meixner et al. (2006) in orange (at 8.0 $\mu$m) and $I$-band time-series photometry from Udalski et al. (2008a, 2008b) in grey. Objects with ESO WFI optical observations considered in this work are bounded by yellow boxes (see Sect. \ref{sec:wfi}).
      }
      \label{fig:lmc}
   \end{figure*}

   Figure \ref{fig:lmc} shows the distribution of objects previously catalogued as PNe and the 6 LMC tiles observed during the first year of VMC operations. Table \ref{tab:location} breaks down the distribution of the 102 objects located within the VMC tiles. In total there are 31 objects from Leisy et al. (1997), 67 from RP2006b and 4 (three bona-fide and one possible) from Miszalski et al. (2011b). We have nominally added the emission line candidate LM2-39 (Lindsay 1963) into the Leisy et al. (1997) sample since it was listed by RP2006b. 

\begin{table}
      \centering
      \caption{Tile location of objects in our sample.} 
      \label{tab:location}
      \begin{tabular}{lrrr|r}
         \hline\hline
         Tile & (1) & (2) & (3) & Total\\
         \hline
         4\_2 &     2             & 0                        & 0                       & 2\\
         4\_3 &     0             & 0                        & 0                       & 0\\
         6\_6 &     20            & 62                          & 3                       & 85\\
         8\_3 &     4             & 5                           & 0                       & 9\\
         8\_8 &     1             & 0                           & 1                         & 2\\
         9\_3 &     4              & 0                          & 0                         & 4\\
         \hline
         Total &    31             & 67                      & 4                       & 102 \\
         \hline
      \end{tabular}
      \tablebib{
      (1) Leisy et al. (1997); (2) Reid \& Parker (2006b); (3) Miszalski et al. (2011b).
      }
   \end{table}

   Table \ref{tab:basic1} lists the names and coordinates of our entire sample along with the parent VMC tile number, the presence of ESO WFI observations (Sect. \ref{sec:wfi}), our new classifications (Sect. \ref{sec:sed}) and remarks based on analysis in subsequent sections. The original RP2006b classifications are also given with some slight modifications to incorporate the updated classifications in RP2010. The 67 objects from RP2006b are made up of 21 `true', 9 `likely' and 37 `possible' PNe, respectively. Five objects in our sample, namely RP218, RP256,  RP698, RP833 and RP1923, were omitted from Table A2 of RP2010 without explanation, presumably because they were deemed not to be PNe. Table 1 of RP2010 reclassified RP641 and RP1933 as HII regions which explains their absence from their Table A2. In these cases we have kept the original RP2006b classification in Table \ref{tab:basic1}. RP774 and RP775 were both reclassified in Table A2 of RP2010 from `true' to `possible'. No other classifications of our sample changed between RP2006b and RP2010.

\newpage

\begin{table*}
\centering
\caption{Basic properties and classifications of the sample.}
\label{tab:basic1}
\begin{tabular}{lrlllll|ll}
\hline\hline
Name & RP & Coords & Tile & WFI & ID & Prop. & RP ID & RP Prop.\\
\hline
LM2-39 & 243 & 05 40 14.82 $-$69 28 49.1 & 6\_6 & Y & Sy? & H,R & T & b e Not PN\\
MG12 & 1897 & 05 01 40.22 $-$66 46 46.0 & 8\_3 & N & PN & H,S & T & e H$\alpha$ only\\
MG13 & - & 05 03 03.23 $-$65 23 02.7 & 9\_3 & N & PN & S & - & -\\
MG16 & - & 05 06 05.17 $-$64 48 49.5 & 9\_3 & N & PN & R? & - & -\\
MG17 & - & 05 06 21.17 $-$64 37 03.8 & 9\_3 & N & PN & S & - & -\\
MG18 & - & 05 07 03.07 $-$65 43 22.6 & 8\_3 & N & PN & R? & - & -\\
MG56 & 657 & 05 31 44.90 $-$69 43 07.4 & 6\_6 & N & PN & H,S? & T & e H$\alpha$ only in HII\\
MG60 & 658 & 05 33 30.81 $-$69 08 13.3 & 6\_6 & Y & PN & H,S & T & c H$\alpha$ only\\
MG65 & 659 & 05 35 10.24 $-$69 39 39.1 & 6\_6 & Y & PN & H,S & T & e s H$\alpha$ only\\
MG68 & 267 & 05 38 19.42 $-$68 58 37.4 & 6\_6 & Y & PN & H,R,B,U & T & c H$\alpha$ only\\
MG73 & 214 & 05 41 36.75 $-$69 27 09.4 & 6\_6 & Y & PN & H,R & T & e poss bp\\
MG75 & 271 & 05 42 15.37 $-$68 48 56.7 & 6\_6 & Y$\dagger$ & ND,PN & H,S & T & s H$\alpha$ only\\
MG76 & 215 & 05 42 23.99 $-$69 53 05.3 & 6\_6 & Y & PN & WH,S & T & c s star\\
MG77 & 272 & 05 43 47.57 $-$68 38 35.1 & 6\_6 & N & PN & H,R & T & e H$\alpha$ only\\
MNC1 & - & 05 42 46.80 $-$69 20 30.0 & 6\_6 & Y & PN & H,R & - & -\\
MNC2 & - & 05 42 43.97 $-$69 35 34.0 & 6\_6 & Y & PN & H,R & - & -\\
MNC3 & - & 05 41 28.35 $-$69 43 53.0 & 6\_6 & Y & PN? & H,R & - & -\\
MNC4 & - & 06 00 59.20 $-$66 36 15.3 & 8\_8 & N & PN & H,S & - & -\\
Mo30 & 662 & 05 31 35.20 $-$69 23 46.7 & 6\_6 & N & PN & H,S & T & e poss bp\\
Mo32 & 664 & 05 32 05.35 $-$69 57 26.6 & 6\_6 & N & PN & H,R & T & b c\\
Mo34 & 663 & 05 35 13.91 $-$70 01 19.5 & 6\_6 & N & PN & H,R & T & e bp\\
Mo36 & 179 & 05 38 53.42 $-$69 57 55.1 & 6\_6 & N & PN & H,S & T & e compact H$\alpha$\\
Mo37 & 152 & 05 39 14.35 $-$70 00 19.1 & 6\_6 & N & PN & H,S & T & e f strong H$\alpha$\\
Mo38 & - & 05 40 32.26 $-$68 44 48.0 & 6\_6 & Y$\dagger$ & PN & H,S & - & -\\
Mo39 & 134 & 05 42 41.01 $-$70 05 49.3 & 6\_6 & N & PN & H,R? & T & c H$\alpha$ only\\
Mo42 & - & 05 55 14.60 $-$66 50 25.1 & 8\_8 & N & PN & R & - & -\\
Sa122 & 650 & 05 34 24.25 $-$69 34 28.6 & 6\_6 & Y & PN & H,R & T & c in HII region\\
SMP4 & - & 04 43 21.88 $-$71 30 08.9 & 4\_2 & N & PN & S & - & -\\
SMP6 & - & 04 47 38.97 $-$72 28 20.6 & 4\_2 & N & PN & S & - & -\\
SMP27 & 1894 & 05 07 54.85 $-$66 57 45.5 & 8\_3 & N & PN & H,S,DH,R & T & c halo mainly N\\
SMP30 & 1552 & 05 09 10.54 $-$66 53 38.4 & 8\_3 & N & PN & H,S & T & c bp\\
SMP35 & - & 05 10 49.90 $-$65 29 30.7 & 9\_3 & N & PN & S & - & -\\
SMP77 & 646 & 05 34 06.24 $-$69 26 18.4 & 6\_6 & Y & PN & H,S & T & poss ds Poss PN\\
SMP78 & 647 & 05 34 21.21 $-$68 58 25.2 & 6\_6 & Y & PN & H,S & T & e$\sim$c poss ds\\
SMP82 & 648 & 05 35 57.55 $-$69 58 17.0 & 6\_6 & N & PN & H,S & T & c compact\\
RP135 & 135 & 05 42 37.77 $-$70 04 36.7 & 6\_6 & N & Em? & WH,S,B & T & c s p\\
RP142 & 142 & 05 39 34.86 $-$70 06 13.1 & 6\_6 & N & FD?,NL & H,R,B & T & e H$\alpha$ only\\
RP143 & 143 & 05 39 31.25 $-$70 06 15.4 & 6\_6 & N & ND,NL & H,R,B & T & irreg b\\
RP162 & 162 & 05 43 17.65 $-$69 56 50.5 & 6\_6 & N & PN & H,S & T & c ds s\\
RP163 & 163 & 05 44 28.78 $-$69 54 44.5 & 6\_6 & N & FD,NL & WH,R? & P & c f H$\alpha$ only diffuse\\
RP178 & 178 & 05 40 28.57 $-$69 54 39.0 & 6\_6 & N & NL & WH?,R?,B & T & c p\\
RP180 & 180 & 05 37 00.55 $-$69 54 32.1 & 6\_6 & N & PN & WH,S & L & c ds\\
RP182 & 182 & 05 39 04.89 $-$69 50 48.0 & 6\_6 & Y & ND,DHII & DH,R,U & L & e diffuse H$\alpha$ only\\
RP187 & 187 & 05 42 36.03 $-$69 40 23.9 & 6\_6 & Y & ND,DHII & DH,R,U & P & irreg f diffuse H$\alpha$ only\\
RP188 & 188 & 05 42 32.73 $-$69 40 23.8 & 6\_6 & Y & ND,DHII & DH,R,B,U & P & irreg VLE hidden\\
RP198 & 198 & 05 44 19.12 $-$69 24 41.9 & 6\_6 & N & HII & H,R,B,U & P & VLE c\\
RP202 & 202 & 05 44 19.15 $-$69 12 07.7 & 6\_6 & Y$\dagger$ & FD,PN & H,R,B & T & e H$\alpha$ only\\
RP203 & 203 & 05 44 17.36 $-$69 11 01.5 & 6\_6 & Y$\dagger$ & ND,DHII & DH,R,U & T & e diffuse H$\alpha$ only\\
RP218 & 218 & 05 39 07.21 $-$69 35 14.6 & 6\_6 & Y & HII & H,R,B,U & P & c b in HII region VLE\\
RP219 & 219 & 05 39 02.83 $-$69 35 09.4 & 6\_6 & Y & FS & S & P & c p in HII region\\
RP223 & 223 & 05 38 26.22 $-$69 32 51.6 & 6\_6 & Y & FS & S & P & c in HII region\\
RP227 & 227 & 05 37 46.82 $-$69 31 55.9 & 6\_6 & Y & LPV/Em? & WH,S & P & e in HII region\\
RP228 & 228 & 05 37 06.76 $-$69 27 09.1 & 6\_6 & Y & FS & S & P & c f p\\
RP231 & 231 & 05 36 49.57 $-$69 23 56.1 & 6\_6 & Y & FS & S & L & e f in HII region\\
RP232 & 232 & 05 36 35.13 $-$69 22 28.6 & 6\_6 & Y & ND,DHII & DH,R,U & L & irreg f diffuse\\
\hline
\end{tabular}
\begin{flushleft}
\textbf{ID} PN (?): true (possible) PN; NL: neutral; ND/FD: VMC non-/faint-detection; FS: field star no H$\alpha$ emission; Em: field star with H$\alpha$ emission; (D)HII: (diffuse) HII region; LPV: long period variable; Sy?: candidate symbiotic star.\\
\textbf{Prop.} H/WH/DH: significant/weak/diffuse H$\alpha$ emission; R: resolved nebula; S: stellar/unresolved; B: close to bright star; U: unusual morphology atypical of PNe.\\
\textbf{RP ID.} T/L/P: True/Likely/Possible PN.\\
\textbf{RP Prop.} b: bright; c: circlular; e: elliptical; f: faint; s: small; p: pt. source; irreg: irregular; ds: double star; bp: bipolar. See RP2006b.\\
$\dagger$WFI observations listed in Table \ref{tab:wfiextra}. All others have WFI observations taken from the ESO reduced data described in Miszalski et al. (2011b).\\
\end{flushleft}
\end{table*}
\begin{table*}
\centering
\caption{Basic properties and classifications of the sample (continued).}
\label{tab:basic2}
\begin{tabular}{lrlllll|ll}
\hline\hline
Name & RP & Coords & Tile & WFI & ID & Prop. & RP ID & RP Prop.\\
\hline
RP234 & 234 & 05 36 41.30 $-$69 22 08.9 & 6\_6 & Y & ND,DHII & DH,R,U & P & e p f\\
RP240 & 240 & 05 40 55.50 $-$69 14 10.0 & 6\_6 & Y & FS & S & P & e f\\
RP241 & 241 & 05 40 20.70 $-$69 13 01.5 & 6\_6 & Y & FS & S & P & c s in HII region\\
RP242 & 242 & 05 40 08.65 $-$68 58 26.8 & 6\_6 & Y & LPV/HII & H,R,B,U & P & c b s in HII\\
RP246 & 246 & 05 38 57.25 $-$69 33 57.1 & 6\_6 & Y & FS & S & P & c in HII region\\
RP247 & 247 & 05 38 48.22 $-$69 34 07.8 & 6\_6 & Y & HII & DH,S & P & c s in HII region\\
RP250 & 250 & 05 44 24.15 $-$69 16 42.1 & 6\_6 & N & HII & DH,R,B,U & P & c ds symb or PN + star\\
RP251 & 251 & 05 44 15.74 $-$69 17 22.7 & 6\_6 & N & HII & S & P & c b s halo VLE\\
RP254 & 254 & 05 43 37.77 $-$69 20 10.3 & 6\_6 & Y & ND,DHII & DH,R,B,U & P & c p diffuse\\
RP256 & 256 & 05 38 51.40 $-$69 44 51.0 & 6\_6 & Y & HII & H,R,B,U & P & c f halo VLE\\
RP259 & 259 & 05 36 48.60 $-$69 26 44.9 & 6\_6 & Y & FS & S & P & c s\\
RP264 & 264 & 05 43 30.35 $-$69 24 46.6 & 6\_6 & Y & Sy? & H,R & P & e s f H$\alpha$ only VLE\\
RP265 & 265 & 05 37 00.70 $-$69 21 29.5 & 6\_6 & Y & FD,PN & H,R,B & T & e p\\
RP266 & 266 & 05 37 27.85 $-$69 08 55.6 & 6\_6 & Y & HII & H,R,B,U & P & c in HII region\\
RP268 & 268 & 05 39 30.10 $-$68 58 57.7 & 6\_6 & Y & FS & S & P & c ds in HII region\\
RP277 & 277 & 05 41 26.74 $-$68 48 00.9 & 6\_6 & Y$\dagger$ & Em & WH,S & P & c f p\\
RP283 & 283 & 05 37 48.27 $-$68 39 54.5 & 6\_6 & Y$\dagger$ & Em & WH,S & P & c s\\
RP312 & 312 & 05 36 19.97 $-$68 55 37.9 & 6\_6 & Y & FS & S & L & c b in HII region\\
RP315 & 315 & 05 36 13.27 $-$68 56 19.5 & 6\_6 & Y & LPV & S & P & c in HII region\\
RP641 & 641 & 05 37 06.39 $-$69 47 17.9 & 6\_6 & Y & HII & H,R,B,U & P & e VLE p\\
RP691 & 691 & 05 35 25.96 $-$69 59 22.2 & 6\_6 & N & NL & S & T & poss c hidden\\
RP698 & 698 & 05 33 30.52 $-$69 52 27.0 & 6\_6 & Y & HII & H,R,B,U & L & poss ds symbiotic-\\
RP700 & 700 & 05 31 29.50 $-$69 50 42.4 & 6\_6 & N & ND,NL & WH,B & T & e diffuse hidden\\
RP701 & 701 & 05 30 57.41 $-$69 49 00.7 & 6\_6 & N & FD,NL & H,R,B,U & T & e 3 stars H$\alpha$ only\\
RP748 & 748 & 05 31 47.13 $-$69 45 44.4 & 6\_6 & N & NL & WH,R & T & e H$\alpha$ only\\
RP774 & 774 & 05 32 39.69 $-$69 30 49.5 & 6\_6 & Y & Sy? & H,S & P & c s b some HII\\
RP775 & 775 & 05 32 44.26 $-$69 30 05.9 & 6\_6 & Y & HII & DH,R,B,U & P & c b half hidden in HII region\\
RP776 & 776 & 05 32 39.24 $-$69 31 53.9 & 6\_6 & Y & Sy? & H,S & T & c s p in HII region\\
RP789 & 789 & 05 32 35.58 $-$69 25 42.1 & 6\_6 & Y & PN & H,R & T & c s H$\alpha$ only\\
RP790 & 790 & 05 32 33.65 $-$69 24 55.6 & 6\_6 & Y & FS & S & L & c s\\
RP791 & 791 & 05 33 07.00 $-$69 29 45.9 & 6\_6 & Y & Em & WH,S & L & c s f weak H$\alpha$\\
RP793 & 793 & 05 34 41.40 $-$69 26 30.7 & 6\_6 & Y & LPV/Mira & H,S & T & p H$\alpha$ only\\
RP828 & 828 & 05 33 40.29 $-$69 12 51.2 & 6\_6 & Y & FS & S & P & c b some HII\\
RP833 & 833 & 05 31 05.78 $-$69 10 41.5 & 6\_6 & N & YSO & WH,S,B & P & p VLE\\
RP883 & 883 & 05 35 56.89 $-$69 00 44.9 & 6\_6 & Y & LPV/Sy? & H,S & P & e p b\\
RP896 & 896 & 05 31 34.35 $-$68 52 45.8 & 6\_6 & N & PN & H,R & T & c b H$\alpha$ only\\
RP907 & 907 & 05 34 48.03 $-$68 48 35.6 & 6\_6 & Y$\dagger$ & PN & H,S & T & c b p in HII region\\
RP908 & 908 & 05 33 23.20 $-$68 39 34.1 & 6\_6 & Y$\dagger$ & Em & WH,S & L & c p s stars N and S\\
RP913 & 913 & 05 32 12.47 $-$68 39 24.6 & 6\_6 & Y$\dagger$ & Em & WH,S & P & c VLE s in HII region\\
RP1018 & 1018 & 05 40 55.06 $-$68 39 54.0 & 6\_6 & Y$\dagger$ & FS & S & P & p s\\
RP1037 & 1037 & 05 37 25.05 $-$69 48 00.1 & 6\_6 & Y & PN & H,R & T & c s H$\alpha$ only\\
RP1040 & 1040 & 05 37 21.05 $-$70 04 08.4 & 6\_6 & N & ND,NL & WH? & T & e f H$\alpha$ only\\
RP1923 & 1923 & 05 04 40.45 $-$66 49 49.2 & 8\_3 & N & FS & S & P & c fading halo\\
RP1930 & 1930 & 04 59 20.54 $-$66 45 59.7 & 8\_3 & N & FD?,NL & WH? & P & e f s H$\alpha$ only\\
RP1933 & 1933 & 05 04 47.17 $-$66 40 30.8 & 8\_3 & N & HII & H,R,B,U & P & e b large\\
RP1934 & 1934 & 05 03 45.66 $-$66 39 17.1 & 8\_3 & N & PN & WH,S? & T & e H$\alpha$ only bp\\
RP1938 & 1938 & 05 01 42.33 $-$66 35 56.9 & 8\_3 & N & PN & WH,R & T & c s f H$\alpha$ only\\
\hline
\end{tabular}
\end{table*}

   The largest contribution to our sample comes from the 6\_6 tile which covers the 30 Doradus star-forming region. Objects in this tile are located within a highly-variable emission-line background that is subject to high levels of interstellar absorption. Identifying emission-line objects in this region is naturally more difficult and it would not be unexpected if a larger fraction of 6\_6 objects were not bona-fide PNe. This adds significant caveats to the results of our study, however the 6\_6 objects also constitute a challenging and thorough training set with which to establish robust multi-wavelength classification criteria that may be applied to other regions as the VMC survey progresses further.

   \section{Observations}
   \label{sec:obs}
   We adopt a multi-wavelength approach to identify non-PNe amongst our sample of 102 objects. A large fraction of the LMC tiles have $UBVi$ photometry from the Magellanic Cloud Photometric Survey (MCPS, Zaritsky et al. 2004) and MIR \emph{Spitzer} IRAC and MIPS photometry from SAGE (Meixner et al. 2006). We also identify variable sources using $I$-band time-series photometry from the OGLE-III microlensing survey (Udalski et al. 2008a, 2008b). Miszalski et al. (2009) first demonstrated the power of this approach for removing strongly variable non-PNe such as symbiotic stars towards the Galactic Bulge. The respective survey footprints cover $\sim$65\% (MCPS), $\sim$50\% (SAGE) and $\sim$30\% (OGLE-III) of the total VMC tiled area (see Cioni et al. 2011). The region surveyed by RP2006a has complete coverage in all surveys except OGLE-III where the N-E corner is not covered.

   Additional observations may also be found in the literature. There is \emph{HST} coverage of RP218, RP232, RP265 and RP268, though only RP265 had a nebula detection (Shaw et al. 2007a). A larger number of PNe from the Leisy et al. (1997) sample have been imaged and resolved with the \emph{HST} that leaves little doubt to their PN nature. These include SMP4, SMP6, MG12, MG13, MG16, SMP27, SMP30, MG60, SMP78, SMP82 and Mo36 (Shaw et al. 2001, 2006; Stanghellini et al. 2002, 2003). Leisy \& Dennefeld (2006) published emission line intensities and chemical abundances for MG17, Mo42, Sa122 and all eight SMP objects in our sample.
   
   The following subsections describe the data used to perform the multi-wavelength analysis. Appendix \ref{sec:phot} describes the derivation of VMC and SAGE magnitudes and tabulates all the photometry. Note that magnitudes may include stellar or nebula contributions, or a combination of both.
   Appendix \ref{sec:images} presents multi-wavelength images of the sample. Objects were identified using finder charts from Leisy et al. (1997) and RP2006b (kindly provided by W. Reid). Note that the large amount of data employed here will not necessarily be available for all other LMC PNe, especially in the outer most tiles. A key aim in our work is to characterise those PNe with the maximal amount of available data, such that objects with only VMC data, or some combination of MCPS, VMC or SAGE data, may be characterised based on that data alone with confidence.

 \subsection{ESO WFI 30 Doradus imaging}
   \label{sec:wfi}
   In any study of PNe it helps to have narrow-band imaging with sufficient depth and resolution to detect and resolve the morphologies of faint nebulae. The RP2006a survey data are adequate for identification, but do not resolve sufficient morphological detail that can assist in the classification process. Fortunately a high proportion of our sample located in the 6\_6 tile (49/85 objects) have deep $B$, $V$, [O~III] and H$\alpha$ imaging taken with the Wide Field Imager (WFI) of the ESO 2.2-m telescope under program ID 076.C-0888.\footnote{http://archive.eso.org/archive/adp/ADP/30\_Doradus} We refer the reader to Miszalski et al. (2011b) for full details of the data products which cover an area of $63\times63$ arcmin$^2$ centred near 30 Doradus ($\alpha_\mathrm{J2000}=05^\mathrm{h}37^\mathrm{m}54.7^\mathrm{s}$, $\delta_\mathrm{J2000}=-69^\circ21'55''$; see Fig. \ref{fig:lmc}). 
   
   A search of the ESO archive found similar but unreduced WFI observations for a further 10 objects obtained under the ESO programs 066.B-0553 and 68.C-0019(A). Table \ref{tab:wfiextra} lists the exposures taken with the Halpha/7, OIII/8, MB (medium-band) 604/21 and MB 485/31 filters whose central wavelengths/FWHMs are 658.8/7.4 nm, 502.4/8.0 nm, 604.3/21.0 nm and 485.8/31.4 nm, respectively. An approximate WCS was applied to the raw frames to allow for raw images to be extracted and stacked when appropriate. No other reduction steps were applied to the data.

  \begin{table}
      \centering
      \caption{Additional ESO WFI archival data.}
      \label{tab:wfiextra}
      \begin{tabular}{clrrcc}
         \hline\hline
         Name & Filter & Exp. & Observed&  PID$\dagger$ & FWHM\\
              &        & (s)      & (YY-MM-DD) & & (\arcsec) \\
              \hline
MG75 & Halpha/7 & 1200 & 01-12-07 & (2)	& 0.90    \\
-- & MB 485/31 & 1200 & 01-12-08 & (2)	& 1.09    \\
-- & OIII/8 & 1200 & 01-12-08 & (2)	&   1.04  \\
Mo38 & Halpha/7 & 1200 & 01-12-07 & (2)	&   1.00  \\
-- & MB 485/31 & 1200 & 01-12-08 & (2)	&   1.12  \\
-- & OIII/8 & 1200 & 01-12-08 & (2)	&   1.08  \\
RP202 & Halpha/7 & 1200 & 01-12-07 & (2)	&  0.96   \\
-- & MB 485/31 & 1200 & 01-12-08 & (2)	&   1.17  \\
-- & OIII/8 & 1200 & 01-12-08 & (2)	&   1.07  \\
RP203 & Halpha/7 & 1200 & 01-12-07 & (2)	&  1.11   \\
-- & MB 485/31 & 1200 & 01-12-08 & (2)	&   0.99  \\
-- & OIII/8 & 1200 & 01-12-08 & (2)	&   1.05  \\
RP277 & Halpha/7 & 1200 & 01-12-07 & (2)	& 0.89    \\
-- & MB 485/31 & 1200 & 01-12-08 & (2)	&  1.04   \\
-- & OIII/8 & 1200 & 01-12-08 & (2)	&   1.07  \\
RP283 & Halpha/7 & 360 & 00-10-18 & (1)	& 1.52    \\
-- & MB 604/21   & 300 & 00-10-18 & (1)	&  1.89   \\
-- & OIII/8 & 480 & 00-10-18 & (1)	&  1.64   \\
RP907 & Halpha/7 & 360 & 00-10-18 & (1)	& 1.59    \\
-- & MB 604/21   & 300 & 00-10-18 & (1)	&  1.94   \\
-- & OIII/8 & 480 & 00-10-18 & (1)	&   1.61  \\
RP908 & Halpha/7 & 360 & 00-10-18 & (1)	&  1.57   \\
-- & MB 604/21   & 300 & 00-10-18 & (1)	& 2.00    \\
-- & OIII/8 & 480 & 00-10-18 & (1)	&  1.75   \\
RP913 & Halpha/7 & 360 & 00-10-18 & (1)	& 1.84    \\
-- & MB 604/21   & 300 & 00-10-18 & (1)	& 1.91    \\
-- & OIII/8 & 480 & 00-10-18 & (1)	&  1.64   \\
RP1018 & Halpha/7 & 1200 & 01-12-07 & (2)	&  0.98   \\
-- & MB 485/31 & 1200 & 01-12-08 & (2)	&   0.99  \\
-- & OIII/8 & 1200 & 01-12-08 & (2)	&   1.07  \\
              \hline
      \end{tabular}
      \tablefoot{
      $\dagger$ESO Program IDs: (1) 066.B-0553; (2) 68.C-0019(A).
      }
   \end{table}

   \subsection{Magellanic Cloud Photometric Survey}
   \label{sec:optical}
   Zaritsky et al. (2004) produced a catalogue of deep Johnson $UBV$ and Gunn $i$ photometry for a large fraction of the LMC (see Fig. \ref{fig:lmc}). To extract the photometry for each object we overlaid the catalogue photometry on the VMC colour-composite image and selected entries corresponding to the VMC object positions. Table \ref{tab:opmags1} lists the extracted magnitudes where the `Status' column indicates no survey coverage (NC) or a non-detection (ND) which occurs for objects too faint or diffuse. As the images are not available it is difficult to judge whether the magnitudes include the central star (CSPN), the nebula, or both. A small number of PNe have photometry that is likely to originate from the hot central star as judged by the faintness of the nebula and the very blue $U-B$ colour (Mo42, MG12, MG13 and MG77), provided there is no strong [O~II] contribution to $U$. Most of the fainter PNe are expected to have evolved CSPN beyond the survey detection limits of $V\sim20$ mag (e.g. Villaver, Stanghellini \& Shaw 2007). In brighter PNe the stronger nebula contributions occur in $B$ and $V$ which have the effect of producing redder $U-B$ and $B-V$ colours than expected for isolated CSPN.

   \subsection{VMC}
   \label{sec:vmc}
   We refer the reader to Cioni et al. (2011) for full details of the VMC data products and their reduction procedures. The strongest emission lines in the NIR for PNe include He I 1.083 $\mu$m in $Y$, Pa $\beta$ in $J$, while $K_s$ contains Br $\gamma$, multiple He I and molecular H$_2$ lines (Hora et al. 1999; Rudy et al. 2001). In the six tiles we have complete $YJ$ observations and a number of additional $K_s$ epochs whose completion status varies between tiles (see Tab. 4 of Cioni et al. 2011). These $K_s$ epochs increase the $K_s$ depth and provide $K_s$ lightcurves over a period of at least 100 days. In the respective tiles we have used data observed up to 31 May 2010 that includes $N$ additional $K_s$ epochs (denoted $TKN$) as follows: $TK5$ (4\_3 and 4\_2), $TK6$ (9\_3), $TK7$ (8\_3) and $TK10$ (6\_6 and 8\_8). Average 5$\sigma$ depths for a single tile are $Y=21.11$ mag, $J=20.53$ mag and $K_s=19.22$ mag (Tab. 7 of Cioni et al. 2011). Artificial star tests on the stacked observations for all tiles give the 5$\sigma$ depth as $Y=22.44$ mag, $J=22.16$ mag and $K_s=21.15$ mag with a completeness level of $\sim$57\% (Rubele et al. in preparation). Small variations in the depth will occur locally depending on the position of objects within a tile and crowding will also affect the completeness. The calculation of VMC magnitudes is described in Appendix \ref{sec:phot}. 

   \subsection{SAGE}
   \label{sec:sage}
   In the Milky Way \emph{Spitzer} IRAC photometry and images are a valuable tool in the removal of non-PNe (Cohen et al. 2007, 2011) and are well suited to the detection of obscured RGB and AGB stars. No similar LMC studies have been published, however Hora et al. (2008) gave IRAC and MIPS photometry for non-RP LMC PNe measured from custom-reduced mosaics that incorporated observations from both epochs of the SAGE survey (Meixner et al. 2006). Here we have largely performed our own aperture photometry following the \emph{Spitzer} science center IRAC and MIPS instrument handbooks with the appropriate IRAC aperture corrections applied. This was necessary because the default catalogues are not optimised for extended sources, though in some cases we adopted catalogue magnitudes for the brightest objects. Brighter objects in our sample overlap with Hora et al. (2008) which serves as an independent check of our procedure and fills in some gaps in their photometry where detections in some bands were sometimes absent. Data products used were from the SAGE data release 3 including IRAC 0.6\arcsec/pixel mosaics (version 2.1), E12 MIPS24 mosaics and less often the \verb|SAGELMCcatalogIRAC| and \verb|SAGELMCcatalogMIPS24| catalogues. Image cutouts were extracted for each object and averaged when objects were included in more than one mosaic sub-frame. Only SMP4 and SMP6 were located outside the SAGE survey (Fig. \ref{fig:lmc}). The calculation of SAGE magnitudes is described in Appendix \ref{sec:phot}.

   \subsection{OGLE-III $I$-band and VMC $K_s$ lightcurves}
   \label{sec:lcs}
   Large-scale photometric monitoring surveys offer a powerful means to identify symbiotic stars which are commonly mistaken for PNe (Miszalski et al. 2009). Their high level of variability makes them particularly conspicuous in IR and NIR lightcurves (Miko{\l}ajewska 2001; Gromadzki et al. 2009). 
   We employed a similar procedure to that described by Miszalski et al. (2009) to extract OGLE-III $I$-band lightcurves for our sample that had survey coverage (Udalski et al. 2008a, 2008b; see Fig. \ref{fig:lmc}) and to search for periodic variability. Objects located outside of the OGLE-III survey footprint are MG13, MG16, MG17, MG18, Mo42, SMP4, SMP6, SMP35, RP1923 and RP1933. 
   VMC $K_s$ lightcurves are more suited to probing variability of obscured symbiotic systems which contain RGB or AGB stars interacting with a white dwarf companion. In some symbiotic stars strong intrinsic reddening created by dust means that variability may only be seen in the NIR out of the reach of OGLE-III (e.g. Miko{\l}ajewska et al. 1999). The sampling of the VMC $K_s$ lightcurves is not suitable for measuring periods, but variability with amplitudes larger than expected for a star of a given magnitude can be detected thanks to the scheme developed by Cross et al. (2009) that is implemented in the VSA archive.
   
   \section{Results}
   \label{sec:results}
   \subsection{Photometric variability}
   \label{sec:variability}
\begin{figure}
      \begin{center}
         \includegraphics[scale=0.60,angle=270]{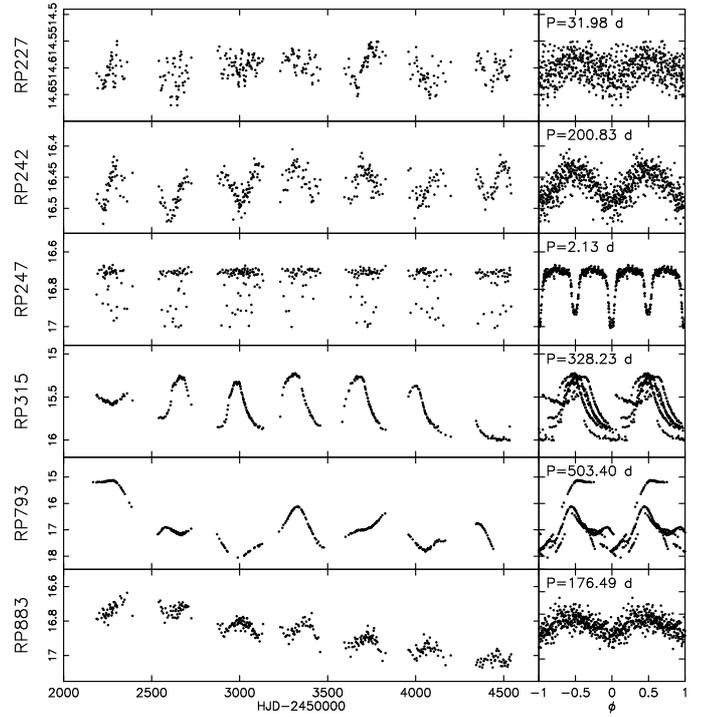}
      \end{center}
      \caption{OGLE-III $I$-band lightcurves of periodic variables in our sample.}
      \label{fig:lpv}
   \end{figure}

   Photometric variability, in combination with corroborating imaging and photometry, is a particularly strong constraint in the classification of variable stars in our sample. Overall we found six periodic variables from OGLE-III (Fig. \ref{fig:lpv}). RP227 and RP793 were previously catalogued by Soszy{\'n}ski et al. (2009) as an OGLE small amplitude red giant and  a Mira, respectively. \emph{None of the six periodic variables are bona-fide central stars of PNe} (see Tab. \ref{tab:basic1} and Sect. \ref{sec:indiv}). Only the HII regions RP242 and RP247, and the symbiotic star candidate RP883, are associated with H$\alpha$ emission. Section \ref{sec:binary} will discuss the influence of these results upon the search for binary central stars in the LMC. 

\begin{figure*}
      \begin{center}
         \includegraphics[scale=0.325,angle=270]{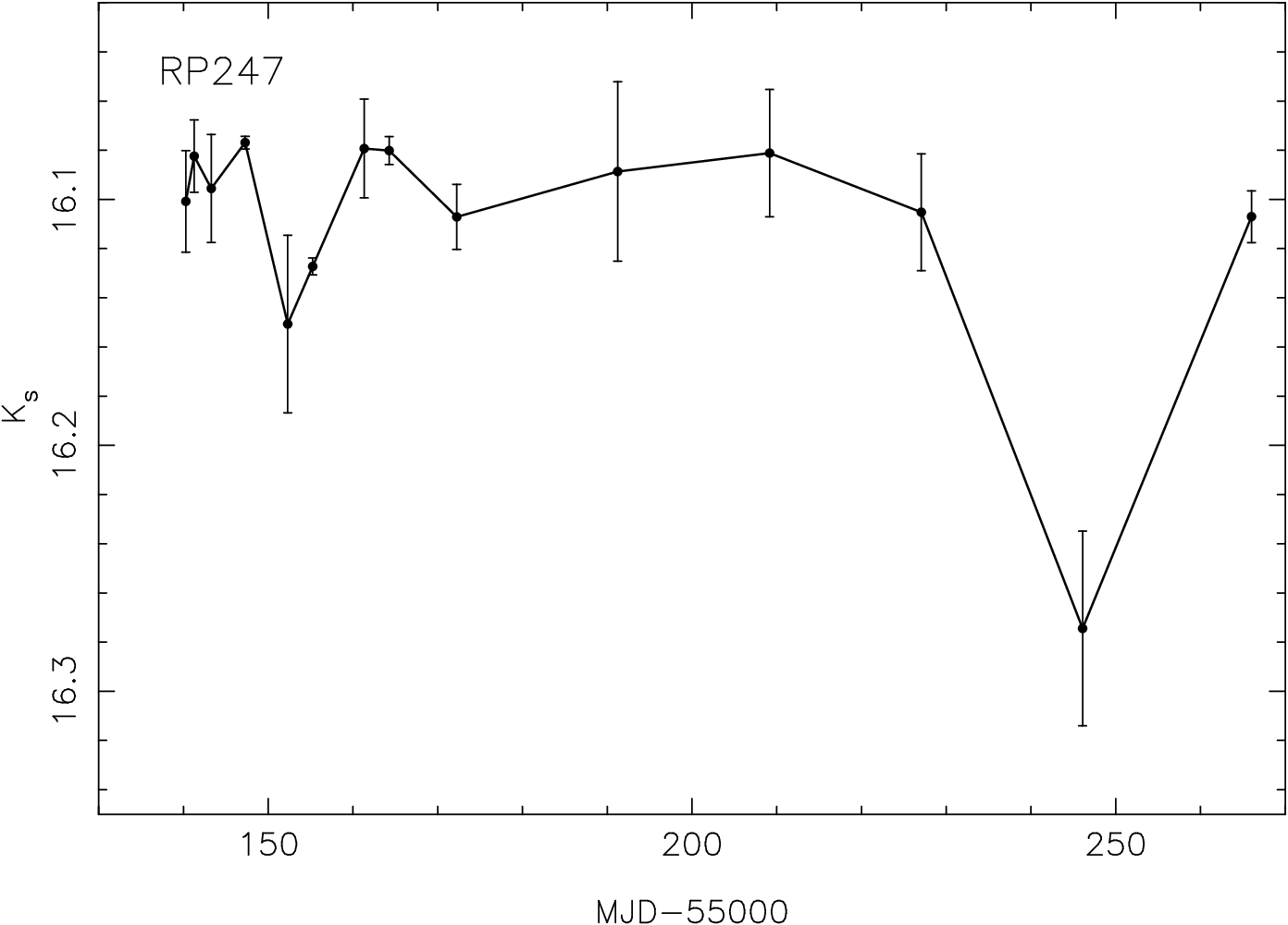}
         \includegraphics[scale=0.325,angle=270]{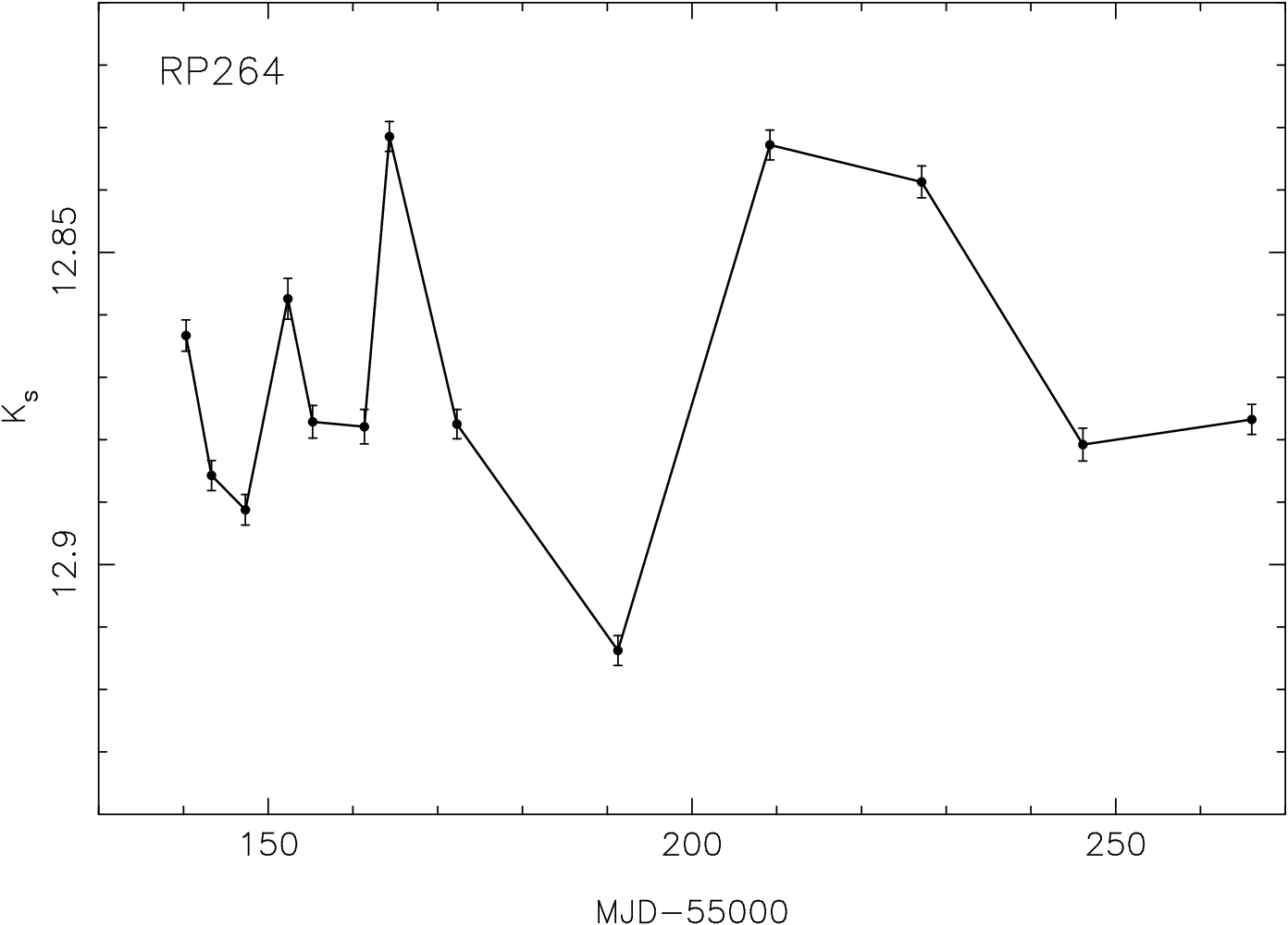}
         \includegraphics[scale=0.325,angle=270]{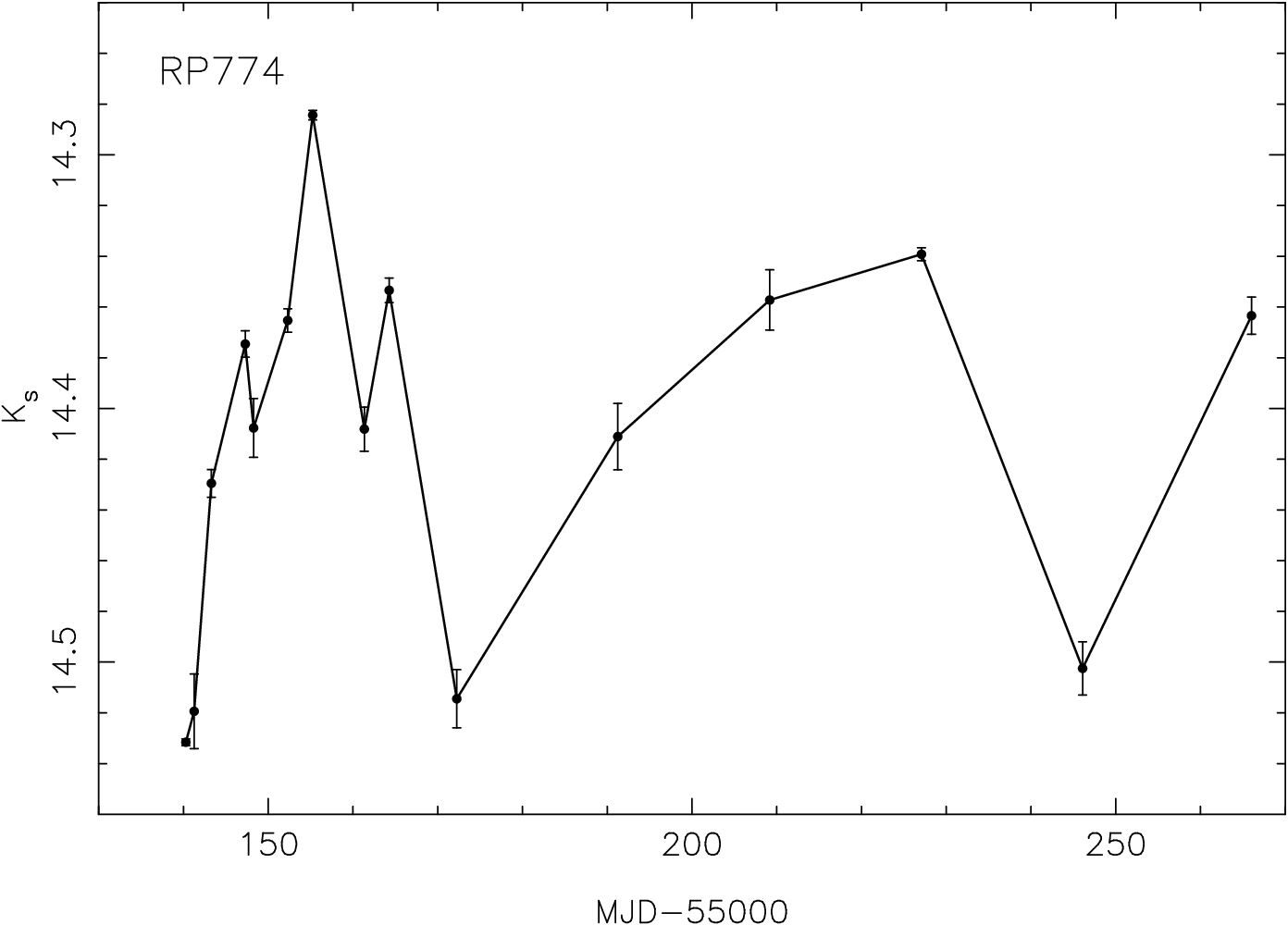}
         \includegraphics[scale=0.325,angle=270]{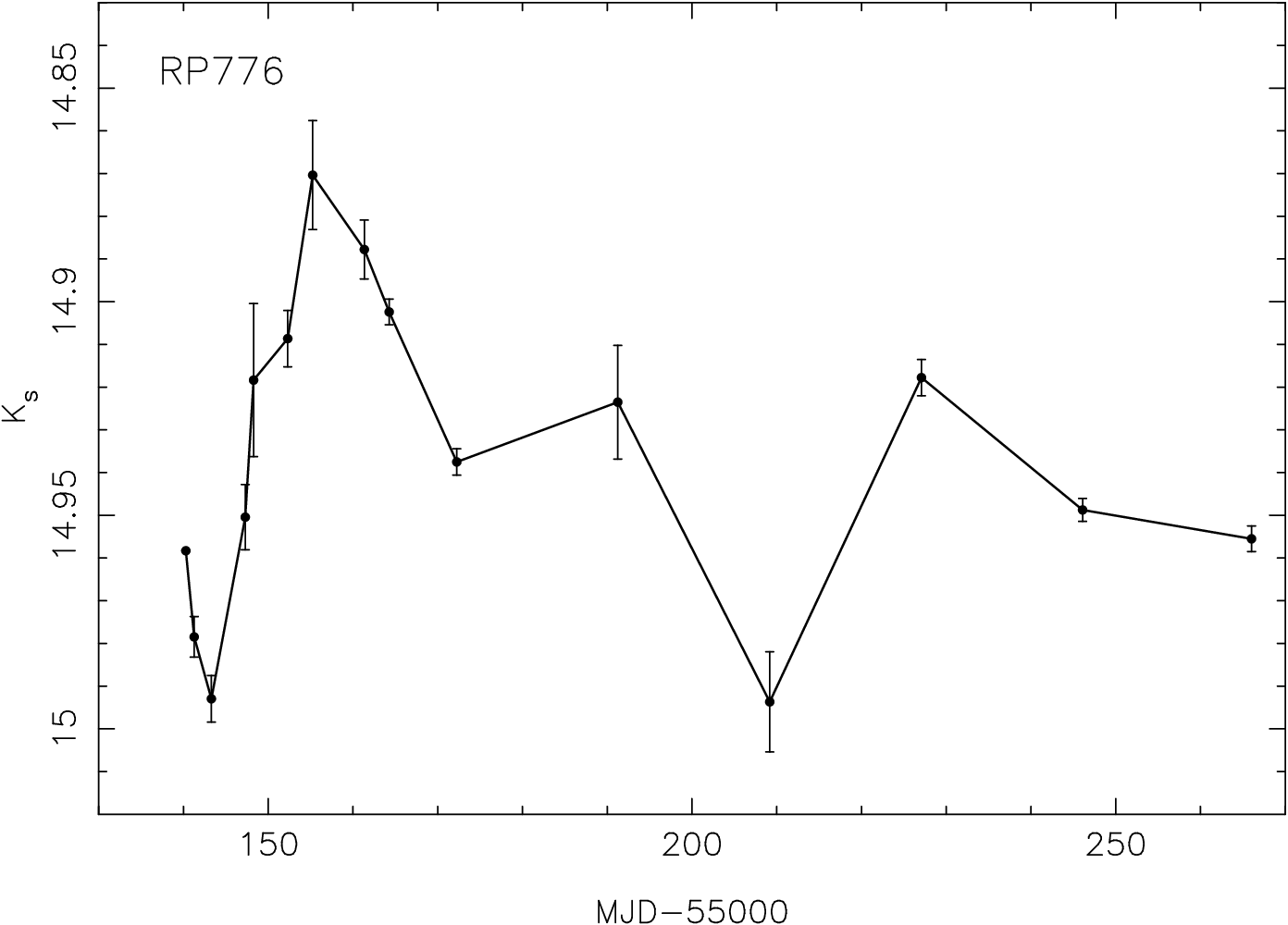}
         \includegraphics[scale=0.325,angle=270]{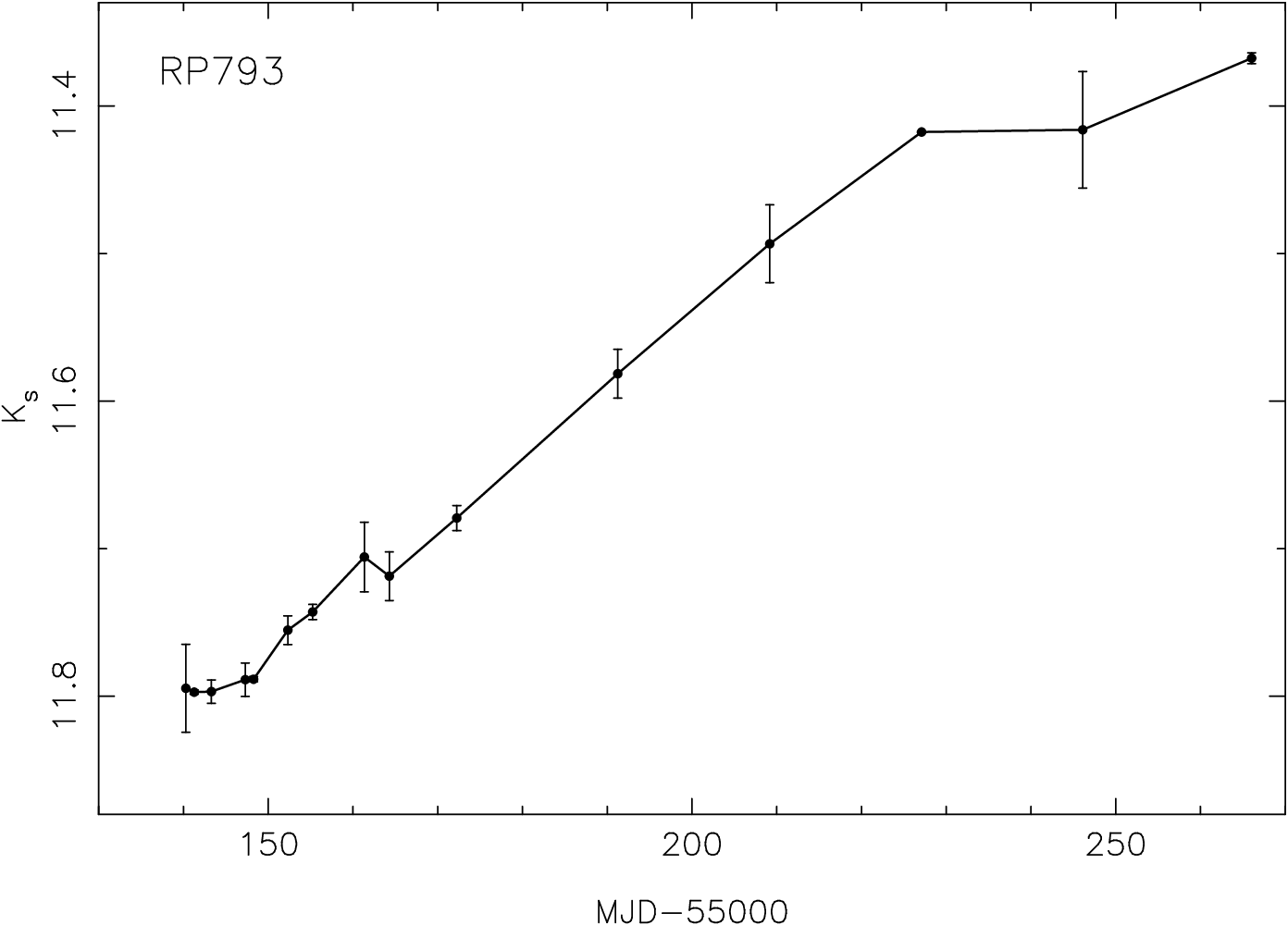}
         \includegraphics[scale=0.325,angle=270]{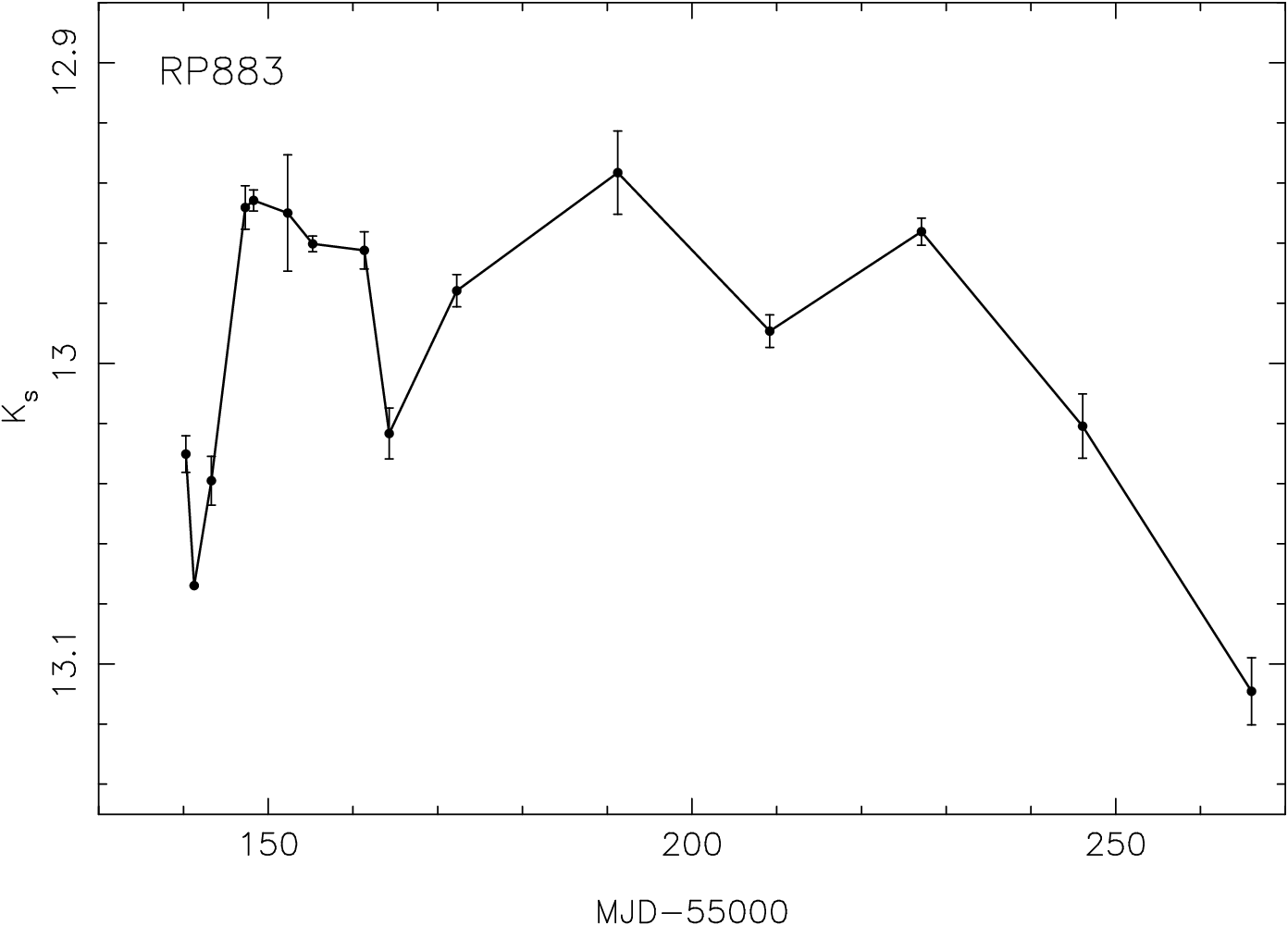}
      \end{center}
      \caption{VMC $K_s$ lightcurves of the variables in Tab. \ref{tab:varstat}. Observations taken on the same night were combined using a weighted mean.}
      \label{fig:ks}
   \end{figure*}

   Table \ref{tab:varstat} lists six variables found to have variable $K_s$ lightcurves which are shown in Fig. \ref{fig:ks}. 
   We have averaged the nightly data using a weighted mean and used the standard deviation of magnitudes on each night as the error.
   Each object has 11--31 good observations for which variability statistics were calculated (Cross et al. 2009). Each median $K_s$ magnitude, $\tilde{K_s}$, has an associated level of expected rms variability for a non-variable point source, $\sigma_\mathrm{exp}$, based on a noise model fitted to the VMC survey data. A measure of the intrinsic variability of the source, $\sigma_\mathrm{int}$, is also calulated and used to form the ratio $\sigma_\mathrm{int}/\sigma_\mathrm{exp}$ which is a measure of the standard deviation above the noise for a given magnitude. A chi-squared statistic is also calculated which is used to determine the probability that a source is variable, $p$, and all objects in Tab. \ref{tab:varstat} have $p=100$\%. Objects are considered to be variable if $p>96$\% and $\sigma_\mathrm{int}/\sigma_\mathrm{exp} \ge 3$. 
   The actual cutoff used to flag variables from $\sigma_\mathrm{int}/\sigma_\mathrm{exp}$ may be slightly lower or higher depending on how the noise model is built from the data and this strategy will be refined as the VMC survey progresses (N. Cross, private communication). For this reason we have also included two objects with $\sigma_\mathrm{int}/\sigma_\mathrm{exp}\ge 2.5$ as variables, both of which are OGLE-III variables. Some other objects of interest may also be variable but have lower, less significant $\sigma_\mathrm{int}/\sigma_\mathrm{exp}$ values of 2.08 (LM2-39 and Mo42) and 2.24 (MNC4). As expected the largest $K_s$ amplitude of 0.4 mag belongs to the Mira RP793 (note there is no evidence to say it is a symbiotic system, see Sect. \ref{sec:indiv}). In the eclipsing RP247 an eclipse seems to be responsible for the variability detection, though the limited amount of data suggests this conclusion should be taken with caution. In RP264, RP774, RP776 and RP883 we attribute the variability to a combination of stellar and dust variability in these symbiotic star candidates. 
   
\begin{table}
      \centering
      \caption{VMC $K_s$ lightcurve variability statistics for $\sigma_\mathrm{int}/\sigma_\mathrm{exp}\ge 2.5$ (Cross et al. 2009).}
      \label{tab:varstat}
      \begin{tabular}{llllllr}
         \hline\hline
         Name & $N$ & $\tilde{K_s}$ & $\sigma_\mathrm{exp}$ & $\sigma_\mathrm{int}$ & $\sigma_\mathrm{int}/\sigma_\mathrm{exp}$\\
         \hline
         RP247 & 29 & 16.03 &  0.018 & 0.050 &    2.755   \\
         RP264 & 14 & 12.87 &  0.009 & 0.024 &    2.539   \\
         RP774 & 31 & 14.39 &  0.011 & 0.074 &    6.900   \\
         RP776 & 31 & 14.82 &  0.012 & 0.035 &    2.978   \\
         RP793 & 11 & 11.77 &  0.009 & 0.145 &   16.091   \\
         RP883 & 31 & 12.90 &  0.009 & 0.053 &    5.664   \\
         \hline
      \end{tabular}
   \end{table}

\begin{figure*}
      \begin{center}
         \includegraphics[scale=0.79,angle=270]{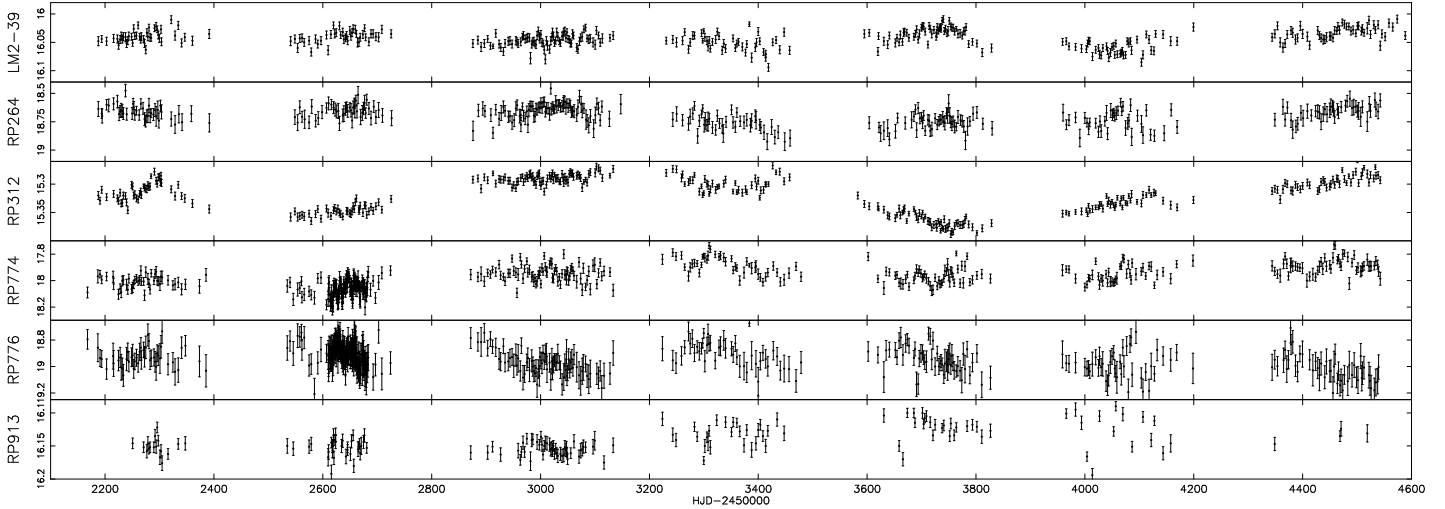}
      \end{center}
      \caption{OGLE-III $I$-band lightcurves for a sample of non-periodic variables.}
      \label{fig:lcother}
   \end{figure*}

   Figure \ref{fig:lcother} shows additional OGLE-III lightcurves for particular objects of interest or non-periodic variables. 
   Four of these, LM2-39, RP264, RP774 and RP776, are probable symbiotic stars, two of which have resolved H$\alpha$ nebulae, for which the variability is small or even negligible in the $I$-band (e.g. Miko{\l}ajewska et al. 1999). Of these four, only RP264 has significant $I$-band variability for its magnitude (in an average magnitude vs $\sigma$ plot). 
RP774 appears to show semi-regular variations in the $I$-band which suggests an RGB star, while LM2-39 and RP776 may show slow variations.

   \subsection{Classification scheme and spectral energy distributions}
   \label{sec:sed}
   The new classifications given in Tab. \ref{tab:basic1} are based on all the available evidence as discussed in Sect. \ref{sec:indiv}. 
   Where there is insufficient evidence to reclassify an object we have set our classification to neutral (NL). This mostly arises when detections are either weak or absent in the VMC data. 
   We have refrained from identifying some diffuse cases when the source in question cannot be a PN, i.e. it is diffuse background HII emission, even though unrelated stars may appear superposed on the diffuse nebula (e.g. RP232 and RP234; see Sect. \ref{sec:indiv}).
   Objects with H$\alpha$ emission as well as a significant continuum contribution are classified as emission line stars. 

\begin{figure*}
      \begin{center}
         \includegraphics[angle=270,scale=0.34]{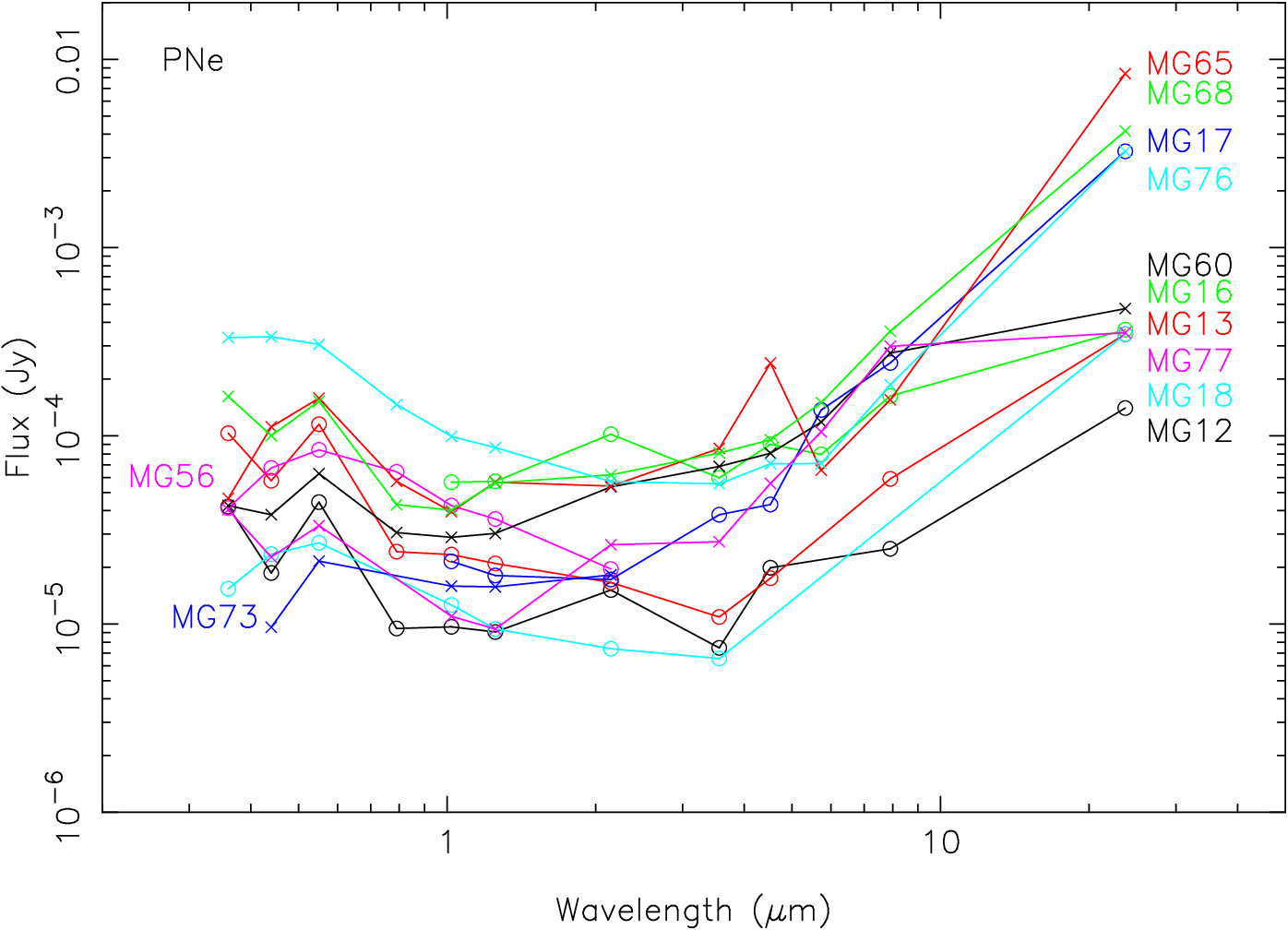}
         \includegraphics[angle=270,scale=0.34]{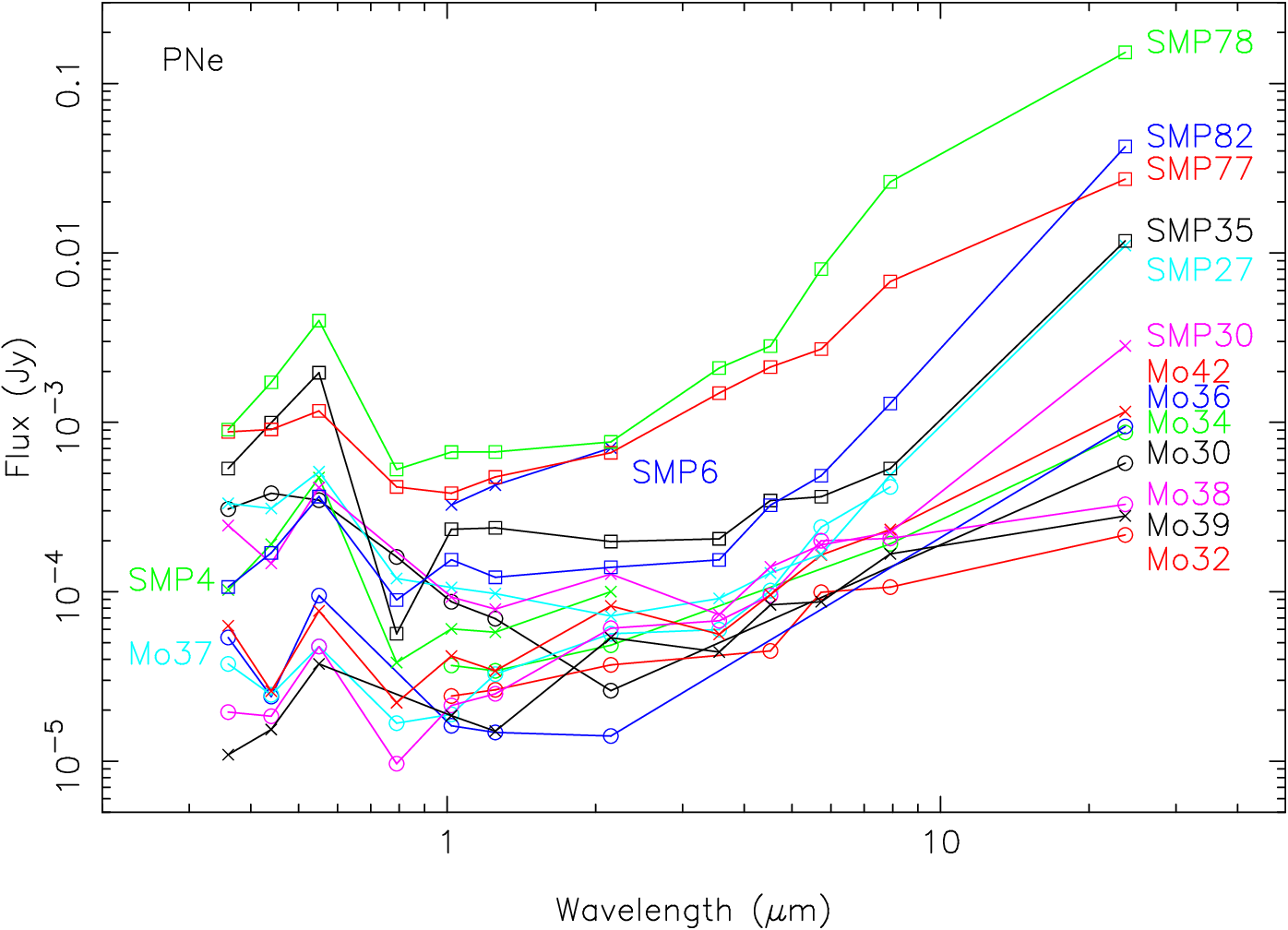}
         \includegraphics[angle=270,scale=0.34]{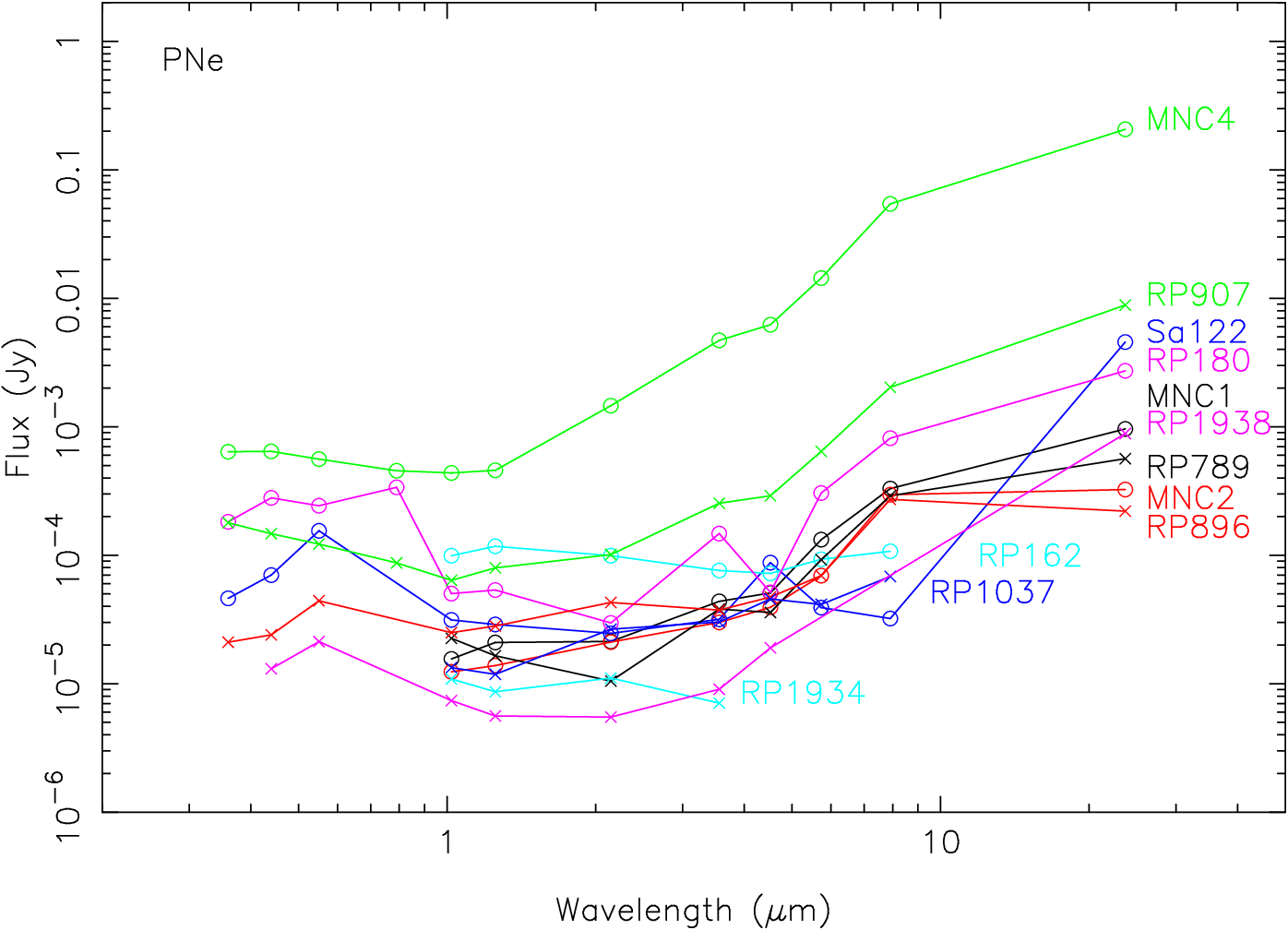}
         \includegraphics[angle=270,scale=0.34]{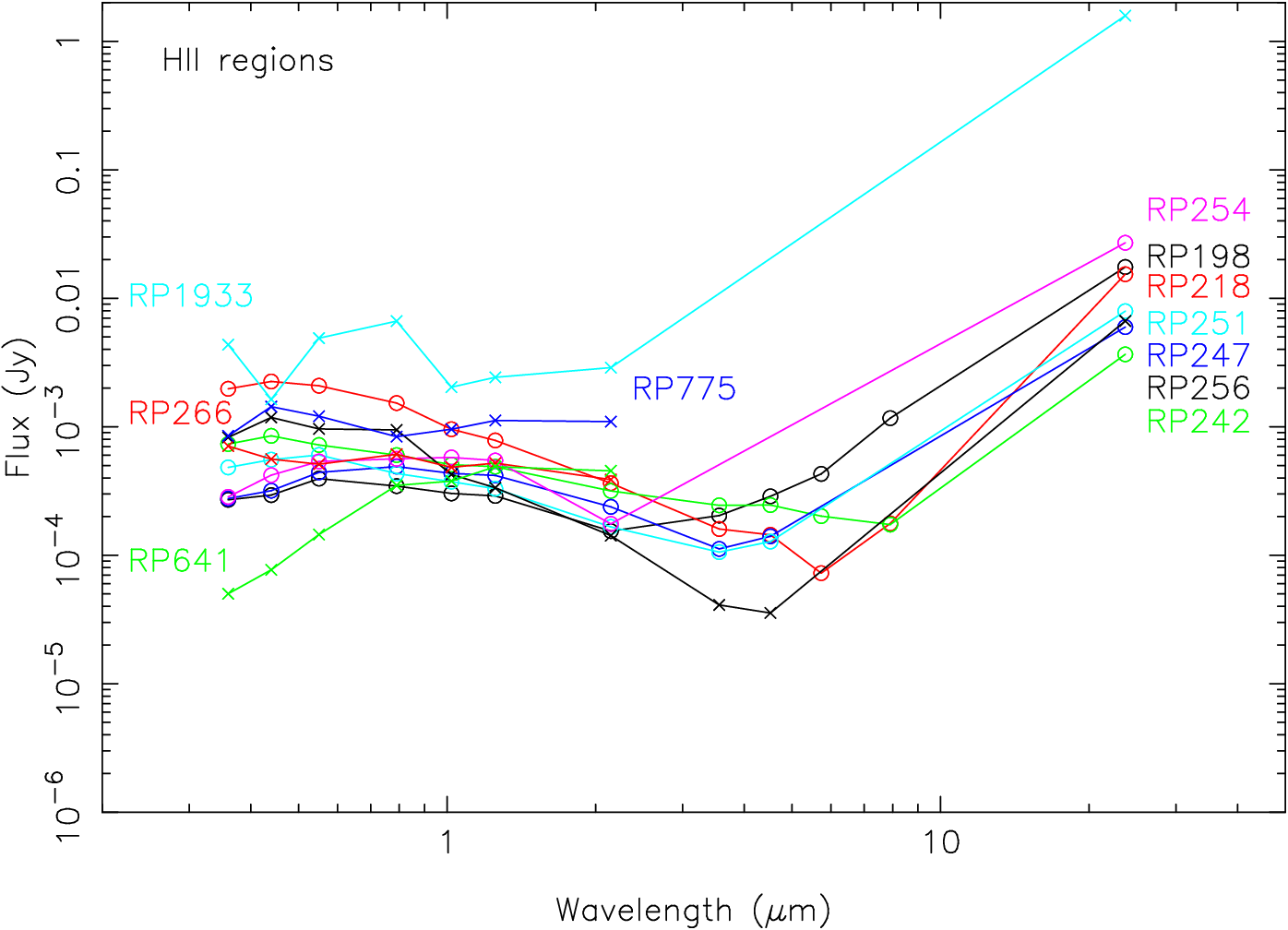}
         \includegraphics[angle=270,scale=0.34]{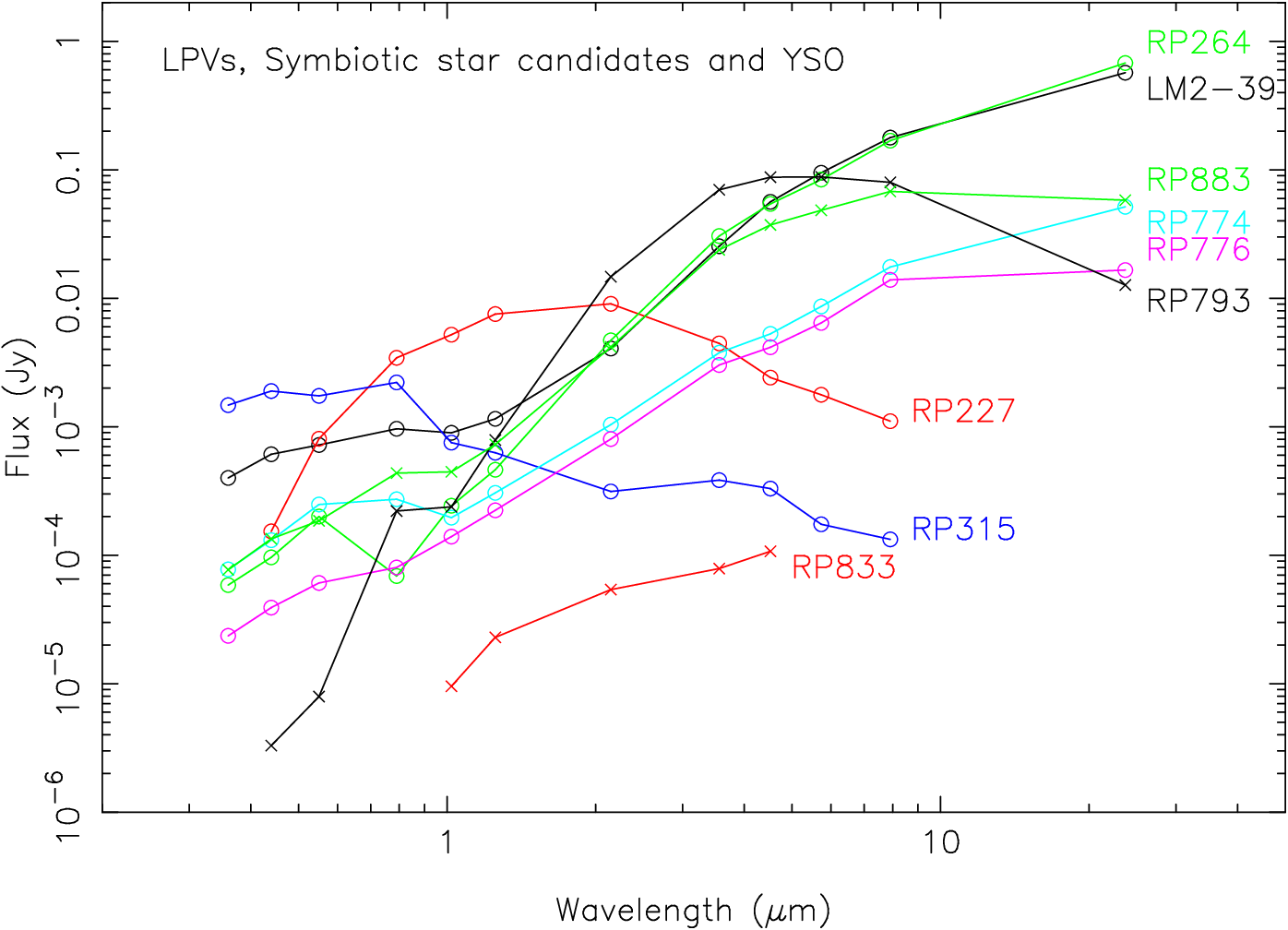}
         \includegraphics[angle=270,scale=0.34]{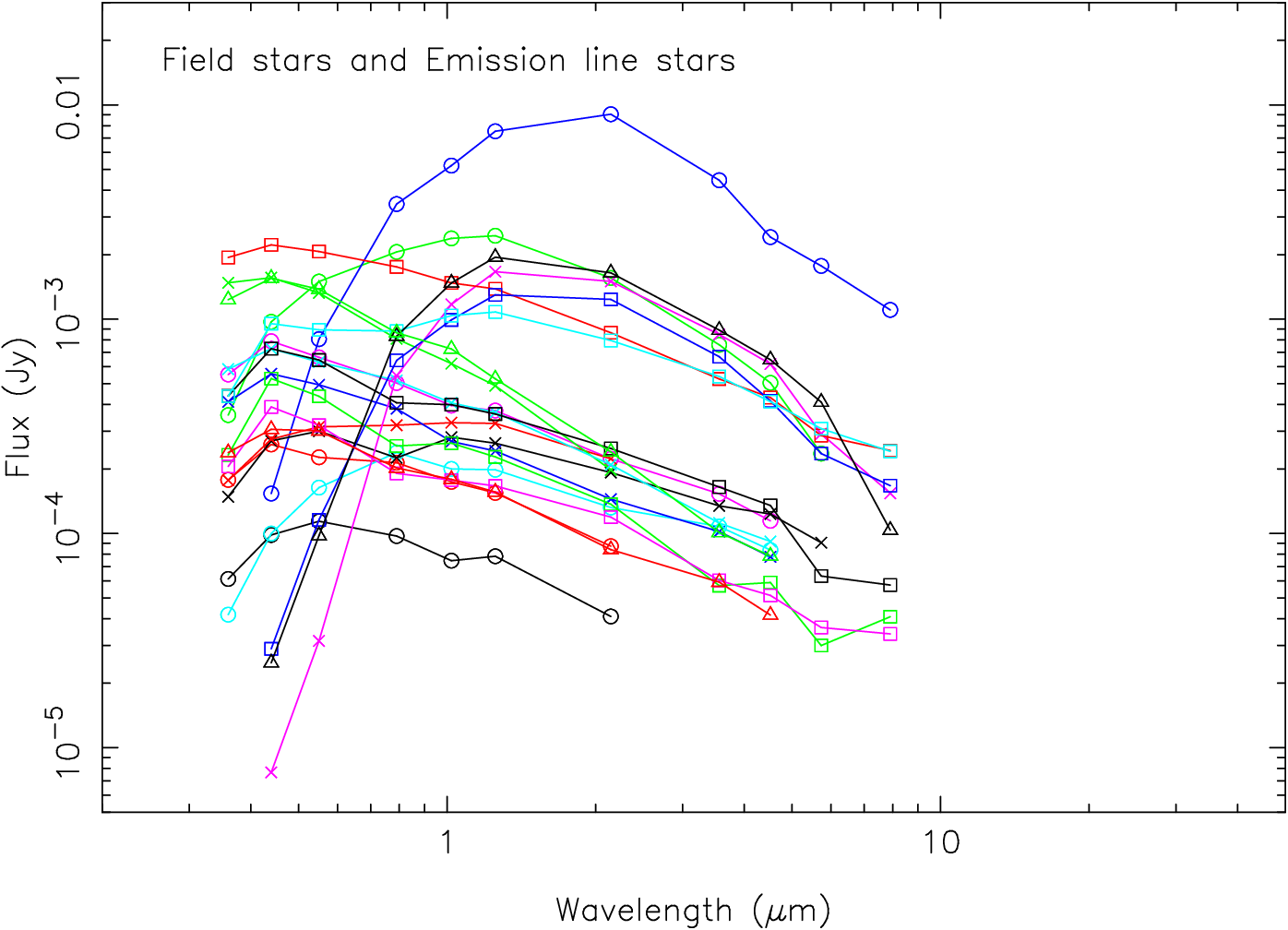}
         \includegraphics[angle=270,scale=0.34]{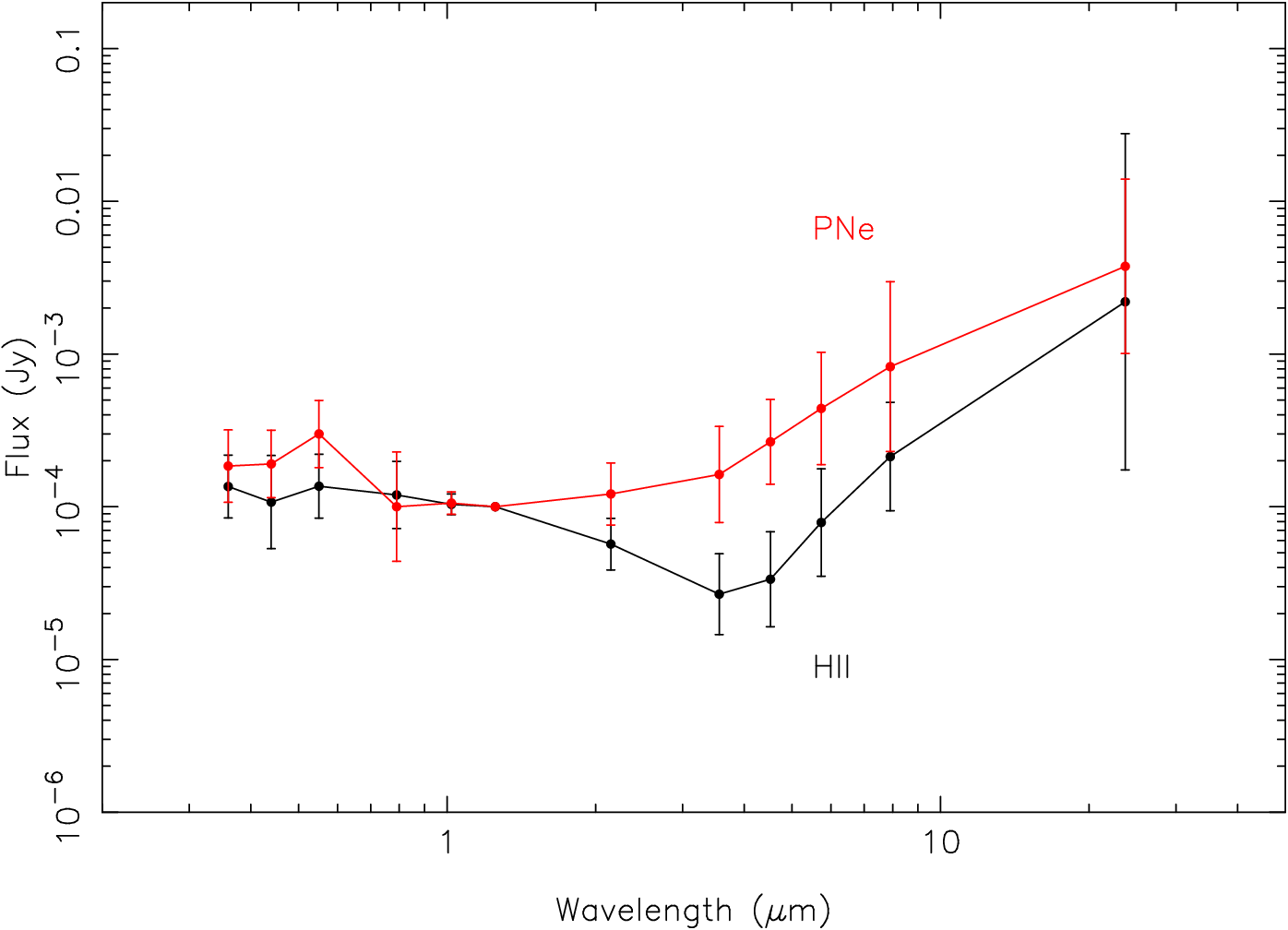}
         \includegraphics[angle=270,scale=0.34]{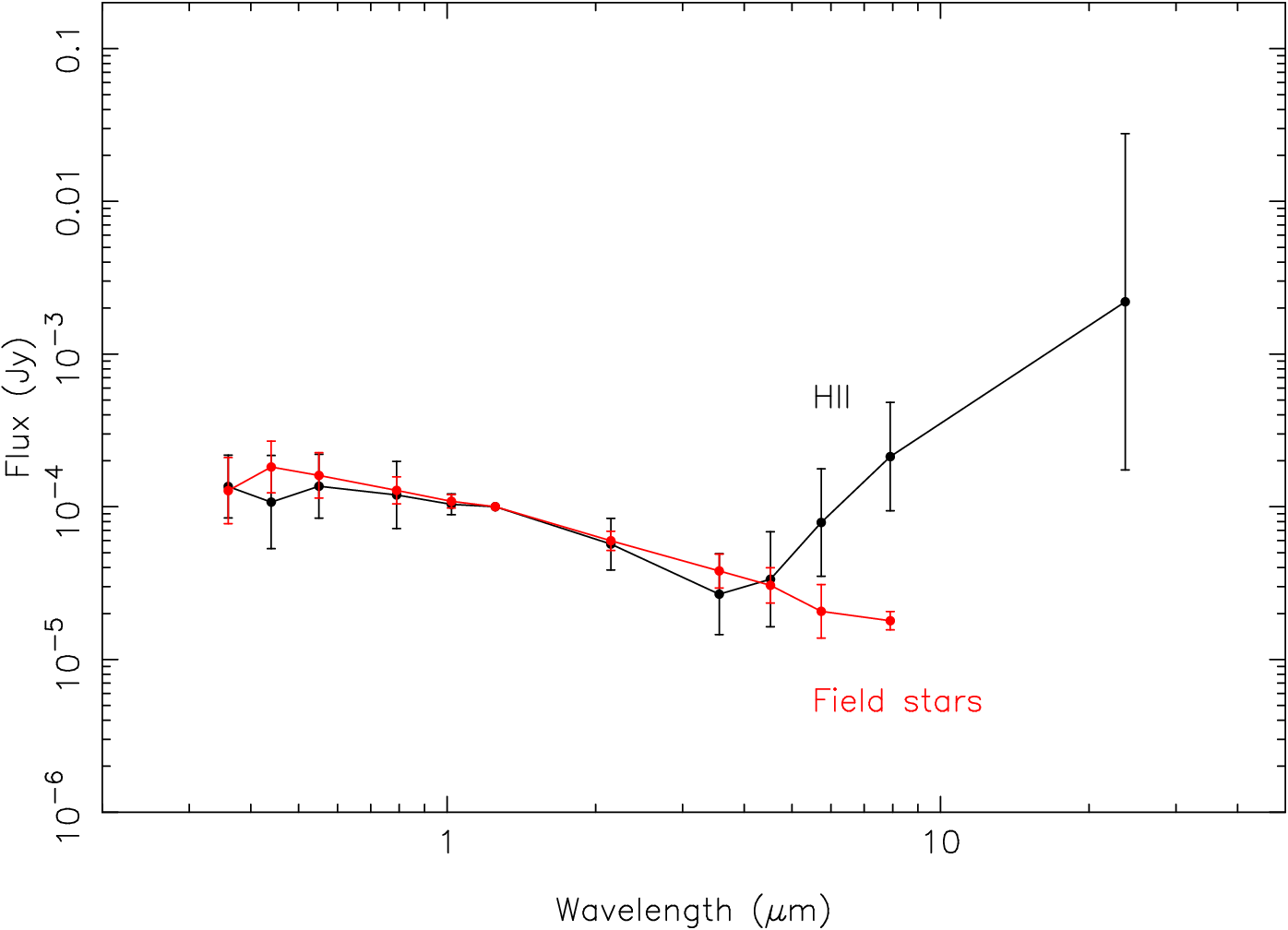}
      \end{center}
      \caption{Spectral energy distributions (SEDs) for objects from different samples (top six panels) and normalised median SEDs (bottom two panels).}
      \label{fig:seds}
   \end{figure*}

   To gain a clear overview of all classifications it is most instructive to look at the spectral energy distributions (SEDs) which are shown in Fig. \ref{fig:seds}. Unlike Hora et al. (2008) which focused on the brightest PNe only in the NIR and MIR, we are able to present the full SED from $U$ to 24 $\mu$m for most of our sample. No dereddening has been applied to the magnitudes before their fluxes were calculated using zeropoints from Bessell (1979) for $UBV$, Fouqu\'e et al. (2000) for $i$, the Cambridge Astronomical Survey Unit (CASU) measured Vega to AB conversions\footnote{http://casu.ast.cam.ac.uk/surveys-projects/vista/technical/filter-set} for $YJK_s$, and those in Sect. \ref{sec:sagephot} for MIR bands. The VMC conversions were derived by CASU following Hewett et al. (2006) and assume the effect of the atmospheric transmission profile on them is very small. 
   
   Figure \ref{fig:seds} is split into six panels grouping objects according to our classifications and two panels contrasting median SEDs normalised to $F_J=10^{-4}$ Jy with 1$\sigma$ error bars. PNe have a general $U$-shaped SED that is due to a large variety of different emission and continuum components. In the optical the nebular and central star continuum dominates with a strong contribution from [O~III] nebula emission at $V$ (see Sect. \ref{sec:optical}). Hot dust is probed by VMC and SAGE IRAC observations, but these wavelengths are also influenced by atomic and molecular emission lines which dominate the SED in fainter PNe. Cool dust is also commonly found in PNe and detected at 24 $\mu$m by SAGE MIPS observations. HII regions have much cooler ionising O/B stars which dominate optical and NIR wavelengths producing a similar median SED to the field stars. These stars are generally not sufficient to produce the hot dust as seen in PNe which leaves only cool dust dominating the MIR SED at 24 $\mu$m. The final two panels contrast HII regions against PNe and field stars. Note in particular the significant gap between HII regions and PNe in the MIR. We also show the periodic variables of Fig. \ref{fig:lpv} (excluding HII regions RP242 and RP247) together with suspected symbiotic stars and the YSO candidate RP833. These sources are  generally much redder than others in the sample with hot dust and high circumstellar reddening that suppresses the SED at optical wavelengths. The field stars form quite a homogeneous group as none have 24 $\mu$m detections, while emission line stars are similar with many being late-type. 
   
   \subsection{Diagnostic diagrams}
\begin{figure*}
          \begin{center}
             \includegraphics[scale=0.69,angle=270]{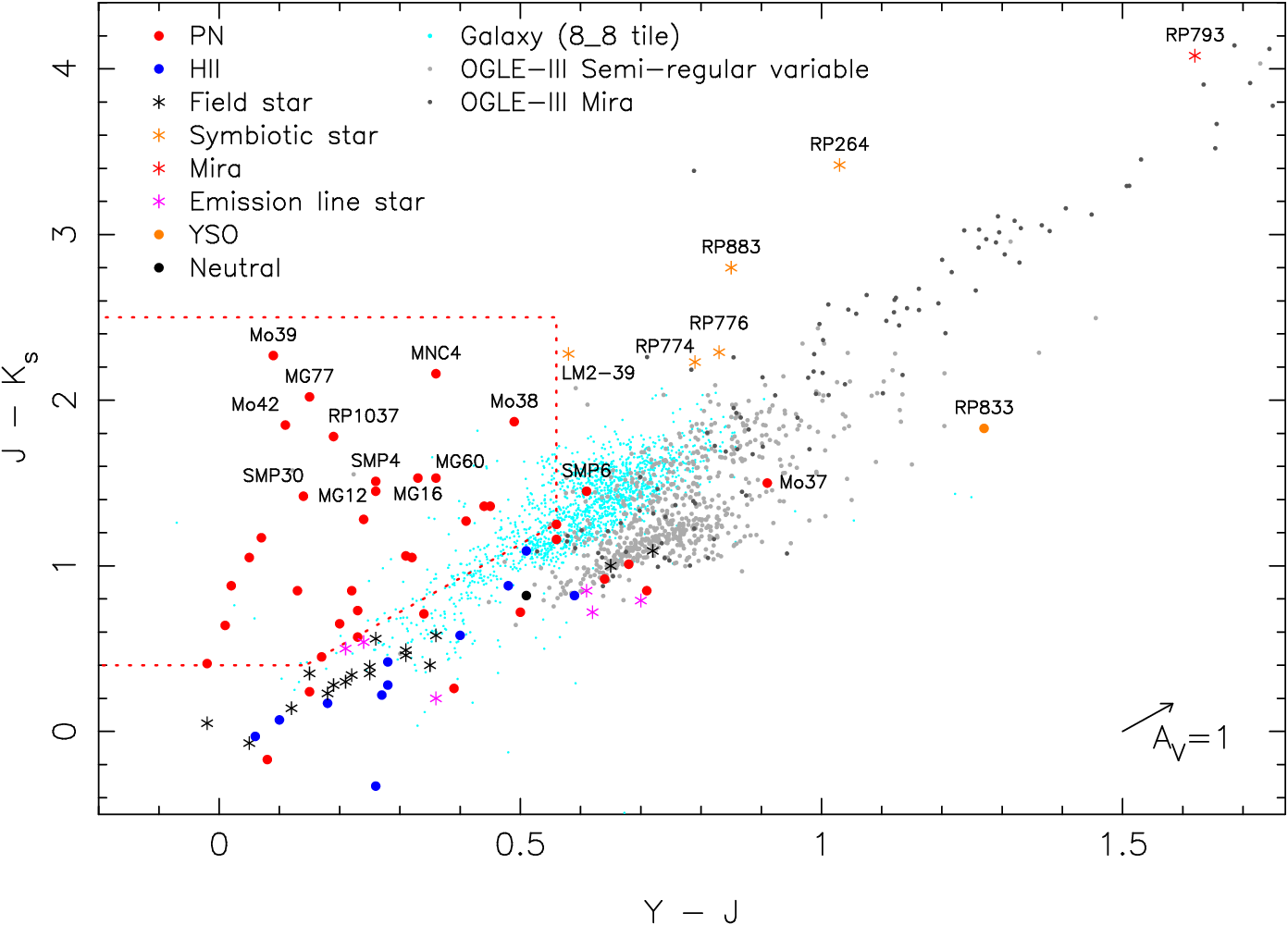}
             \includegraphics[scale=0.69,angle=270]{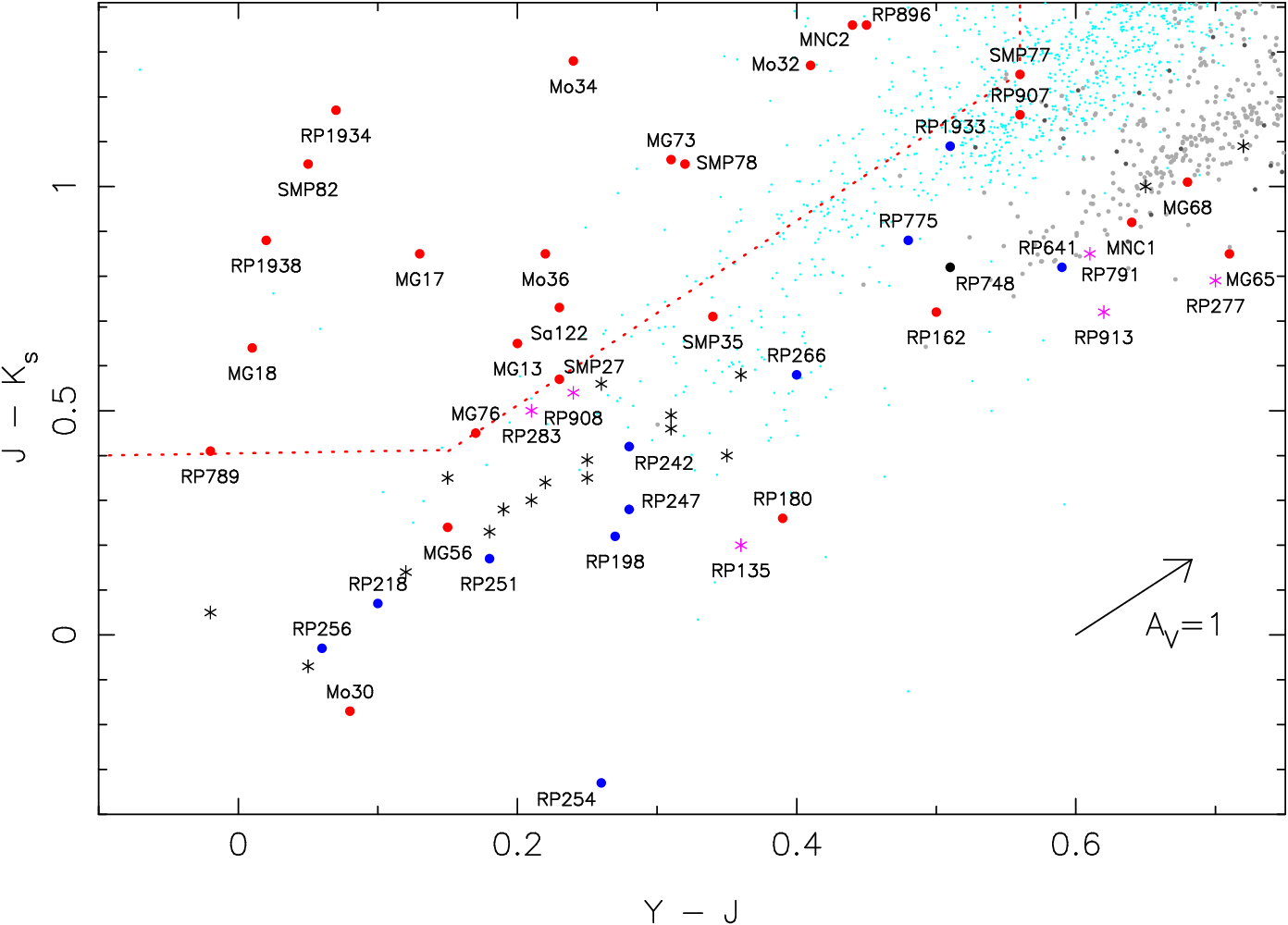}
          \end{center}
          \caption{VMC ant diagram showing the position of objects in our sample. A PN-rich region is bounded by the red dashed lines. The lower panel zooms into the more crowded region of the upper panel.}
          
          \label{fig:vmc}
       \end{figure*}

   With the wealth of optical, NIR and MIR magnitudes available we focus here on 
   diagnostic colour-colour diagrams which best isolate PNe from non-PNe. 
   Figure \ref{fig:vmc} shows PNe in the VMC colour-colour plane or `ant diagram' (see Fig. 9 of Cioni et al. 2011). PNe mostly lie within a region demarcated by red dotted lines in Fig. \ref{fig:vmc} which is defined by $0.4\le J-K_s\le2.5$ [$Y-J\le0.15$] and $J-K_s\ge2.05\,(Y-J-0.17)+0.45$ [$0.15\le Y-J\le0.56$, $J-K_s\le2.5$]. A small number of PNe with either hot dust (e.g. MG68) or perhaps stronger Pa$\beta$ nebular emission (e.g. MNC1 and Mo37) have redder $Y-J$ colours. Also shown in Fig. \ref{fig:vmc} are semi-regular and Mira variables (Soszy\'nski et al. 2009), and a representative sample of extended galaxies selected from the 8\_8 tile with Petrosian $K_s<17$ mag. The data show PNe are quite well separated from the majority of other sources in the fields. Symbiotic star candidates have much redder $J-K_s$ colours per $Y-J$ colour than PNe, while RP793 lies in the typical colour space expected for non-symbiotic Miras. 

\begin{figure*}
             \begin{center}
                \includegraphics[scale=0.69,angle=270]{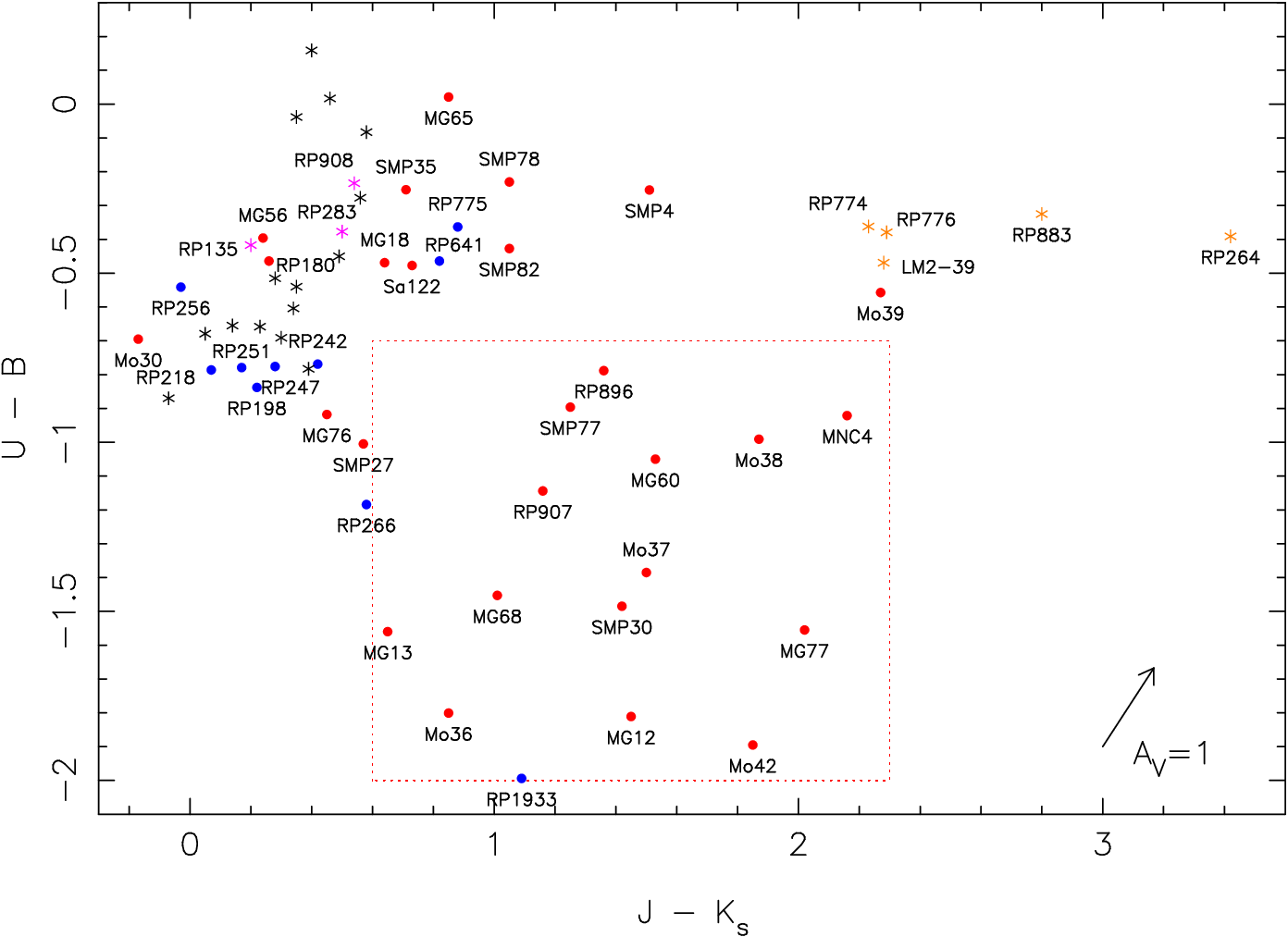}
                \includegraphics[scale=0.69,angle=270]{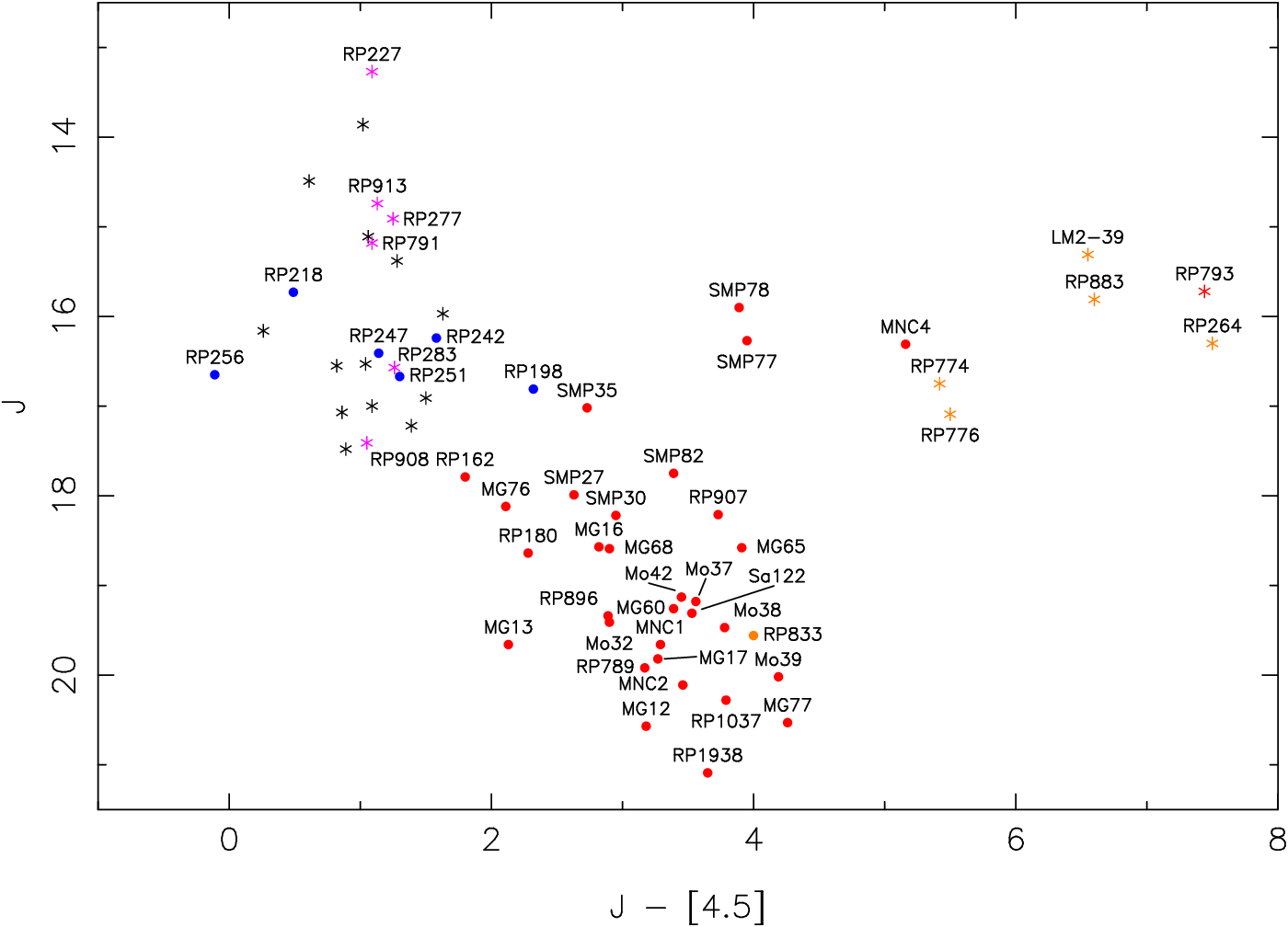}
             \end{center}
             \caption{As in Fig. \ref{fig:vmc} but for $U-B$ vs. $J-K_s$ (top) and $J$ vs. $J$ - [4.5] (bottom).}
             \label{fig:cmd1}
          \end{figure*}

\begin{figure*}
            \begin{center}
               \includegraphics[scale=0.69,angle=270]{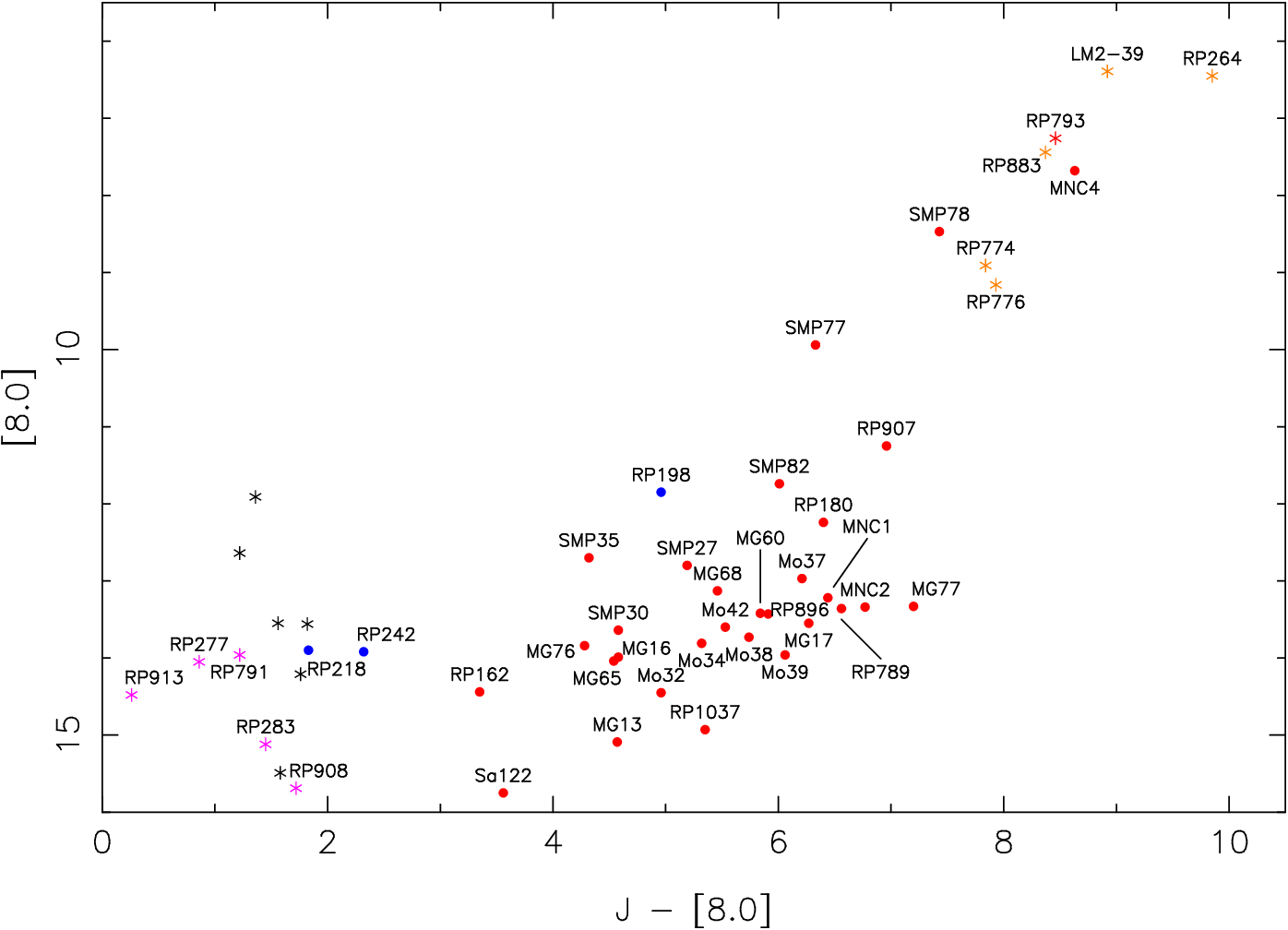}
               \includegraphics[scale=0.69,angle=270]{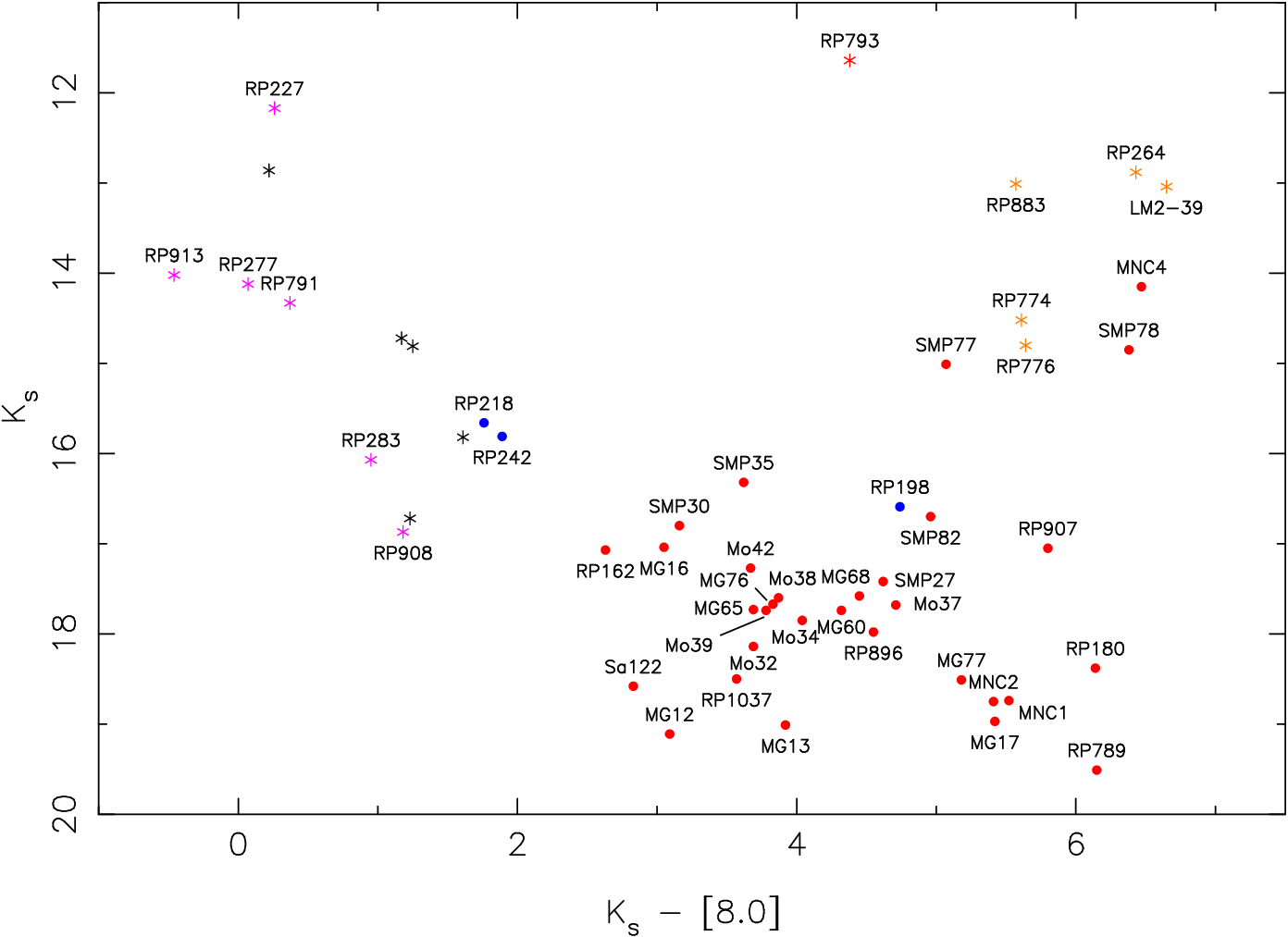}
            \end{center}
             \caption{As in Fig. \ref{fig:vmc} but for [8.0] vs. $J$ - [8.0] (top) and $K_s$ vs. $K_s$ - [8.0] (bottom).}
            \label{fig:cmd2}
         \end{figure*}

   The VMC ant diagram alone appears to be a very useful tool to isolate most PNe. Additional wavelengths can further improve the isolation of PNe. In the $U-B$ vs $J-K_s$ plane many PNe fall within a fairly isolated patch defined by $0.6\le J-K_s \le 2.3$ and $-2 \le U-B \le -0.7$ (Fig. \ref{fig:cmd1}). An HII region falls within this patch (RP1933) but morphology should allow these to be easily identified. Similar congregations of bona-fide PNe can also be found in a combination of various VMC and SAGE magnitudes and colours (Figs. \ref{fig:cmd1} and \ref{fig:cmd2}). The depth provided by the VMC survey allows redder $J-[8.0]$ and $K_s-[8.0]$ colours to be probed than previously possible and earlier predictions of Hora et al. (2008) are largely verified to this effect. Note the large $K_s$ and MIR luminosities of the symbiotic star candidates which form a well separated group from most PNe. MNC4 and other PNe with hot dust may also fall into this area. We found the $J$ - [4.5] colour to give the cleanest separation between HII regions ($J$ - [4.5] $\la2$), PNe ($2\la J$ - [4.5] $\la$ 4.5) and the obscured RGB or AGB stars of symbiotic stars and Miras ($J$ - [4.5] $\ga$ 5). 

       
\section{Notes on individual objects}
\label{sec:indiv}
\hspace{0.4cm}
\textbf{LM2-39}.--- RP2006b remarked this object as a late-type star. The resolved H$\alpha$ nebula, NIR and MIR colours, and slow $I$-band variations all suggest LM2-39 to be a strong symbiotic star candidate. 

\textbf{MG13}.--- We suspect our VMC magnitudes mostly reflect the CSPN. The knot immediately to NW may be resolved nebula emission from the enhanced rim of diameter 1.55\arcsec\ identified by Stanghellini et al. (2003). We measure a diameter of 1.95\arcsec\ that is consistent with this value within a VMC pixel.

\textbf{MG16}.--- There is a hint of extended emission in the VMC images as two lobes towards the NE and E. These could be extensions of the faint bipolar lobes detected by Shaw et al. (2006) or alternatively they could be evidence of ISM interaction.

\textbf{MG68}.--- The WFI images reveal a highly asymmetric nebula that is shaped by strong ISM interaction from the SE (e.g. Wareing et al. 2007). The [O~III] emission is concentrated towards the SE suggesting the presence of a bow-shock, while H$\alpha$ emission appears separated into multiple tails before being obscured behind a nearby dark cloud. The high spatial variation of [O~III] emission may explain the anomalous [O~III] flux published by RP2010 and revised by Miszalski et al. (2011b).

\textbf{Mo42, MG73 and RP202}.--- These are all bipolar PNe whose nebulae are resolved for the first time. The VMC image of Mo42 reveals a central star that is brightest in $Y$ and an inner bipolar nebula with point-symmetric extensions that resemble those in SMP10 (Shaw et al. 2001). In MG73 the WFI images resolve the bipolar nebula, while the VMC $K_s$ filament detected traces the inner torus aligned with PA=0--5$^\circ$. The brightest part of RP202 is a `waist' oriented at PA=45$^\circ$ and thin extensions extend either side along PA=0$^\circ$. RP2006b suspected MG73 was bipolar based on their unpublished spectroscopy (see properties in Tab. \ref{tab:basic1}). The morphology of Mo42 is consistent with the Type-I (Peimbert \& Torres-Peimbert 1983) chemical abundances of He/H=11.21 and N/O=$+$0.34 measured by Leisy \& Dennefeld (2006). 

\textbf{SMP27}.--- Shaw et al. (2001) describe \emph{HST} observations of this PN which shows an inner compact nebula (the PN proper) surrounded by an extended arc of emission. Here we have only measured magnitudes for the compact nebula. The arc was later confirmed to be an asymmetric shell that extends around the compact nebula from H$\alpha$ (RP2006a) and MIR observations (Cohen et al. 2009). We can now add a NIR $K_s$ detection to the list which subtends a diameter of 15.4\arcsec\ (3.7 pc across at 49 kpc). The shell is often mentioned in the context of AGB haloes (Corradi et al. 2003), extremely faint outer nebulae created by previous mass-loss episodes on the AGB. However, this interpretation is inconsistent with the atypically large diameter and high surface brightness of the shell whose peak $K_s$ intensity is 0.14 that of the inner nebula (most haloes are $10^{-3}$--$10^{-4}$ times fainter than their inner nebulae). Instead, the shell is most likely just ionised ISM as appears to be the case for NGC 3242 and NGC 6751 (Corradi et al. 2003).

\textbf{SMP30}.--- The Zaritsky et al. (2004) $i$ magnitude of $14.125\pm0.013$ is erroneously bright and was removed from Table \ref{tab:opmags1}.

\textbf{RP135}.--- The NIR and MIR colours of this object suggest a non-PN classification, perhaps an emission line star if there is indeed weak H$\alpha$ emission.

\textbf{RP203}.--- The WFI images reveal the RP2006b identification to be a knot of diffuse HII. Without a NIR or MIR counterpart it is difficult to classify this object as a PN. 

\textbf{RP232 and RP234}.--- The WFI images reveal an ionisation structure in RP232 opposite to that expected for a PN with H$\alpha$ in the centre surrounded by [O~III]. Both nebulae are part of extensive background emission in the vicinity of 30 Doradus. Superposed stars near the centre of the purported nebulae are therefore unrelated and in lower resolution imaging may have unduly amplified any apparent H$\alpha$ excess. 

\textbf{RP242}.--- There is no intrinsic [O~III] emission detected in the WFI image contrary to the spectrum published by RP2006b which has [O~III] $\lambda$5007/H$\alpha=0.20$. This greatly reduces the chance that this object is a PN, but it could also be symptomatic of a very low metallicity (e.g. Jacoby et al. 2002). Furthermore, RP242 is placed amongst objects lacking hot dust (stars and HII regions) in the diagnostic colour-colour planes. All our evidence points to an HII region with a central ionising O/B binary with $P=200.83$ days (Fig. \ref{fig:lpv}). In this sense RP242 seems to be a longer period version of RP247 (see below). Spectroscopy of the central stars is required to check our classification. If indeed a hot post-AGB star were found to be present, then RP242 would host the first binary central star discovered in the LMC.

\textbf{RP247}.--- The OGLE-III lightcurve reveals an eclipsing binary with $P=2.133$ days and minima with similar depths indicating similar $T_\mathrm{eff}$ components. Our main reason for reclassifying RP247 is the extremely weak H$\alpha$ emission in the WFI image compared to even other HII regions in our sample (e.g. RP254). Furthermore, the similar colours to RP242 and the SED with a 24 $\mu$m detection point to a weak Stromgren sphere surrounding a close O/B binary. 

\textbf{RP254}, \textbf{RP266}, \textbf{RP641} and \textbf{RP698}.--- The VMC and SAGE data clearly reveal the dusty environment of these HII regions. Soszy\'nski et al. (2009) found an unrelated semi-regular variable $\sim$7\arcsec\ NE of the RP2006b position for RP698 which has a period of 97.02 days. RP698 itself measures $\sim$10\arcsec\ across as an elongated enhancement of a larger region of nebulosity visible in the MIR.

\textbf{RP264}.--- An elliptical H$\alpha$ nebula measuring 1.9\arcsec$\times$3.9\arcsec\ appears in the WFI data. The very red $J-K_s=3.42$ colour and very high MIR luminosity points to an obscured AGB star for which we detect significant variability in the OGLE-III $I$-band lightcurve (Fig. \ref{fig:lcother}) and the VMC $K_s$ lightcurve (Fig. \ref{fig:ks}). The combination of nebula and obscured AGB star make RP264 an excellent symbiotic star candidate.

\textbf{RP265}.--- Is a resolved bipolar nebula in the H$\alpha$ WFI images that is also seen in broadband \emph{HST} observations (Shaw et al. 2007a). Only a weak detection was found in the VMC images and severe crowding meant no attempt was made to measure integrated magnitudes. The WFI images did not detect any [O~III] emission, but this is not entirely unexpected for bipolar PNe.

\textbf{RP277, RP283, RP791, RP908 and RP913}.--- All these objects appear to be emission line stars. A strong continuum component is found in the NIR alongside a small to moderate H$\alpha$ excess in WFI images. 

\textbf{RP691}.--- This may be a case of superposition with the purported PN lying just North of the star identified in Fig. \ref{fig:dbl1} (Reid \& Parker, private communication). No detection in the NIR or MIR of such a PN was found, so we leave the classification of RP691 as neutral. The magnitudes of the bright star are recorded for reference. 

\textbf{RP701}.--- A diffuse nebula is detected in the VMC images but severe crowding precludes measurement of its properties. 

\textbf{RP774 and RP776}.--- Both RP774 and RP776 are catalogued as `True' PNe by RP2006b and Woods et al. (2011) classified RP774 as a YSO. They appear to have significant nebular H$\alpha$ emission in the WFI images and the NIR and MIR properties of both objects suggest either an obscured RGB (or possibly AGB) star or the presence of hot dust. It is difficult to accept the latter explanation given the significant $K_s$ variability, unless this variability was attributed to line-of-sight dust obscuration events. This is however unlikely since these events are extremely rare in PNe and we would have expected to see more dramatic $I$-band variability if this were the case (Miszalski et al. 2011a). The OGLE-III lightcurve of RP774 in Fig. \ref{fig:lcother} appears to show low-amplitude semi-regular variability on the order of $\sim$40--80 days and this could be attributed to an RGB star. 

\textbf{RP775}.--- van Loon et al. (2010) classified RP775 as a possible HII region. With the high resolution of the WFI data we can see the ionised front powered by an embedded massive object. This leaves no doubt that RP775 is an HII region.

\textbf{RP793}.--- The very large 0.4 mag $K_s$ amplitude in the VMC lightcurve supports the C-rich Mira classification made in the LPV catalogue of Soszy\'nski et al. (2009). We have replaced the $I$ and $V$ magnitudes in Tab. \ref{tab:opmags1} with the average OGLE-III values of $I=17.63$ and $V=21.65$ mag (Soszy\'nski et al. 2009). The very red optical colours in Tab. \ref{tab:opmags1} and the extremely weak H$\alpha$ source in the WFI image essentially rule out the presence of a symbiotic nebula in RP793. 

\textbf{RP833}.--- There is no OGLE-III $I$-band detection of this object indicating a very high level of obscuration. It is most likely a YSO based on its NIR colours, the presence of other similar condensations directly to the North and South and the dusty MIR HII environment. 

\textbf{RP883}.--- As for RP774 and RP776, RP883 appears to have a strong nebular component in the optical combined with NIR and MIR colours typical of an obscured RGB or AGB star. We found a period of 176.49 days from the OGLE-III lightcurve after fitting and subtracting the general decline. The significant $K_s$ variability in the VMC lightcurve further supports a classification as a candidate symbiotic star with the measured period lying at the lower range for symbiotic stars (Miko{\l}ajewska 2003; Gromadzki et al. 2009). 

\textbf{RP1933}.--- Discussed by van Loon et al. (2010) and also reclassified by RP2010, the bright extended emission clearly points to an HII region powered by a very hot central source ($U-B=-1.99$).

\section{Discussion}
\label{sec:discussion}
\subsection{Reclassifications and selection effects}
\label{sec:reclass}
Table \ref{tab:removed} summarises the 45 objects we have reclassified as non-PNe from 3 `true', 8 `likely' and 34 `possible' PNe classified by RP2006b. Apart from the symbiotic star candidates we have not added to Tab. \ref{tab:removed} any objects with uncertain classifications (those with `?' in Tab. \ref{tab:basic1}). Periodic variables were included in the table as field stars except for when they are surrounded by HII regions (e.g. RP242). Out of the non-RP sample only LM2-39 was reclassified as a symbiotic star candidate.

\begin{table}
   \centering
   \caption{Objects reclassified as non-PNe from the RP2006b catalogue.}
   \label{tab:removed}
   \begin{tabular}{lrrr}
      \hline\hline
      Object Type & True & Likely & Possible\\
      \hline
      HII regions & - & 1 & 11\\
      Diffuse HII & 1 & 2 & 4\\ 
      Field stars & 1 & 3 & 12\\ 
      Emission line stars &- & 2 & 3 \\
      Symbiotic stars & 1 & - & 3\\ 
      YSO         & - & - & 1 \\   
      \hline
      Reclassified/Total & 3/21 & 8/9 & 34/37\\
      Percentage & 14\% & 89\% & 92\%\\
      \hline
   \end{tabular}
\end{table}

The contamination fraction of 45/67 objects or 67\% is rather high. Field stars and non-diffuse HII regions make up the bulk of the contaminants, most of which were designated as possible PNe by RP2006b. Figure \ref{fig:lmc} shows that most of the RP objects classified as possible PNe are located around 30 Doradus. This suggests that the strong emission-line background in this area played a large part in their inclusion in the RP survey. The dominant background complicates multi-object spectroscopic followup whose sky subtraction is generally less accurate than longslit spectroscopy. The variable spectrograph point-spread function of 2dF prior to the AAOmega spectrograph upgrade (Sharp et al. 2006) and use of only $\sim$40 sky fibres over the large 2.1 degree field-of-view (RP2010), may have indeed resulted in residual emission lines being mistaken for genuine emission. This could explain the inclusion of some field stars such as RP1018 for which we cannot identify any H$\alpha$ emission in the WFI images. Although this does not seem widespread as we were able to calibrate the ESO WFI [O~III] images based on spectroscopic [O~III] fluxes from RP2010 (Miszalski et al. 2011b). In a few cases intrinsic variability (e.g. RP312, RP315) may also have given the false appearance of H$\alpha$ emission in the RP survey material. However, not all field stars could be explained in this fashion as many have negligibly small OGLE-III $I$-band variability.

Perhaps the largest contributing factor towards the high contamination fraction is the lower resolution of the RP survey. The increased resolution and depth of the multi-wavelength data presented here have resolved some cases of superpositions between field stars and diffuse background HII (e.g. RP232, RP234) and have been particularly helpful in identifying smaller HII regions. Actual superpositions may also occur between bona-fide PNe and field stars, synonymous with our emission line star classification, but should be rare based on the rarity of real superpositions in the Galactic Bulge PN sample (Miszalski et al. in preparation). 

\subsection{Implications for binary central star surveys} 
\label{sec:binary}
There has been some interest in searching for binary central stars in the LMC population (Shaw et al. 2007c; Shaw, Rest \& Damke 2009). Miszalski et al. (2009) used OGLE-III to more than double the known Galactic population of close binary CSPN but to date no such binaries have been found in the LMC. LMC PNe offer a potentially very useful platform to measure the binary central star fraction for a large population, but the extreme faintness of LMC CSPN are a major obstacle to this task (Villaver et al. 2007). A major repercussion of our multi-wavelength analysis is that unless an LMC PN can be shown to be a bona-fide PN, then any claims of variability, even if periodic, cannot be used to claim a binary central star. An interesting example is RP247 whose orbital period of 2.13 days (Fig. \ref{fig:lpv}) is not entirely inconsistent with the observed Galactic period distribution (Miszalski et al. 2009, 2011c), however the weak trace of H$\alpha$ emission is insufficient for a PN classification of this HII region. 

There is also a danger that field stars are mistaken for secondary star companions. Shaw et al. (2009) found a large proportion of `PNe' with NIR colours of giants. Even though there are only a few Galactic PNe with bona-fide luminous companions (e.g. Tab. 4 of De Marco 2009), the fraction claimed by Shaw et al. (2009) is anomalously high and readily explained by the high numbers of field stars and emission line stars revealed in this work. Many of these stars will have slowly-variable lightcurves (e.g. RP312 and RP913 in Fig. \ref{fig:lcother}) and periodic lightcurves of an extremely large variety (e.g. Kiss \& Bedding 2003). Inspection of lightcurves in the remaining RP2006b catalogue not analysed in this work has revealed additional examples of variability due to field star contaminants. As yet we have not found any genuine periodic variability that can be ascribed to a binary CSPN. 

   \section{Conclusions}
   \label{sec:conclusion}
   We have presented multi-wavelength data for a sample of 102 objects previously classified as PNe that were observed during the first year of the VMC survey. The six LMC tiles constituting the first VMC observations sample a range of LMC environments, but our results are dominated by the 6\_6 tile centred on 30 Doradus. This complex region serves as an excellent training ground for developing multi-wavelength criteria to maximise the exploitation of the VMC survey for PNe studies and to identify non-PNe contaminating extant MCPNe catalogues. The main conclusions are as follows:

\begin{itemize}
   \item PNe are well-separated from other objects in multiple colour-colour diagnostic diagrams. Their typical colours in the VMC ant diagram are defined by the region $0.4\le J-K_s\le2.5$ [$Y-J\le0.15$] and $J-K_s\ge2.05\,(Y-J-0.17)+0.45$ [$0.15\le Y-J\le0.56$, $J-K_s\le2.5$] which shares minimal overlap with galaxies and evolved variable stars. Distinction between a variety of non-PNe was well catered for by the incorporation of MCPS and SAGE photometry. 
   \item A large non-PNe contamination fraction of 67\% was identified in the RP subset of our study based on an overall assessment of all the multi-wavelength data. This is made up of 3 `true', 8 `likely' and 34 `possible' PNe from RP2006b. The RP2006b reclassifications were made up of 16 field stars, 5 emission line stars, 19 HII regions, 4 symbiotic star candidates and 1 YSO candidate. Factors contributing to these results may include the complex emission-line background near 30 Doradus, the lower resolution of RP survey and residual emission-lines following sky-subtraction. 
   \item Six periodic variables were discovered amongst the RP2006b sample from OGLE-III lightcurves and were reclassified as non-PNe. These discoveries emphasise the importance of a clean PN sample before searches for binary central stars of LMC PNe are conducted. Furthermore, the large proportion of field stars and emission line stars we have reclassified in the sample disproves previous claims of a large population of giant secondary companions to LMC PNe (Shaw et al. 2009). The success of incorporating time-series photometry into multi-wavelength studies of PNe, first employed by Miszalski et al. (2009) and reaffirmed here, should where possible be incorporated into standardised schemes for identifying non-PNe (e.g. Frew \& Parker 2010).
   \item The high resolution of the presented imaging has resolved the morphologies of some PNe for the first time. These include the bipolar nebulae of MG73, RP202 and RP265 (confirming prior \emph{HST} imaging), the asymmetric, bow-shocked nebula of MG68, and the exquisite point-symmetric extensions to the inner bipolar nebula of Mo42. 
   \item We have identified 5 new symbiotic star candidates in the LMC based primarily on their H$\alpha$ emission, NIR and MIR colours, and $I$-band and $K_s$ variability. This is a dramatic boost to only 8 previously catalogued LMC symbiotics (Belczy\'nski et al. 2000; J. Miko{\l}ajewska, private communication). 
\end{itemize}

\begin{acknowledgements}
   We kindly thank Warren Reid and Quentin Parker for sharing their finder charts of the RP sample and comments on an early version of this paper. BM thanks Eduardo Gonz\'alez-Solares for helpful discussions concerning the VMC data products and Nicholas Cross for discussions regarding variability of sources in the VSA archive. We also thank J. Emerson, L. Girardi and V. Ivanov for helpful comments and the anonymous referee for constructive comments that helped refine the paper. We thank the UK team responsible for the realisation of VISTA, and the ESO team who have been operating and maintaining this new facility. The VISTA Data Flow System comprising the VISTA pipeline at CASU and the VISTA Science Archive at WFAU has been crucial in providing us with calibrated data products for this paper, and is supported by the UK Science and Technology Facilities Council. The OGLE project has received funding from the European Research Council under the European Community's Seventh Framework Programme (FP7/2007-2013) / ERC grant agreement no. 246678. This research has made use of SAOImage DS9, developed by Smithsonian Astrophysical Observatory.
\end{acknowledgements}

\clearpage
\begin{appendix}
\section{Photometry}
\label{sec:phot}
\subsection{VMC}
   The mixture of stellar and extended sources in our sample required different approaches to obtaining their magnitudes. As a first step we created a stacked image per waveband (0.339\arcsec/pixel) for each object by averaging individual paw-print observations (Irwin et al. 2004). As described above, the depth in these images is uniform across all tiles with some small variation. The catalogue photometry hosted by the VISTA Science Archive (VSA) is adopted when the \verb|mergedClass| flag indicates a stellar or probably stellar object. In these instances we adopt the catalogued magnitudes calculated with a small aperture of diameter 2.0\arcsec\ (AperMag3) which is corrected for flux outside the aperture assuming a stellar profile. This is appropriate for compact, unresolved PNe and of course for stars. For some extended objects we can make use of uncorrected magnitudes (NoAperCorr) calculated using apertures of diameter 2.0\arcsec\ (AperMag3), 2.8\arcsec (AperMag4) or 5.7\arcsec\ (AperMag6), where the aperture is visually selected to enclose the most object flux. The VSA magnitudes are based on the previously mentioned level of $TKN$ completeness. 

   When catalogue magnitudes were not available for fainter extended sources we have performed our own aperture photometry on our stacked image cutouts. A specially developed plugin for \textsc{ds9} (Joye \& Mandel 2003) served as a wrapper for the \textsc{ds9} funtools program \textsc{funcnts}\footnote{http://hea-www.harvard.edu/RD/funtools/ds9.html} which calculates the total number of counts within a given \textsc{ds9} region (e.g. a circle). 
   A circular aperture of maximal radius is chosen for each object along with multiple nearby sky apertures of the same radius. 
   The magnitudes were then calculated using:
   \begin{equation}
      m = \mathrm{NZP} - 2.5\, \mathrm{log}_{10}\, [(O-S)/T]
      \label{eqn:mag}
   \end{equation}
   where $S$ is the average of the total sky counts in all sky apertures, $O$ is the total counts in the object aperture, $T$ is the normalised exposure time (5 s for $K_s$, 10 s for $J$ and 20 s for $Y$) and NZP is the nightly photometric zeropoint. We assigned a $1\sigma$ error of 0.20 mag based on the comparison between our measurements and the catalogued photometry. As no errors were given by the VSA for NoAperCorr magnitudes we also assigned them these errors. Some fields were too crowded for aperture photometry and we have remarked in Tab. \ref{tab:basic1} where this occurred. In one case an elliptical aperture seemed most suitable (RP789). No aperture corrections were applied to our aperture photometry. 
   
\subsection{SAGE}
\label{sec:sagephot}
   To perform the aperture photometry we developed a similar \textsc{ds9} plugin as for the VMC aperture photometry that would allow for the efficient calculation of all magnitudes per object once the IRAC and MIPS images were loaded alongside the VMC colour-composite image for guidance. The total number of sky subtracted counts per object were converted to Jy following the instrument handbooks and converted to magnitudes using the zeropoints given in the accompanying SAGE data release documentation, i.e. 280.9, 179.7, 115.0, 64.13 and 7.14 Jy for the 3.6, 4.5, 5.8, 8.0 and 24.0 $\mu$m bands, respectively. Table \ref{tab:mir1} contains the measured magnitudes for which we have assigned $1\sigma$ errors of 0.25, 0.25, 0.30, 0.35 and 0.40 mag, respectively, based on a comparison with Hora et al. (2008) and catalogue magnitudes. 
   These errors are larger than the 0.1--0.2 mag estimated by Hora et al. (2008) and more realistically reflect the inherent difficulty in selecting optimal object and sky apertures under sometimes very challenging circumstances. This is particularly true at 8.0 $\mu$m where diffuse PAH emission in the field creates a highly variable background. An excellent example is Mo38 which has negligible background at shorter wavelengths. If our sky aperture is placed in a background minimum SE of the PN we reproduce [8.0]=12.9 mag of Hora et al. (2008). However, a placement immediately West of the PN on a similar background level as the PN results in [8.0]=13.7 mag -- a difference of 0.8 mag! In the case of a few bright objects we used the catalogued magnitudes and their errors. Very extended HII regions were not measured. 

\begin{table*}
\centering
\caption{Optical magnitudes from Zaritsky et al. (2004).}
\label{tab:opmags1}
\begin{tabular}{lllllllllllrrr}
\hline\hline
Name & ID & Status & $U$ & $U$e & $B$ & $B$e & $V$ & $V$e & $i$ & $i$e & $U-B$ & $B-V$ & $V-i$ \\
\hline
LM2-39 & Sy? & - & 16.64 & 0.04 & 17.11 & 0.04 & 16.76 & 0.04 & 16.03 & 0.04 & $-$0.47 & 0.35 & 0.72\\
MG12 & PN & - & 19.09 & 0.15 & 20.90 & 0.10 & 19.78 & 0.08 & 21.05 & 0.22 & $-$1.81 & 1.11 & $-$1.27\\
MG13 & PN & - & 18.11 & 0.16 & 19.67 & 0.11 & 18.75 & 0.04 & 20.04 & 0.13 & $-$1.56 & 0.92 & $-$1.28\\
MG16 & PN & NC & - & - & - & - & - & - & - & - & - & - & -\\
MG17 & PN & NC & - & - & - & - & - & - & - & - & - & - & -\\
MG18 & PN & - & 20.18 & 0.18 & 20.65 & 0.14 & 20.32 & 0.09 & - & - & $-$0.47 & 0.32 & -\\
MG56 & PN & - & 19.11 & 0.17 & 19.50 & 0.19 & 19.09 & 0.08 & 18.97 & 0.12 & $-$0.40 & 0.41 & 0.12\\
MG60 & PN & - & 19.07 & 0.10 & 20.12 & 0.07 & 19.41 & 0.05 & 19.78 & 0.09 & $-$1.05 & 0.72 & $-$0.37\\
MG65 & PN & - & 18.98 & 0.09 & 18.96 & 0.07 & 18.40 & 0.07 & 19.10 & 0.12 & 0.02 & 0.55 & $-$0.70\\
MG68 & PN & - & 17.62 & 0.13 & 19.07 & 0.10 & 18.44 & 0.08 & 19.41 & 0.10 & $-$1.45 & 0.63 & $-$0.97\\
MG73 & PN & - & - & - & 21.62 & 0.16 & 20.57 & 0.18 & - & - & - & 1.05 & -\\
MG75 & ND,PN & ND & - & - & - & - & - & - & - & - & - & - & -\\
MG76 & PN & - & 16.84 & 0.11 & 17.76 & 0.04 & 17.69 & 0.05 & 18.08 & 0.12 & $-$0.92 & 0.07 & $-$0.39\\
MG77 & PN & - & 19.13 & 0.14 & 20.68 & 0.16 & 20.10 & 0.12 & - & - & $-$1.56 & 0.58 & -\\
MNC1 & PN & ND & - & - & - & - & - & - & - & - & - & - & -\\
MNC2 & PN & ND & - & - & - & - & - & - & - & - & - & - & -\\
MNC3 & PN? & ND & - & - & - & - & - & - & - & - & - & - & -\\
MNC4 & PN & - & 16.13 & 0.37 & 17.05 & 0.02 & 17.03 & 0.03 & 16.85 & 0.04 & $-$0.92 & 0.02 & 0.18\\
Mo30 & PN & - & 16.93 & 0.04 & 17.62 & 0.05 & 17.55 & 0.13 & 17.98 & 0.06 & $-$0.69 & 0.07 & $-$0.43\\
Mo32 & PN & ND & - & - & - & - & - & - & - & - & - & - & -\\
Mo34 & PN & ND & - & - & - & - & - & - & - & - & - & - & -\\
Mo36 & PN & - & 18.82 & 0.11 & 20.62 & 0.16 & 18.96 & 0.16 & - & - & $-$1.80 & 1.66 & -\\
Mo37 & PN & - & 19.21 & 0.10 & 20.59 & 0.07 & 19.71 & 0.05 & 20.43 & 0.19 & $-$1.39 & 0.88 & $-$0.72\\
Mo38 & PN & - & 19.92 & 0.13 & 20.91 & 0.09 & 19.71 & 0.05 & 21.03 & 0.29 & $-$0.99 & 1.20 & $-$1.32\\
Mo39 & PN & - & 20.55 & 0.29 & 21.11 & 0.09 & 19.97 & 0.07 & - & - & $-$0.56 & 1.14 & -\\
Mo42 & PN & - & 18.64 & 0.10 & 20.54 & 0.26 & 19.19 & 0.11 & 20.13 & 0.10 & $-$1.90 & 1.35 & $-$0.95\\
Sa122 & PN & - & 18.98 & 0.20 & 19.46 & 0.09 & 18.43 & 0.18 & - & - & $-$0.48 & 1.03 & -\\
SMP4 & PN & - & 18.12 & 0.06 & 18.37 & 0.04 & 17.22 & 0.03 & 19.54 & 0.07 & $-$0.25 & 1.16 & $-$2.32\\
SMP6 & PN & NC & - & - & - & - & - & - & - & - & - & - & -\\
SMP27 & PN & - & 16.84 & 0.04 & 17.84 & 0.12 & 17.13 & 0.03 & 18.30 & 0.09 & $-$1.00 & 0.71 & $-$1.17\\
SMP30 & PN & - & 17.17 & 0.06 & 18.65 & 0.05 & 17.37 & 0.05 & - & - & $-$1.48 & 1.29 & -\\
SMP35 & PN & - & 16.33 & 0.20 & 16.58 & 0.04 & 15.67 & 0.07 & 19.11 & 0.34 & $-$0.25 & 0.91 & $-$3.44\\
SMP77 & PN & - & 15.78 & 0.04 & 16.68 & 0.10 & 16.23 & 0.04 & 16.95 & 0.10 & $-$0.90 & 0.45 & $-$0.71\\
SMP78 & PN & - & 15.75 & 0.03 & 15.98 & 0.03 & 14.90 & 0.02 & 16.69 & 0.03 & $-$0.23 & 1.08 & $-$1.79\\
SMP82 & PN & - & 18.07 & 0.06 & 18.50 & 0.04 & 17.50 & 0.05 & 18.62 & 0.06 & $-$0.43 & 1.00 & $-$1.12\\
RP135 & Em? & - & 18.68 & 0.08 & 19.09 & 0.06 & 18.76 & 0.10 & 18.53 & 0.12 & $-$0.42 & 0.33 & 0.23\\
RP142 & FD?,NL & ND & - & - & - & - & - & - & - & - & - & - & -\\
RP143 & ND,NL & ND & - & - & - & - & - & - & - & - & - & - & -\\
RP162 & PN & ND & - & - & - & - & - & - & - & - & - & - & -\\
RP163 & FD,NL & ND & - & - & - & - & - & - & - & - & - & - & -\\
RP178 & NL & ND & - & - & - & - & - & - & - & - & - & - & -\\
RP180 & PN & - & 17.49 & 0.09 & 17.96 & 0.10 & 17.94 & 0.05 & 17.17 & 0.16 & $-$0.46 & 0.02 & 0.77\\
RP182 & ND,DHII & ND & - & - & - & - & - & - & - & - & - & - & -\\
RP187 & ND,DHII & ND & - & - & - & - & - & - & - & - & - & - & -\\
RP188 & ND,DHII & ND & - & - & - & - & - & - & - & - & - & - & -\\
RP198 & HII & - & 17.07 & 0.06 & 17.90 & 0.11 & 17.41 & 0.07 & 17.15 & 0.05 & $-$0.84 & 0.49 & 0.26\\
RP202 & FD,PN & ND & - & - & - & - & - & - & - & - & - & - & -\\
RP203 & ND,DHII & ND & - & - & - & - & - & - & - & - & - & - & -\\
RP218 & HII & - & 14.91 & 0.04 & 15.69 & 0.03 & 15.60 & 0.04 & 15.53 & 0.07 & $-$0.79 & 0.09 & 0.07\\
RP219 & FS & - & 17.52 & 0.09 & 18.04 & 0.04 & 18.02 & 0.25 & 17.67 & 0.05 & $-$0.52 & 0.02 & 0.35\\
RP223 & FS & - & 16.76 & 0.04 & 16.60 & 0.07 & 15.96 & 0.03 & 15.21 & 0.04 & 0.16 & 0.64 & 0.75\\
RP227 & LPV/Em? & - & - & - & 18.61 & 0.04 & 16.64 & 0.05 & 14.65 & 0.02 & - & 1.97 & 1.98\\
RP228 & FS & - & 19.09 & 0.11 & 19.08 & 0.05 & 18.37 & 0.06 & 17.55 & 0.05 & 0.02 & 0.71 & 0.82\\
RP231 & FS & - & 16.29 & 0.04 & 16.83 & 0.07 & 16.85 & 0.08 & 16.74 & 0.05 & $-$0.54 & $-$0.02 & 0.11\\
RP232 & ND,DHII & ND & - & - & - & - & - & - & - & - & - & - & -\\
RP234 & ND,DHII & ND & - & - & - & - & - & - & - & - & - & - & -\\
RP240 & FS & - & 17.72 & 0.05 & 17.99 & 0.06 & 17.71 & 0.03 & 17.61 & 0.06 & $-$0.28 & 0.28 & 0.10\\
RP241 & FS & - & 17.52 & 0.06 & 17.97 & 0.04 & 17.66 & 0.05 & 17.23 & 0.05 & $-$0.45 & 0.32 & 0.43\\
RP242 & LPV/HII & - & 15.98 & 0.04 & 16.75 & 0.02 & 16.76 & 0.03 & 16.55 & 0.04 & $-$0.77 & $-$0.01 & 0.21\\
RP246 & FS & - & 15.22 & 0.03 & 16.09 & 0.04 & 16.10 & 0.03 & 16.22 & 0.04 & $-$0.87 & $-$0.01 & $-$0.13\\
RP247 & HII & - & 17.04 & 0.05 & 17.82 & 0.08 & 17.29 & 0.04 & 16.77 & 0.04 & $-$0.78 & 0.53 & 0.52\\
RP250 & HII & ND & - & - & - & - & - & - & - & - & - & - & -\\
RP251 & HII & - & 16.43 & 0.04 & 17.21 & 0.05 & 16.95 & 0.05 & 16.90 & 0.05 & $-$0.78 & 0.26 & 0.05\\
RP254 & ND,DHII & - & 17.01 & 0.04 & 17.52 & 0.09 & 17.08 & 0.06 & 16.62 & 0.07 & $-$0.51 & 0.44 & 0.46\\
RP256 & HII & - & 15.85 & 0.04 & 16.39 & 0.03 & 16.44 & 0.03 & 16.06 & 0.09 & $-$0.54 & $-$0.05 & 0.39\\
\hline
\end{tabular}
\end{table*}
\begin{table*}
\centering
\caption{Optical mags (continued).}
\label{tab:opmags2}
\begin{tabular}{lllllllllllrrr}
\hline\hline
Name & ID & Status & $U$ & $U$e & $B$ & $B$e & $V$ & $V$e & $i$ & $i$e & $U-B$ & $B-V$ & $V-i$ \\
\hline
RP259 & FS & - & 16.61 & 0.04 & 17.21 & 0.05 & 17.17 & 0.05 & 17.04 & 0.05 & $-$0.60 & 0.04 & 0.13\\
RP264 & Sy? & - & 18.72 & 0.10 & 19.11 & 0.07 & 18.15 & 0.13 & 18.90 & 0.12 & $-$0.39 & 0.96 & $-$0.75\\
RP265 & FD,PN & ND & - & - & - & - & - & - & - & - & - & - & -\\
RP266 & HII & - & 16.02 & 0.14 & 17.20 & 0.07 & 17.13 & 0.11 & 16.53 & 0.07 & $-$1.18 & 0.07 & 0.60\\
RP268 & FS & - & 16.23 & 0.04 & 16.92 & 0.02 & 16.90 & 0.03 & 16.71 & 0.03 & $-$0.69 & 0.01 & 0.19\\
RP277 & Em & - & - & - & 21.86 & 0.14 & 20.16 & 0.07 & 16.66 & 0.03 & - & 1.71 & 3.50\\
RP283 & Em & - & 16.54 & 0.04 & 16.92 & 0.02 & 16.88 & 0.03 & 16.97 & 0.07 & $-$0.38 & 0.04 & $-$0.09\\
RP312 & FS & - & 14.92 & 0.06 & 15.71 & 0.02 & 15.61 & 0.03 & 15.38 & 0.03 & $-$0.78 & 0.09 & 0.23\\
RP315 & LPV & - & 15.22 & 0.03 & 15.88 & 0.02 & 15.80 & 0.09 & 15.13 & 0.03 & $-$0.66 & 0.08 & 0.67\\
RP641 & HII & - & 18.89 & 0.09 & 19.36 & 0.07 & 18.50 & 0.08 & 17.14 & 0.06 & $-$0.46 & 0.86 & 1.36\\
RP691 & NL & - & 21.05 & 0.40 & 18.37 & 0.07 & 16.65 & 0.06 & 15.15 & 0.04 & 2.68 & 1.72 & 1.50\\
RP698 & HII & ND & - & - & - & - & - & - & - & - & - & - & -\\
RP700 & ND,NL & ND & - & - & - & - & - & - & - & - & - & - & -\\
RP701 & FD,NL & ND & - & - & - & - & - & - & - & - & - & - & -\\
RP748 & NL & - & - & - & 22.36 & 0.26 & 21.67 & 0.22 & 21.44 & 0.44 & - & 0.70 & 0.23\\
RP774 & Sy? & - & 18.42 & 0.06 & 18.78 & 0.07 & 17.92 & 0.11 & 17.40 & 0.08 & $-$0.36 & 0.86 & 0.51\\
RP775 & HII & - & 15.82 & 0.05 & 16.18 & 0.07 & 16.20 & 0.07 & 16.19 & 0.05 & $-$0.36 & $-$0.01 & 0.01\\
RP776 & Sy? & - & 19.71 & 0.18 & 20.09 & 0.06 & 19.44 & 0.09 & 18.73 & 0.08 & $-$0.38 & 0.65 & 0.71\\
RP789 & PN & ND & - & - & - & - & - & - & - & - & - & - & -\\
RP790 & FS & - & 17.23 & 0.04 & 17.27 & 0.04 & 17.30 & 0.05 & 17.47 & 0.05 & $-$0.04 & $-$0.04 & $-$0.17\\
RP791 & Em & - & - & - & 20.42 & 0.09 & 18.75 & 0.07 & 16.48 & 0.05 & - & 1.68 & 2.27\\
RP793 & LPV/Mira & - & - & - & 22.78 & 0.45 & 21.65$^*$ & - & 17.64$^*$ & - & - & 1.12 & 4.02\\
RP828 & FS & - & 16.54 & 0.04 & 16.62 & 0.05 & 16.53 & 0.04 & 16.13 & 0.04 & $-$0.08 & 0.10 & 0.39\\
RP833 & YSO & ND & - & - & - & - & - & - & - & - & - & - & -\\
RP883 & LPV/Sy? & - & 18.43 & 0.09 & 18.75 & 0.03 & 18.23 & 0.19 & 16.89 & 0.03 & $-$0.33 & 0.52 & 1.34\\
RP896 & PN & - & 19.84 & 0.22 & 20.62 & 0.09 & 19.79 & 0.09 & - & - & $-$0.79 & 0.83 & -\\
RP907 & PN & - & 17.51 & 0.04 & 18.66 & 0.05 & 18.68 & 0.06 & 18.64 & 0.06 & $-$1.14 & $-$0.03 & 0.04\\
RP908 & Em & - & 17.36 & 0.04 & 17.60 & 0.03 & 17.64 & 0.04 & 17.79 & 0.04 & $-$0.23 & $-$0.04 & $-$0.15\\
RP913 & Em & - & - & - & 20.58 & 0.06 & 18.93 & 0.04 & 16.19 & 0.03 & - & 1.65 & 2.74\\
RP1018 & FS & - & 17.20 & 0.04 & 17.86 & 0.03 & 17.71 & 0.04 & 17.73 & 0.03 & $-$0.66 & 0.15 & $-$0.02\\
RP1037 & PN & ND & - & - & - & - & - & - & - & - & - & - & -\\
RP1040 & ND,NL & ND & - & - & - & - & - & - & - & - & - & - & -\\
RP1923 & FS & - & 15.42 & 0.13 & 16.10 & 0.03 & 16.05 & 0.04 & 16.15 & 0.04 & $-$0.68 & 0.04 & $-$0.10\\
RP1930 & FD?,NL & ND & - & - & - & - & - & - & - & - & - & - & -\\
RP1933 & HII & - & 14.05 & 0.14 & 16.04 & 0.13 & 14.68 & 0.12 & 13.94 & 0.12 & $-$1.99 & 1.36 & 0.74\\
RP1934 & PN & ND & - & - & - & - & - & - & - & - & - & - & -\\
RP1938 & PN & - & - & - & 21.28 & 0.18 & 20.58 & 0.13 & - & - & - & - & -\\
\hline
\end{tabular}
\tablefoot{
$^*$ Replaced with mean values from Soszy\'nski et al. (2009)
}
\end{table*}

\begin{table*}
\centering
\caption{VMC magnitudes.}
\label{tab:mags1}
\begin{tabular}{llrcllllllrr}
\hline\hline
Name & ID & Class & Aperture & $Y$ & $Y$e & $J$ & $J$e & $K_s$ & $K_{s}$e & $J-K_s$ & $Y-J$ \\
\hline
LM2-39 & Sy? & $-$1 & AperMag3 & 15.90 & 0.01 & 15.31 & 0.01 & 13.04 & 0.01 & 2.28 & 0.58 \\
MG12 & PN & 1 & NoAperCorr3 & 20.82 & 0.20 & 20.57 & 0.20 & 19.11 & 0.20 & 1.45 & 0.26 \\
MG13 & PN & 1 & NoAperCorr4 & 19.86 & 0.20 & 19.66 & 0.20 & 19.01 & 0.20 & 0.65 & 0.20 \\
MG16 & PN & 1 & NoAperCorr6 & 18.90 & 0.20 & 18.57 & 0.20 & 17.04 & 0.20 & 1.53 & 0.33 \\
MG17 & PN & $-$1 & AperMag3 & 19.95 & 0.05 & 19.82 & 0.07 & 18.97 & 0.09 & 0.85 & 0.13 \\
MG18 & PN & 1 & NoAperCorr4 & 20.53 & 0.20 & 20.53 & 0.20 & 19.89 & 0.20 & 0.64 & 0.01 \\
MG56 & PN & 1 & NoAperCorr3 & 19.21 & 0.20 & 19.07 & 0.20 & 18.83 & 0.20 & 0.24 & 0.15 \\
MG60 & PN & 1 & NoAperCorr4 & 19.63 & 0.20 & 19.26 & 0.20 & 17.74 & 0.20 & 1.53 & 0.36 \\
MG65 & PN & 1 & 2.5\arcsec & 19.29 & 0.20 & 18.58 & 0.20 & 17.73 & 0.20 & 0.85 & 0.71 \\
MG68 & PN & 1 & NoAperCorr4 & 19.27 & 0.20 & 18.59 & 0.20 & 17.58 & 0.20 & 1.01 & 0.68 \\
MG73 & PN & - & 2.1\arcsec & 20.28 & 0.20 & 19.97 & 0.20 & 18.91 & 0.20 & 1.06 & 0.31 \\
MG75 & ND,PN & - & - & - & - & - & - & - & - & - & - \\
MG76 & PN & $-$1 & AperMag3 & 18.29 & 0.03 & 18.12 & 0.04 & 17.67 & 0.04 & 0.45 & 0.17 \\
MG77 & PN & - & 3.3\arcsec & 20.68 & 0.20 & 20.53 & 0.20 & 18.51 & 0.20 & 2.02 & 0.15 \\
MNC1 & PN & - & 2.8\arcsec & 20.30 & 0.20 & 19.66 & 0.20 & 18.74 & 0.20 & 0.92 & 0.64 \\
MNC2 & PN & 1 & 2.4\arcsec & 20.55 & 0.20 & 20.11 & 0.20 & 18.75 & 0.20 & 1.36 & 0.44 \\
MNC3 & PN? & - & - & - & - & - & - & - & - & - & - \\
MNC4 & PN & $-$1 & AperMag3 & 16.68 & 0.01 & 16.31 & 0.01 & 14.15 & 0.01 & 2.16 & 0.36 \\
Mo30 & PN & 1 & NoAperCorr3 & 18.43 & 0.20 & 18.36 & 0.20 & 18.52 & 0.20 & $-$0.17 & 0.08 \\
Mo32 & PN & - & 2.3\arcsec & 19.82 & 0.20 & 19.41 & 0.20 & 18.14 & 0.20 & 1.27 & 0.41 \\
Mo34 & PN & - & 1.8\arcsec & 19.37 & 0.20 & 19.13 & 0.20 & 17.85 & 0.20 & 1.28 & 0.24 \\
Mo36 & PN & - & 1.5\arcsec & 20.26 & 0.20 & 20.04 & 0.20 & 19.19 & 0.20 & 0.85 & 0.22 \\
Mo37 & PN & 1 & 3.2\arcsec & 20.09 & 0.20 & 19.18 & 0.20 & 17.68 & 0.20 & 1.50 & 0.91 \\
Mo38 & PN & 1 & 3.3\arcsec & 19.96 & 0.20 & 19.47 & 0.20 & 17.60 & 0.20 & 1.87 & 0.49 \\
Mo39 & PN & 1 & NoAperCorr4 & 20.11 & 0.20 & 20.02 & 0.20 & 17.74 & 0.20 & 2.27 & 0.09 \\
Mo42 & PN & 1 & NoAperCorr6 & 19.23 & 0.20 & 19.13 & 0.20 & 17.27 & 0.20 & 1.85 & 0.11 \\
Sa122 & PN & - & 2.5\arcsec & 19.54 & 0.20 & 19.31 & 0.20 & 18.58 & 0.20 & 0.73 & 0.23 \\
SMP4 & PN & 1 & NoAperCorr6 & 18.83 & 0.20 & 18.56 & 0.20 & 17.06 & 0.20 & 1.51 & 0.26 \\
SMP6 & PN & $-$1 & AperMag3 & 17.00 & 0.01 & 16.39 & 0.01 & 14.94 & 0.01 & 1.45 & 0.61 \\
SMP27 & PN & $-$1 & AperMag3 & 18.22 & 0.02 & 17.99 & 0.02 & 17.42 & 0.03 & 0.57 & 0.23 \\
SMP30 & PN & 1 & NoAperCorr6 & 18.36 & 0.20 & 18.22 & 0.20 & 16.80 & 0.20 & 1.42 & 0.14 \\
SMP35 & PN & 1 & NoAperCorr6 & 17.36 & 0.20 & 17.02 & 0.20 & 16.32 & 0.20 & 0.71 & 0.34 \\
SMP77 & PN & $-$1 & AperMag3 & 16.83 & 0.01 & 16.27 & 0.01 & 15.01 & 0.01 & 1.25 & 0.56 \\
SMP78 & PN & $-$1 & AperMag3 & 16.22 & 0.01 & 15.90 & 0.01 & 14.85 & 0.01 & 1.05 & 0.32 \\
SMP82 & PN & $-$1 & AperMag3 & 17.81 & 0.03 & 17.75 & 0.04 & 16.70 & 0.03 & 1.05 & 0.05 \\
RP135 & Em? & 1 & AperMag3 & 18.60 & 0.04 & 18.23 & 0.04 & 18.03 & 0.06 & 0.20 & 0.36 \\
RP142 & FD?,NL & - & - & - & - & - & - & - & - & - & - \\
RP143 & ND,NL & - & - & - & - & - & - & - & - & - & - \\
RP162 & PN & 1 & NoAperCorr4 & 18.29 & 0.20 & 17.79 & 0.20 & 17.07 & 0.20 & 0.72 & 0.50 \\
RP163 & FD,NL & - & - & - & - & - & - & - & - & - & - \\
RP178 & NL & - & - & - & - & - & - & - & - & - & - \\
RP180 & PN & - & 1.4\arcsec & 19.03 & 0.20 & 18.64 & 0.20 & 18.38 & 0.20 & 0.26 & 0.39 \\
RP182 & ND,DHII & - & - & - & - & - & - & - & - & - & - \\
RP187 & ND,DHII & - & - & - & - & - & - & - & - & - & - \\
RP188 & ND,DHII & - & - & - & - & - & - & - & - & - & - \\
RP198 & HII & 1 & AperMag3 & 17.08 & 0.01 & 16.81 & 0.01 & 16.59 & 0.02 & 0.22 & 0.27 \\
RP202 & FD,PN & - & - & - & - & - & - & - & - & - & - \\
RP203 & ND,DHII & - & - & - & - & - & - & - & - & - & - \\
RP218 & HII & $-$1 & AperMag3 & 15.83 & 0.01 & 15.73 & 0.01 & 15.66 & 0.01 & 0.07 & 0.10 \\
RP219 & FS & $-$1 & AperMag3 & 17.68 & 0.02 & 17.49 & 0.02 & 17.21 & 0.03 & 0.28 & 0.19 \\
RP223 & FS & $-$1 & AperMag3 & 14.84 & 0.01 & 14.49 & 0.01 & 14.08 & 0.01 & 0.40 & 0.35 \\
RP227 & LPV/Em? & $-$1 & AperMag3 & 13.99 & 0.01 & 13.27 & 0.01 & 12.17 & 0.01 & 1.09 & 0.72 \\
RP228 & FS & $-$2 & AperMag3 & 17.53 & 0.02 & 17.22 & 0.02 & 16.76 & 0.02 & 0.46 & 0.31 \\
RP231 & FS & $-$1 & AperMag3 & 16.79 & 0.01 & 16.53 & 0.01 & 16.19 & 0.01 & 0.35 & 0.25 \\
RP232 & ND,DHII & - & - & - & - & - & - & - & - & - & - \\
RP234 & ND,DHII & - & - & - & - & - & - & - & - & - & - \\
RP240 & FS & $-$1 & AperMag3 & 17.16 & 0.01 & 16.91 & 0.01 & 16.35 & 0.01 & 0.56 & 0.26 \\
RP241 & FS & 1 & AperMag3 & 16.99 & 0.01 & 16.68 & 0.01 & 16.19 & 0.01 & 0.49 & 0.31 \\
RP242 & LPV/HII & $-$1 & AperMag3 & 16.52 & 0.01 & 16.24 & 0.01 & 15.81 & 0.01 & 0.42 & 0.28 \\
RP246 & FS & $-$1 & AperMag3 & 16.30 & 0.01 & 16.24 & 0.01 & 16.32 & 0.02 & $-$0.07 & 0.05 \\
RP247 & HII & $-$1 & AperMag3 & 16.69 & 0.01 & 16.41 & 0.01 & 16.12 & 0.01 & 0.28 & 0.28 \\
RP250 & HII & - & - & - & - & - & - & - & - & - & - \\
RP251 & HII & $-$1 & AperMag3 & 16.85 & 0.01 & 16.67 & 0.01 & 16.51 & 0.01 & 0.17 & 0.18 \\
RP254 & ND,DHII & $-$1 & AperMag3 & 16.38 & 0.01 & 16.12 & 0.01 & 16.45 & 0.01 & $-$0.33 & 0.26 \\
RP256 & HII & $-$1 & AperMag3 & 16.71 & 0.01 & 16.65 & 0.01 & 16.68 & 0.02 & $-$0.03 & 0.06 \\
\hline
\end{tabular}
\end{table*}
\begin{table*}
\centering
\caption{VMC magnitudes (continued).}
\label{tab:mags2}
\begin{tabular}{llrcllllllrr}
\hline\hline
Name & ID & Class & Aperture & $Y$ & $Y$e & $J$ & $J$e & $K_s$ & $K_{s}$e & $J-K_s$ & $Y-J$ \\
\hline
RP259 & FS & $-$1 & AperMag3 & 17.21 & 0.01 & 17.00 & 0.02 & 16.66 & 0.02 & 0.34 & 0.22 \\
RP264 & Sy? & 1 & AperMag3 & 17.32 & 0.01 & 16.30 & 0.01 & 12.88 & 0.01 & 3.42 & 1.03 \\
RP265 & FD,PN & - & - & - & - & - & - & - & - & - & - \\
RP266 & HII & 1 & AperMag3 & 16.57 & 0.01 & 16.17 & 0.01 & 15.59 & 0.01 & 0.58 & 0.40 \\
RP268 & FS & $-$1 & AperMag3 & 16.76 & 0.01 & 16.55 & 0.01 & 16.26 & 0.01 & 0.30 & 0.21 \\
RP277 & Em & $-$1 & AperMag3 & 15.61 & 0.01 & 14.91 & 0.01 & 14.12 & 0.01 & 0.79 & 0.70 \\
RP283 & Em & $-$1 & AperMag3 & 16.78 & 0.01 & 16.57 & 0.01 & 16.07 & 0.01 & 0.50 & 0.21 \\
RP312 & FS & $-$1 & AperMag3 & 15.36 & 0.01 & 15.11 & 0.01 & 14.72 & 0.01 & 0.39 & 0.25 \\
RP315 & LPV & $-$1 & AperMag3 & 16.09 & 0.01 & 15.97 & 0.01 & 15.82 & 0.01 & 0.14 & 0.12 \\
RP641 & HII & 1 & AperMag3 & 16.84 & 0.01 & 16.24 & 0.01 & 15.42 & 0.01 & 0.82 & 0.59 \\
RP691 & NL & $-$1 & AperMag3 & 14.50 & 0.01 & 13.86 & 0.01 & 12.86 & 0.01 & 1.00 & 0.65 \\
RP698 & HII & - & - & - & - & - & - & - & - & - & - \\
RP700 & ND,NL & - & - & - & - & - & - & - & - & - & - \\
RP701 & FD,NL & - & - & - & - & - & - & - & - & - & - \\
RP748 & NL & - & 2.6\arcsec & 20.06 & 0.20 & 19.55 & 0.20 & 18.73 & 0.20 & 0.82 & 0.51 \\
RP774 & Sy? & 1 & NoAperCorr4 & 17.55 & 0.20 & 16.75 & 0.20 & 14.52 & 0.20 & 2.23 & 0.79 \\
RP775 & HII & 1 & NoAperCorr6 & 15.83 & 0.20 & 15.34 & 0.20 & 14.46 & 0.20 & 0.88 & 0.48 \\
RP776 & Sy? & 1 & NoAperCorr6 & 17.92 & 0.20 & 17.09 & 0.20 & 14.80 & 0.20 & 2.29 & 0.83 \\
RP789 & PN & - & ell(1.68,3.64)\arcsec & 19.90 & 0.20 & 19.92 & 0.20 & 19.51 & 0.20 & 0.41 & $-$0.02 \\
RP790 & FS & $-$1 & AperMag3 & 17.23 & 0.01 & 17.07 & 0.02 & 16.72 & 0.02 & 0.35 & 0.15 \\
RP791 & Em & $-$1 & AperMag3 & 15.79 & 0.01 & 15.18 & 0.01 & 14.33 & 0.01 & 0.85 & 0.61 \\
RP793 & LPV/Mira & $-$1 & AperMag3 & 17.34 & 0.01 & 15.72 & 0.01 & 11.64 & 0.01 & 4.08 & 1.62 \\
RP828 & FS & $-$1 & AperMag3 & 15.74 & 0.01 & 15.38 & 0.01 & 14.81 & 0.01 & 0.58 & 0.36 \\
RP833 & YSO & 1 & 2.4\arcsec & 20.83 & 0.20 & 19.56 & 0.20 & 17.73 & 0.20 & 1.83 & 1.27 \\
RP883 & LPV/Sy? & $-$1 & AperMag3 & 16.66 & 0.01 & 15.81 & 0.01 & 13.01 & 0.01 & 2.80 & 0.85 \\
RP896 & PN & 1 & 3.8\arcsec & 19.79 & 0.20 & 19.34 & 0.20 & 17.98 & 0.20 & 1.36 & 0.45 \\
RP907 & PN & $-$1 & AperMag3 & 18.77 & 0.04 & 18.21 & 0.03 & 17.05 & 0.02 & 1.16 & 0.56 \\
RP908 & Em & $-$1 & AperMag3 & 17.66 & 0.01 & 17.41 & 0.02 & 16.87 & 0.02 & 0.54 & 0.24 \\
RP913 & Em & $-$1 & AperMag3 & 15.36 & 0.01 & 14.74 & 0.01 & 14.02 & 0.01 & 0.72 & 0.62 \\
RP1018 & FS & $-$1 & AperMag3 & 17.65 & 0.01 & 17.48 & 0.01 & 17.25 & 0.02 & 0.23 & 0.18 \\
RP1037 & PN & 1 & 2.7\arcsec & 20.47 & 0.20 & 20.28 & 0.20 & 18.50 & 0.20 & 1.78 & 0.19 \\
RP1040 & ND,NL & - & - & - & - & - & - & - & - & - & - \\
RP1923 & FS & $-$1 & AperMag3 & 16.13 & 0.01 & 16.16 & 0.01 & 16.11 & 0.01 & 0.05 & $-$0.02 \\
RP1930 & FD?,NL & - & - & - & - & - & - & - & - & - & - \\
RP1933 & HII & 1 & NoAperCorr6 & 15.01 & 0.20 & 14.50 & 0.20 & 13.41 & 0.20 & 1.09 & 0.51 \\
RP1934 & PN & 1 & NoAperCorr4 & 20.69 & 0.20 & 20.62 & 0.20 & 19.45 & 0.20 & 1.17 & 0.07 \\
RP1938 & PN & - & 2.5\arcsec & 21.11 & 0.20 & 21.09 & 0.20 & 20.21 & 0.20 & 0.88 & 0.02 \\
\hline
\end{tabular}
\end{table*}

\begin{table*}
\centering
\caption{SAGE MIR magnitudes.}
\label{tab:mir1}
\begin{tabular}{llllrrrrrrrrrrr}
\hline\hline
Name & ID & IRAC & MIPS & $J-K_s$ & [3.6] & [3.6]e & [4.5] & [4.5]e & [5.8] & [5.8]e & [8.0] & [8.0]e & [24] & [24]e \\
\hline
LM2-39 & Sy? & CAT & CAT & 2.28 & 10.11 & 0.05 & 8.76 & 0.03 &7.71 & 0.02 &6.39 & 0.03 & 2.75 & 0.01\\
MG12 & PN & AP & AP & 1.45 & 18.94 & 0.25 & 17.39 & 0.25 &- & - &16.02 & 0.35 & 11.77 & 0.40\\
MG13 & PN & AP & AP & 0.65 & 18.53 & 0.25 & 17.53 & 0.25 &- & - &15.09 & 0.35 & 10.79 & 0.40\\
MG16 & PN & AP & AP & 1.53 & 16.68 & 0.25 & 15.75 & 0.25 &15.40 & 0.30 &13.99 & 0.35 & 10.73 & 0.40\\
MG17 & PN & AP & AP & 0.85 & 17.17 & 0.25 & 16.55 & 0.25 &14.81 & 0.30 &13.55 & 0.35 & 8.36 & 0.40\\
MG18 & PN & AP & AP & 0.64 & 19.08 & 0.25 & - & - &- & - &- & - & 10.79 & 0.40\\
MG56 & PN & ND & ND & 0.24 & - & - & - & - &- & - &- & - & - & -\\
MG60 & PN & AP & AP & 1.53 & 16.53 & 0.25 & 15.87 & 0.25 &14.97 & 0.30 &13.42 & 0.35 & 10.45 & 0.40\\
MG65 & PN & AP & AP & 0.85 & 16.29 & 0.25 & 14.67 & 0.25 &15.61 & 0.30 &14.04 & 0.35 & 7.33 & 0.40\\
MG68 & PN & AP & AP & 1.01 & 16.35 & 0.25 & 15.69 & 0.25 &14.71 & 0.30 &13.13 & 0.35 & 8.09 & 0.40\\
MG73 & PN & ND & ND & 1.06 & - & - & - & - &- & - &- & - & - & -\\
MG75 & ND,PN & ND & ND & - & - & - & - & - &- & - &- & - & - & -\\
MG76 & PN & AP & AP & 0.45 & 16.76 & 0.25 & 16.01 & 0.25 &15.52 & 0.30 &13.84 & 0.35 & 8.36 & 0.40\\
MG77 & PN & AP & AP & 2.02 & 17.53 & 0.25 & 16.27 & 0.25 &15.10 & 0.30 &13.33 & 0.35 & 10.77 & 0.40\\
MNC1 & PN & AP & AP & 0.92 & 17.02 & 0.25 & 16.37 & 0.25 &14.85 & 0.30 &13.22 & 0.35 & 9.68 & 0.40\\
MNC2 & PN & AP & AP & 1.36 & 17.43 & 0.25 & 16.65 & 0.25 &15.55 & 0.30 &13.34 & 0.35 & 10.86 & 0.40\\
MNC3 & PN? & ND & ND & - & - & - & - & - &- & - &- & - & - & -\\
MNC4 & PN & AP & CAT & 2.16 & 11.94 & 0.25 & 11.15 & 0.25 &9.76 & 0.30 &7.68 & 0.35 & 3.85 & 0.01\\
Mo30 & PN & ND & AP & $-$0.17 & - & - & - & - &- & - &- & - & 10.24 & 0.40\\
Mo32 & PN & AP & AP & 1.27 & - & - & 16.51 & 0.25 &15.16 & 0.30 &14.45 & 0.35 & 11.30 & 0.40\\
Mo34 & PN & AP & AP & 1.28 & - & - & - & - &- & - &13.81 & 0.35 & 9.79 & 0.40\\
Mo36 & PN & ND & AP & 0.85 & - & - & - & - &- & - &- & - & 9.70 & 0.40\\
Mo37 & PN & AP & ND? & 1.50 & 16.68 & 0.25 & 15.62 & 0.25 &14.20 & 0.30 &12.97 & 0.35 & - & -\\
Mo38 & PN & AP & AP & 1.87 & 16.55 & 0.25 & 15.69 & 0.25 &14.40 & 0.30 &13.73 & 0.35 & 10.85 & 0.40\\
Mo39 & PN & AP & AP & 2.27 & 17.01 & 0.25 & 15.83 & 0.25 &15.30 & 0.30 &13.96 & 0.35 & 11.02 & 0.40\\
Mo42 & PN & AP & AP & 1.85 & 16.75 & 0.25 & 15.68 & 0.25 &14.61 & 0.30 &13.60 & 0.35 & 9.48 & 0.40\\
Sa122 & PN & AP & AP & 0.73 & 17.37 & 0.25 & 15.78 & 0.25 &16.17 & 0.30 &15.75 & 0.35 & 7.99 & 0.40\\
SMP4 & PN & NC & NC & 1.51 & - & - & - & - &- & - &- & - & - & -\\
SMP6 & PN & NC & NC & 1.45 & - & - & - & - &- & - &- & - & - & -\\
SMP27 & PN & AP & AP & 0.57 & 16.22 & 0.25 & 15.36 & 0.25 &14.60 & 0.30 &12.80 & 0.35 & 7.03 & 0.40\\
SMP30 & PN & AP & AP & 1.42 & 16.46 & 0.25 & 15.27 & 0.25 &14.46 & 0.30 &13.64 & 0.35 & 8.51 & 0.40\\
SMP35 & PN & AP & AP & 0.71 & 15.34 & 0.25 & 14.29 & 0.25 &13.75 & 0.30 &12.70 & 0.35 & 6.96 & 0.40\\
SMP77 & PN & AP & AP & 1.25 & 13.19 & 0.25 & 12.32 & 0.25 &11.57 & 0.30 &9.94 & 0.35 & 6.05 & 0.40\\
SMP78 & PN & AP & AP & 1.05 & 12.82 & 0.25 & 12.01 & 0.25 &10.39 & 0.30 &8.47 & 0.35 & 4.18 & 0.40\\
SMP82 & PN & AP & AP & 1.05 & 15.65 & 0.25 & 14.36 & 0.25 &13.44 & 0.30 &11.74 & 0.35 & 5.57 & 0.40\\
RP135 & Em? & ND & ND & 0.20 & - & - & - & - &- & - &- & - & - & -\\
RP142 & FD?,NL & ND & ND & - & - & - & - & - &- & - &- & - & - & -\\
RP143 & ND,NL & ND & ND & - & - & - & - & - &- & - &- & - & - & -\\
RP162 & PN & AP & ND & 0.72 & 16.42 & 0.25 & 15.99 & 0.25 &15.23 & 0.30 &14.44 & 0.35 & - & -\\
RP163 & FD,NL & ND & ND & - & - & - & - & - &- & - &- & - & - & -\\
RP178 & NL & AP & AP & - & - & - & 16.40 & 0.25 &16.27 & 0.30 &14.71 & 0.35 & 10.62 & 0.40\\
RP180 & PN & CAT & AP & 0.26 & 15.70 & 0.07 & 16.36 & 0.16 &13.94 & 0.06 &12.24 & 0.06 & 8.55 & 0.40\\
RP182 & ND,DHII & ND & ND & - & - & - & - & - &- & - &- & - & - & -\\
RP187 & ND,DHII & ND & ND & - & - & - & - & - &- & - &- & - & - & -\\
RP188 & ND,DHII & ND & ND & - & - & - & - & - &- & - &- & - & - & -\\
RP198 & HII & AP & AP & 0.22 & 15.35 & 0.25 & 14.49 & 0.25 &13.57 & 0.30 &11.85 & 0.35 & 6.53 & 0.40\\
RP202 & FD,PN & ND & ND & - & - & - & - & - &- & - &- & - & - & -\\
RP203 & ND,DHII & ND & ND & - & - & - & - & - &- & - &- & - & - & -\\
RP218 & HII & AP & AP & 0.07 & 15.61 & 0.25 & 15.24 & 0.25 &15.50 & 0.30 &13.90 & 0.35 & 6.67 & 0.40\\
RP219 & FS & ND & ND & 0.28 & - & - & - & - &- & - &- & - & - & -\\
RP223 & FS & CAT & ND & 0.40 & 13.92 & 0.04 & 13.88 & 0.05 &14.22 & 0.11 &- & - & - & -\\
RP227 & LPV/Em? & CAT & ND & 1.09 & 12.00 & 0.04 & 12.18 & 0.04 &12.03 & 0.04 &11.91 & 0.04 & - & -\\
RP228 & FS & CAT & ND & 0.46 & 16.04 & 0.07 & 15.83 & 0.05 &- & - &- & - & - & -\\
RP231 & FS & CAT & ND & 0.35 & 15.66 & 0.05 & 15.49 & 0.06 &- & - &- & - & - & -\\
RP232 & ND,DHII & ND & ND & - & - & - & - & - &- & - &- & - & - & -\\
RP234 & ND,DHII & ND & ND & - & - & - & - & - &- & - &- & - & - & -\\
RP240 & FS & AP & ND & 0.56 & 15.80 & 0.25 & 15.41 & 0.25 &15.26 & 0.30 &- & - & - & -\\
RP241 & FS & ND & ND & 0.49 & - & - & - & - &- & - &- & - & - & -\\
RP242 & LPV/HII & AP & AP & 0.42 & 15.15 & 0.25 & 14.66 & 0.25 &14.39 & 0.30 &13.92 & 0.35 & 8.23 & 0.40\\
RP246 & FS & ND & ND & $-$0.07 & - & - & - & - &- & - &- & - & - & -\\
RP247 & HII & CAT & AP & 0.28 & 16.00 & 0.09 & 15.27 & 0.11 &- & - &- & - & 7.70 & 0.40\\
RP250 & HII & ND & ND & - & - & - & - & - &- & - &- & - & - & -\\
RP251 & HII & AP & AP & 0.17 & 16.06 & 0.25 & 15.37 & 0.25 &- & - &- & - & 7.39 & 0.40\\
RP254 & ND,DHII & EX & AP & $-$0.33 & - & - & - & - &- & - &- & - & 6.06 & 0.40\\
RP256 & HII & AP & AP & $-$0.03 & 17.09 & 0.25 & 16.76 & 0.25 &- & - &- & - & 7.58 & 0.40\\
\hline
\end{tabular}
\end{table*}
\begin{table*}
\centering
\caption{SAGE MIR magnitudes (continued).}
\label{tab:mir2}
\begin{tabular}{llllrrrrrrrrrrr}
\hline\hline
Name & ID & IRAC & MIPS & $J-K_s$ & [3.6] & [3.6]e & [4.5] & [4.5]e & [5.8] & [5.8]e & [8.0] & [8.0]e & [24] & [24]e \\
\hline
RP259 & FS & CAT & ND & 0.34 & 16.10 & 0.05 & 15.91 & 0.06 &- & - &- & - & - & -\\
RP264 & Sy? & CAT & CAT & 3.42 & 9.91 & 0.03 & 8.80 & 0.03 &7.84 & 0.02 &6.45 & 0.02 & 2.56 & 0.01\\
RP265 & FD,PN & ND & ND & - & - & - & - & - &- & - &- & - & - & -\\
RP266 & HII & EX & ND & 0.58 & - & - & - & - &- & - &- & - & - & -\\
RP268 & FS & AP & ND & 0.30 & 16.00 & 0.25 & 15.73 & 0.25 &- & - &- & - & - & -\\
RP277 & Em & AP & ND & 0.79 & 13.80 & 0.25 & 13.66 & 0.25 &13.99 & 0.30 &14.05 & 0.35 & - & -\\
RP283 & Em & AP & ND & 0.50 & 15.58 & 0.25 & 15.31 & 0.25 &15.65 & 0.30 &15.12 & 0.35 & - & -\\
RP312 & FS & CAT & ND & 0.39 & 14.32 & 0.03 & 14.05 & 0.05 &14.01 & 0.06 &13.55 & 0.10 & - & -\\
RP315 & LPV & AP & ND & 0.14 & 14.66 & 0.25 & 14.34 & 0.25 &14.55 & 0.30 &14.21 & 0.35 & - & -\\
RP641 & HII & EX & EXTD & 0.82 & - & - & - & - &- & - &- & - & - & -\\
RP691 & NL & CAT & ND & 1.00 & 12.72 & 0.03 & 12.84 & 0.03 &12.75 & 0.05 &12.64 & 0.09 & - & -\\
RP698 & HII & EX & EXTD & - & - & - & - & - &- & - &- & - & - & -\\
RP700 & ND,NL & ND & ND & - & - & - & - & - &- & - &- & - & - & -\\
RP701 & FD,NL & ND & ND & - & - & - & - & - &- & - &- & - & - & -\\
RP748 & NL & ND & ND & 0.82 & - & - & - & - &- & - &- & - & - & -\\
RP774 & Sy? & CAT & AP & 2.23 & 12.18 & 0.05 & 11.33 & 0.03 &10.31 & 0.05 &8.91 & 0.05 & 5.36 & 0.40\\
RP775 & HII & EX & EXTD & 0.88 & - & - & - & - &- & - &- & - & - & -\\
RP776 & Sy? & CAT & AP & 2.29 & 12.42 & 0.05 & 11.59 & 0.03 &10.63 & 0.04 &9.16 & 0.04 & 6.59 & 0.40\\
RP789 & PN & AP & AP & 0.41 & 17.17 & 0.25 & 16.75 & 0.25 &15.24 & 0.30 &13.36 & 0.35 & 10.26 & 0.40\\
RP790 & FS & AP & ND & 0.35 & 16.73 & 0.25 & 16.21 & 0.25 &16.46 & 0.30 &15.49 & 0.35 & - & -\\
RP791 & Em & CAT & ND & 0.85 & 14.06 & 0.03 & 14.09 & 0.03 &14.22 & 0.06 &13.96 & 0.08 & - & -\\
RP793 & LPV/Mira & CAT & AP & 4.08 & 9.01 & 0.04 & 8.28 & 0.03 &7.79 & 0.02 &7.26 & 0.02 & 6.88 & 0.40\\
RP828 & FS & CAT & ND & 0.58 & 14.29 & 0.03 & 14.10 & 0.04 &13.93 & 0.07 &13.56 & 0.06 & - & -\\
RP833 & YSO & AP & ND & 1.83 & 16.38 & 0.25 & 15.56 & 0.25 &- & - &- & - & - & -\\
RP883 & LPV/Sy? & CAT & AP & 2.80 & 10.18 & 0.02 & 9.21 & 0.01 &8.44 & 0.02 &7.44 & 0.01 & 5.23 & 0.40\\
RP896 & PN & AP & AP & 1.36 & 17.19 & 0.25 & 16.45 & 0.25 &15.55 & 0.30 &13.43 & 0.35 & 11.28 & 0.40\\
RP907 & PN & CAT & AP & 1.16 & 15.11 & 0.04 & 14.48 & 0.04 &13.13 & 0.04 &11.25 & 0.03 & 7.27 & 0.40\\
RP908 & Em & AP & ND & 0.54 & 16.67 & 0.25 & 16.36 & 0.25 &16.25 & 0.30 &15.69 & 0.35 & - & -\\
RP913 & Em & AP & ND & 0.72 & 13.75 & 0.25 & 13.61 & 0.25 &13.62 & 0.30 &14.48 & 0.35 & - & -\\
RP1018 & FS & CAT & ND & 0.23 & 16.69 & 0.05 & 16.59 & 0.07 &- & - &- & - & - & -\\
RP1037 & PN & AP & ND & 1.78 & 17.42 & 0.25 & 16.49 & 0.25 &16.11 & 0.30 &14.93 & 0.35 & - & -\\
RP1040 & ND,NL & ND & ND & - & - & - & - & - &- & - &- & - & - & -\\
RP1923 & FS & CAT & EXTD & 0.05 & 16.11 & 0.11 & 15.90 & 0.14 &- & - &- & - & - & -\\
RP1930 & FD?,NL & ND & ND & - & - & - & - & - &- & - &- & - & - & -\\
RP1933 & HII & EX & CAT & 1.09 & - & - & - & - &- & - &- & - & 1.64 & 0.01\\
RP1934 & PN & AP & ND & 1.17 & 19.00 & 0.25 & - & - &- & - &- & - & - & -\\
RP1938 & PN & AP & AP & 0.88 & 18.73 & 0.25 & 17.44 & 0.25 &- & - &- & - & 9.77 & 0.40\\
\hline
\end{tabular}
\end{table*}

\clearpage

\section{Images}
\label{sec:images}
We present colour-composite images for our sample in two sets of figures. 
Figures \ref{fig:multi1}--\ref{fig:multi10} contain the VMC images made from stacked $K_s$ (red), $J$ (green) and $Y$ (blue) images, WFI images made from either H$\alpha$ (red), [O~III] (green) and $B$ (blue) images or for objects in Tab. \ref{tab:wfiextra} H$\alpha$ (red), MB 485/31 or MB 604/21 (green) and [O~III] (blue), and \emph{Spitzer} SAGE images made from [5.8] (red), [4.5] (green) and [3.6] (blue) images.
Figures \ref{fig:dbl1}--\ref{fig:dbl4} contain VMC and SAGE images for objects lacking ESO WFI coverage.

In each set of figures the objects are ordered following Tab. \ref{tab:basic1}. VMC images include circles with the RP2006b measured H$\alpha$ radius.
Depending on the ratio of H$\alpha$ to [O~III] emission, PNe may appear red as in RP265 (no [O~III] indicative of a cool CSPN), yellow as in MG60 (H$\alpha$/[O~III]$\sim$1) or green as in SMP78 ([O~III]$>$H$\alpha$). Exceptions include MG76, where [O~III] was replaced by $V$ due to missing coverage, and objects in Tab. \ref{tab:wfiextra}, in which case PNe may appear red (H$\alpha$ only) or pink ([O~III] and H$\alpha$).

\begin{figure*}
\begin{center}
\end{center}
\caption{(left column) VMC $K_s$ (red), $J$ (green) and $Y$ (blue) colour-composite; (middle column) \emph{Spitzer} SAGE [5.8] (red), [4.5] (green) and [3.6] (blue); (right column) Optical H$\alpha$ (red), [O~III] (green) and $B$ (blue) excluding MG76 and objects in Tab. \ref{tab:wfiextra} which are H$\alpha$ (red), continuum (green) and [O~III] (blue). Each image is $30\times$30 arcsec$^2$ with North up and East to left.}
\label{fig:multi1}
\end{figure*}

\begin{figure*}
\begin{center}
\end{center}
\caption{Figure \ref{fig:multi1} (continued).}
\label{fig:multi2}
\end{figure*}

\begin{figure*}
\begin{center}
\end{center}
\caption{Figure \ref{fig:multi2} (continued).}
\label{fig:multi3}
\end{figure*}

\begin{figure*}
\begin{center}
\end{center}
\caption{Figure \ref{fig:multi3} (continued).}
\label{fig:multi4}
\end{figure*}

\begin{figure*}
\begin{center}
\end{center}
\caption{Figure \ref{fig:multi4} (continued).}
\label{fig:multi5}
\end{figure*}

\begin{figure*}
\begin{center}
\end{center}
\caption{Figure \ref{fig:multi5} (continued).}
\label{fig:multi6}
\end{figure*}

\begin{figure*}
\begin{center}
\end{center}
\caption{Figure \ref{fig:multi6} (continued).}
\label{fig:multi7}
\end{figure*}

\begin{figure*}
\begin{center}
\end{center}
\caption{Figure \ref{fig:multi7} (continued).}
\label{fig:multi8}
\end{figure*}

\begin{figure*}
\begin{center}
\end{center}
\caption{Figure \ref{fig:multi8} (continued).}
\label{fig:multi9}
\end{figure*}

\begin{figure*}
\begin{center}
\end{center}
\caption{Figure \ref{fig:multi9} (continued).}
\label{fig:multi10}
\end{figure*}

\clearpage

\begin{figure*}
\begin{center}
\end{center}
\caption{Similar to Fig. \ref{fig:multi1} but for objects without WFI coverage in two sets of two columns: (left columns) VMC colour-composite; (right columns) SAGE colour-composite. SMP 4 and SMP 6 have no SAGE coverage.}
\label{fig:dbl1}
\end{figure*}

\begin{figure*}
\begin{center}
\end{center}
\caption{Figure \ref{fig:dbl1} (continued).}
\label{fig:dbl2}
\end{figure*}

\begin{figure*}
\begin{center}
\end{center}
\caption{Figure \ref{fig:dbl2} (continued).}
\label{fig:dbl3}
\end{figure*}

\begin{figure*}
\begin{center}
\end{center}
\caption{Figure \ref{fig:dbl3} (continued).}
\label{fig:dbl4}
\end{figure*}

\end{appendix}
\end{document}